# Community landscapes: an integrative approach to determine overlapping network module hierarchy, identify key nodes and predict network dynamics

**István A. Kovács[1,2], Robin Palotai[1], Máté S. Szalay[1], Peter Csermely[1,*]**

[1]Department of Medical Chemistry, Semmelweis University, Budapest, Hungary
[2]Department of Physics, Loránd Eötvös University, Budapest, Hungary
*Corresponding author; Department of Medical Chemistry, Semmelweis University, P.O.Box 260, H-1444 Budapest 8, Hungary; Tel: +36-1-459-1500/60130; E-mail: csermely@eok.sote.hu

**Background:** Network communities help the functional organization and evolution of complex networks. However, the development of a method, which is both fast and accurate, provides modular overlaps and partitions of a heterogeneous network, has proven to be rather difficult.
**Methodology/Principal Findings:** Here we introduce the novel concept of ModuLand, an integrative method family determining overlapping network modules as hills of an influence function-based, centrality-type community landscape, and including several widely used modularization methods as special cases. As various adaptations of the method family, we developed several algorithms, which provide an efficient analysis of weighted and directed networks, and (1) determine pervasively overlapping modules with high resolution; (2) uncover a detailed hierarchical network structure allowing an efficient, zoom-in analysis of large networks; (3) allow the determination of key network nodes and (4) help to predict network dynamics.
**Conclusions/Significance:** The concept opens a wide range of possibilities to develop new approaches and applications including network routing, classification, comparison and prediction.

## Introduction

In real networks, module or community structure plays a central role in the understanding of topology and dynamics. Numerous module determination methods are based on the intuitive picture identifying the network communities as dense groups of the network, whose nodes have a much stronger influence on each other than on the rest of the network. The development of a method, which translates this intuitive definition of modules into a practically applicable, fast, accurate and widely usable algorithm turned out to be a very challenging problem. So far a wide variety of great ideas and powerful approaches based on very different physical or algorithmic grounds were applied in order to solve this problem. At the moment there is no 'best method' available to find network modules, and even the widely used algorithms may suffer from serious problems (see Figure S1.1, and Tables S1.1 and S1.2 in the Electronic Supplementary Material S1) [1-7], although they usually provide useful and clear dissections of networks.

In 2002 Girvan and Newman published a seminal paper [2] using an algorithm for detecting communities by iteratively removing edges of high betweenness centrality values from the network, and defining communities as the connected components of the network after these edge removals. Later they [8] introduced the modularity measure, Q with which the optimal number of removed edges could be determined. In a short time the Q function evaluating the goodness of partitioning a graph into given clusters became an essential element of a wide range of clustering methods. Different kind of approaches have also been developed, including ones utilizing spectral functions of the graphs [9,10], dynamic algorithms like random walks [5,11], spin models (e.g. the Potts model) [12] or synchronization models [13]. The most popular method to find overlapping communities is the Clique Percolation Method described by Palla et al. in [7], but other excellent methods optimizing overlapping quality functions such as that of Nepusz et al. [4] or the link-based method resulting in pervasive overlaps [14] also exist. Although the field of community detection is quite diversified, we tried to collect the main algorithms in Table S1.2 in the Electronic Supplementary Material S1, and we also recommend a current extensive review of the field by Santo Fortunato [1].

In this paper we introduce an integrative network module determination method family, called ModuLand (see Box 1 for the glossary of novel terms). This module determination method family is based on the novel concept of understanding the overlapping modules as hills of an influence function-based, centrality-type community landscape. The ModuLand method family gives a common framework for the development and comparison of a large variety of modularization methods resulting in network modules with variable overlaps, requiring different computational speed and providing a different level of accuracy.

# Results

## Description of the ModuLand method family

Keeping in mind the emerging needs for an integrative approach for the determination of network modules, we have developed the ModuLand method family (Figure 1 and Figure S1.2 in the Electronic Supplementary Material S1). All members of the ModuLand method family are based on the following common steps:

1) Determination of influence functions: If a node lies in a module, than its influence on the links of the given module is typically larger than on more distant links of the network. As a first step, we determine the influence function, $f_i$ of each node $i$ of the network on the links of the entire network. After perturbation-flow type calculations detailed below starting from each node $i$ of the network, we acquire a set of $f_i(j,k) \geq 0$ values over all links *(j,k)* of the network for node *i*.
2) Construction of a community landscape: The influence functions of different nodes in the same module are generally different. Nevertheless, the module is the set of nodes, which mutually have a large influence on each other. In order to take this mutuality into account, we summarize the influence function values of Step 1 over each link of the network: $c(j,k) = \sum_i f_i(j,k)$. The resulting *c(j,k)* values represent a smooth, centrality-type quantity, which is larger for the module cores and smaller for the surrounding regions. We represent these non-negative *c* values as a vertical measure forming a community landscape over the links of the 2D visualization of the original network (Figure S1.3 in the Electronic Supplementary Material S1).

3) Determination of hills of the community landscape: Modules are determined as the 'mountains and hills' of the community landscape of Step 2. We present two different approaches:
    a. Modules are the connected components above a chosen centrality-threshold.
    b. Modules are determined by local maxima of the community landscape and their surrounding region (Figures S1.4 and S1.5 in the Electronic Supplementary Material S1).
4) Determination of a hierarchy of higher level networks: We note that a higher level network of the modules of Step 3 can also be constructed, where each former module is a node of this higher level network. If the higher level networks are re-assessed with the ModuLand method again and again, a set of hierarchical layers of modules can be defined until the giant component of the whole network coalesces to a single node (Figure S1.6 in the Electronic Supplementary Material S1).

In the followings we will describe the four major steps of the ModuLand method in detail.

**Step 1: Determination of influence functions.** In principle, the determination of the influence functions (or indirect impact of a node or link) requires a network-dependent perturbation-flow simulation on the network (as an example, see our PerturLand algorithm in Section IV.2. in the Electronic Supplementary Material S1), which is a challenging problem in itself. However, the details of the influence functions usually average out during the community landscape construction, which justifies the use of less specific, faster approximations. Here we present our simplest influence function calculation algorithm, the NodeLand algorithm in detail, which can be applied on weighted, undirected networks.

NodeLand algorithm: starting from a given node *s*, the NodeLand algorithm iteratively determines the set of nodes *A*, which is strongly influenced by node *s*. For any given *A* set, we define the density of the set as

$$d = \frac{\sum_{(i,j)\in A} w_{i,j}}{|A|},$$

where $w_{ij}$ is the weight of the link between node *i* and *j*, and $|A|$ is the number of nodes in *A*. Initially, *A* consists of the staring node only, thus $A = \{s\}$. In each iterative step we will expand *A*. For each neighboring node $k \notin A$ we determine the potential new density value, including node *k* in *A*:

$$d' = \frac{\sum_{(i,j)\in A\cup\{k\}} w_{i,j}}{|A|+1}.$$

If the density of *A* can be increased this way, then we add the nodes with maximal $d'$ value to the set *A*, and start a new step of the iteration. We stop the process, when the density can not be further increased with the addition of single neighboring nodes. At this point, we will have a final set $A$ containing the nodes strongly influenced by the starting node, $s$ (including $s$ itself). The influence function $f_s$ over the links is defined as $f_s(i, j) = w_{i,j}$ , if $(i, j) \in A$ and zero otherwise.

LinkLand algorithm: the LinkLand algorithm, used in our module determinations of the main text below, differs from the NodeLand algorithm in two points.

- In the LinkLand algorithm the influence functions are assigned to starting links instead of starting nodes, thus initially $A=\{k,l\}$ containing the two end-nodes of the starting link $(k,l)$.
- In contrast to the NodeLand algorithm, while calculating the influence function the weight of the starting link $(k,l)$ is also taken into account: $f_s(i, j) = w_{i,j} w_{k,l}$, if $(i, j) \in A$, and zero otherwise.

On undirected networks we prefer to use the LinkLand algorithm, which is found to provide an acceptable compromise between precision and speed. Identification of the influence function of a node or link in the case of NodeLand and LinkLand algorithms is structurally similar to a breadth-first search, therefore the worst-case runtime complexity of the two algorithms for all nodes or links is $O(n(n+e))$ and $O(e(n+e))$, respectively, where $n$ is the number of nodes and $e$ is the number of links in the network. However, in practice these algorithms are very fast as the influence zone of any given starting node or link rarely covers the whole network. For downloading the ModuLand program package including the NodeLand, LinkLand influence function calculation algorithms and their User Guide, see our homepage: http://www.linkgroup.hu/modules.php.

As an example for the results obtained by the LinkLand algorithm, Figure 1A shows three influence functions defined over the links of the 'network science' co-authorship network [15]. All the three starting links highlighted by the arrows belong to widely collaborating, key players of the field, resulting extended influence zones.

**Step 2: Construction of a community landscape.** In order to find the regions with nodes mutually having a relatively large influence on each other, we calculate the sum of the individual influence functions on a given network link resulting in the $c(j,k) = \sum_i f_i(j,k) \geq 0$ centrality-type value. From the centrality of the links the centrality of the nodes can be derived by a summation: $c(i) = \sum_j c(i, j)$. When mentioning 'centrality', throughout the paper we refer to these definitions. The centrality values can be plotted on a vertical scale resulting in a community landscape over a 2D representation of the links of the network (see Figure 1B and Figure S1.3 in the Electronic Supplementary Material S1). Now we can see the 'hills and mountains' of the community landscape, consisting of the nodes influencing each other stronger than the rest of the network. This is exactly the intuitive definition of modules given in the first section of this paper. The precise definition of these hills will be the subject of the following section.

**Step 3: Determination of hills of the community landscape.** Here we present two main approaches of hill-determination suitable for the determination of modules.

Centrality threshold based hills: as a natural choice, hills may be identified as the connected components of the community landscape above a given threshold. This approach results in distinct network modules without overlaps, like in case of the widely used Girvan and Newman method [2]. Generally, it is a rather difficult problem to choose the most appropriate value for the threshold (see Figure S1.4 and Table S1.2 in the Electronic Supplementary Material S1). On one hand, if we raise the detection limit too high, we will find only the largest communities. On the other hand, if we set the detection limit too low in order to be able to see the smaller modules, then most of the large communities would merge together. This is the manifestation of the well known giant-component problem [6,16-18]. As the centrality threshold-based approach is very general, most of the former methods yielding non-overlapping modules can be interpreted in the ModuLand method family as the application of the threshold-based hill determination method over an appropriate community landscape.

Local maxima based hills: in this method we start the identification of the modules by finding the module centers, which are identical with the hill-tops or local maxima of the community landscape, defined as follows:

- Undirected networks: A hill-top of the community landscape contains all connected links having the same, locally maximal centrality value, while having all of their neighboring links with lower centrality values.
- Directed networks: The definition of hill-tops is more complicated in directed networks, but we also show it here for clarity. Let the outbound links of a link *(i,j)* denote the outbound links of its end-point, node *j*. Then a hill-top is either a single link with all of its outbound links having a lower centrality, or a strongly connected component (meaning that every node of the component is reachable from every other node by a directed walk on this component) consisting of all links of the same centrality with all of their outbound links having lower centralities.

By this definition the number of local maxima automatically yields the number of modules, and at the same time all small and large modules are identified simultaneously. This is in strong contrast to the previously described threshold-based approach, which often needs special criteria to determine the threshold value.

At this stage only the central links or plateaus of the modules have been identified. In the next step, the modules will be extended towards lower regions of the community landscape. We have developed several methods for this extension process detailed in Section V.2. in the Electronic Supplementary Material S1. We suggest the use of the ProportionalHill method described below, determining the module-membership values of the links proportionally to the membership values of their neighbors located higher in the community landscape. Using this method the hills will naturally overlap, resulting in links which are assigned to multiple modules simultaneously.

ProportionalHill method for the determination of network modules: here we present the algorithm of the ProportionalHill method for undirected networks, while the analogous directed version can be found in Section V.2.b. in the Electronic Supplementary Material S1. As the first step, the community landscape is divided into hills (corresponding to network modules) defined by the local maxima of the community landscape described above. For a given link (*i,j*), $H_m(i,j)$ is the hill (or module) membership value of the link in the *m*-th hill. The module membership values are normalized to the centrality of the given link, $\sum_m H_m(i,j) = c(i,j)$. If link *(i,j)* is part of the *module center* of module *m*, then let $H_k(i,j) = c_{i,j}$, if $k = m$, and $H_k(i,j) = 0$ otherwise. For the other links we apply the following rule:

$$H_k(i,j) = c_{i,j} \frac{\sum_{(a,b)\in N} H_k(a,b)}{\sum_{(a,b)\in N} c_{a,b}},$$

where *N* is the set of neighboring links of link *(i,j)* having larger or equal centrality than that of link *(i,j)*. Now the community landscape is divided into multiple hills identified as the modules of the network. Thus, $H_m(i,j)$ readily gives the module membership value of link *(i,j)* in module *m*. Finally, the module membership values of a given node *i* is given by $H_k(i) = \sum_{j\in S} H_k(i,j)$, where *S* is the set of the neighboring nodes of node *i* and *k* is the module considered. The presented ProportionalHill method has a runtime complexity of *O(edM)*, with *e* being the number of links, *d* being the average node degree and *M* being the number of the identified modules.

If we need smaller or larger overlaps between the modules, than those obtained with the ProportionalHill method, we may use the GradientHill or TotalHill methods, respectively, as described in Section V.2. in the Electronic Supplementary Material S1 and Figure S1.5 in the Electronic Supplementary Material S1. While for practical purposes we suggest the use the ProportionalHill method, the most detailed module overlap information is acquirable with the computationally more expensive TotalHill method. The TotalHill method also takes into account the neighboring links of lower centrality during the module-extension step. The TotalHill approach requires the solution of *M* appropriate linear equation systems of size *n* by *n*, with *n* being the number of nodes. Results obtained using the TotalHill method can be seen on Figure 1C and Figure S1.3 in the Electronic Supplementary Material S1, where large segments of the network belong to at least two modules. For downloading the ModuLand program package including the ProportionalHill and TotalHill method algorithms and their User Guide, see our homepage: http://www.linkgroup.hu/modules.php.

**Step 4: Determination of a hierarchy of higher level networks.** Optionally, a higher level hierarchical representation of the network can also be created, where the nodes of the higher level correspond to the modules of the original network, and the links of the higher level correspond to the overlaps between the respective modules (Figure 1D, Figures S1.2 and S1.6 in the Electronic Supplementary Material S1).

In the description of the calculation of the higher hierarchical level let us consider here the undirected case only (the directed case is described in Section VII. in the Electronic Supplementary Material S1). Let the strength of the overlap (equaling with the weight of the link at the one level higher hierarchy) between modules $i$ and $j$ be the sum of the node-wise calculated overlap values $O_{ij}(n)$:

$$W_{(i,j)\in L'} = \sum_n O_{ij}(n),$$

where $O_{ij}(n)$ is proportional to the module membership values $H_i(n)$ and $H_j(n)$ and being normalized to the centrality as:

$$O_{ij}(n) = 2\frac{H_i(n)H_j(n)}{c(n)},$$

where $c(n)$ is the centrality of node $n$, and the factor of $2$ refers to that both directions between the modules have been taken into account.

The steps leading to a higher level hierarchical representation can be applied repetitively until the giant component of the whole original network is represented by a single node allowing a fast, zoom-in type analysis of large networks (Section VII. in the Electronic Supplementary Material S1).

A simple case illustrating this scenario can be seen on Figure S1.6 in the Electronic Supplementary Material S1 showing the hierarchical levels of the network science collaboration network [15]. It can be seen that the modules of higher and higher hierarchical levels correspond to larger and larger groups (e.g. the modules of the modules etc.) of the original network nodes.

## Characterization of the overlapping modules identified by the ModuLand method family

The ModuLand method family, even with its simplest NodeLand influence function calculation method correctly identified the observed split of the gold-standard Zachary karate club network [19], while uncovering a third, previously identified module and several club-members in modular overlaps (Figure S1.7 in the Electronic Supplementary Material S1).

Application of the LinkLand influence function calculation method to the University of South Florida word association network [20] resulted in a set of modules having a highly heterogeneous degree, module size and module overlap distribution (Figure S1.8 in the Electronic Supplementary Material S1), which is in agreement with earlier data (see Supplementary Discussion in the Electronic Supplementary Material S1) [3,7].

The application of the ModuLand method on the benchmark graphs of Lancichinetti et al. [21] generated over a range of parameter settings showed (Figure S1.13. and Section VI.2. in the Electronic Supplementary Material S1) that the identified ModuLand modules corresponded consistently to the original modules, while modules can be defined in the strong sense (where ‘strong sense’ means, at least the half of the neighboring nodes are assigned to the same module as the given node, see ref. [21] and Table S1.3 in the Electronic Supplementary Material S1).

To obtain a more detailed picture we directly compared the method-pair of NodeLand, or LinkLand influence function calculation algorithm and the ProportionalHill hillfinder method with the InfoMap method [5] Louvain method [22] and the CFinder method using k=4 cliques [7]. Figure 2A shows the accuracy of the 4 methods in terms of generalized normalized mutual information [23] on the non-overlapping benchmark graphs of Lancichinetti et al. [21]. Our method compares well to the other 3 methods, however, at less defined modules it is not as accurate as the InfoMap method. It is worth to note, that the benchmark graphs used [21] have been developed for the comparison of not overlapping modularization methods, which may – in part – explain why the two overlapping methods did not perform extremely well on them.

Benchmark graphs have been criticized recently due to their limited capacity to reflect the complexity of real-world networks [24]. Therefore, we decided to compare the accuracy of the 4 methods mentioned before on a high-fidelity yeast protein-protein interaction network [25]. The analysis of the modular distribution of a central regulator of yeast cells, the cAMP-dependent kinase family [26] and their neighbors is shown on Figure 2B. While the NodeLand method identified 10 highly overlapping modules (and functions) of the 3 catalytic subunits and their regulatory subunit, the InfoMap, Louvain and CFinder methods all assigned these family members to a single module. These single modules were signaling modules in case of the InfoMap and CFinder methods, while the Louvain method assigned the whole cAMP-dependent kinase family to a vesicular traffic module, which reflects only a small part of their biological regulatory function (Fig 2B and Table S1.4 in the Electronic Supplementary Material S1). The modular assignment of the 16 neighbors of the cAMP-dependent kinase family enriched the modular composition by 4, 11, 8, and 6 modules in case of the NodeLand, InfoMap, Louvain and CFinder methods, respectively. In contrast to the other 3 methods, which assigned all 16 neighbors to various modules, the CFinder found modules for 5 neighbors only (Fig 2B and Table S1.4 in the Electronic Supplementary Material S1). The overlap of the neighbor-enriched cAMP kinase-related modules had similar values between all methods tested (25% to 75% and 33% to 50% agreement between the NodeLand and the other 3 methods vs. in between the other three methods, respectively; Table S1.5 in the Electronic Supplementary Material S1). Similarly, the number of modules, whose functional association with the cAMP kinase family was supported by experimental data, was also in the same range (71%, 58%, 89% and 57% for NodeLand, Infomap, Louvain and CFinder methods, respectively; Table S1.5 in the Electronic Supplementary Material S1).

In conclusion, both i.) the comparison of ModuLand-derived modules with those obtained by other methods and ii.) the experimental data of the literature showed that the pervasive overlaps of the ModuLand method give an adequate representation of the functional multiplicity of protein-protein interaction network nodes. It is important to note that, in contrast to the other methods tested, the ModuLand method gives this rich background of functional information at the single node level as opposed to the subnetwork level of other methods. Moreover, the Moduland-based, different modular assignment strengths of related nodes (such as those of the 3 cAMP-kinase catalytic subunits; Table S1.4 in the Electronic Supplementary Material S1) give a detailed suggestion on the nodes' functional specificity.

## Variable overlaps of modules surrounding heteronym and antagonym words in a word association network

Extending the analysis of the gold-standard Zachary karate club network, we examined the much larger University of South Florida word association network having 10,617 nodes and 63,788 links [20], which was a target of a successful previous modularization study yielding overlapping modules [7]. This detailed analysis took 10 minutes on a computer with a 3 GHz Intel CPU. Figure 3 shows the modular environment of the antagonym word, “terrific” and that of the heteronym, “content”. The mingling colors indicate a high overlap between the modules. Importantly, the overlap of the modules with alternative meanings of the two words is much greater in the case of “terrific” than in case of “content”, which is a reasonable consequence of the fact that variations of antagonistic meanings (“terrific”) are often amongst our associations, leading to a large joint module containing words with both positive (like “good”, “better” and “great”) and negative (like “bad”, “awful” and “worse”) meanings. We note that the word “well” has multiple meanings, and therefore it is also the member of other distant modules (like the module of “water” or the module connected to “health”). On the contrary to the word “terrific”, the associations between differently pronounced meanings (“content”) are much more seldom. Overlap between the multiple meanings of the words “bright” and “focus” (Figure S1.9 in the Electronic Supplementary Material S1) is closer to that of “terrific” than that of “content”. However, in case of these latter, multiple meaning words the similarly pronounced meanings are not divided into two major sections as in case of the heteronyms, which is again in agreement with our common knowledge.

## Modular hierarchy of a social network

The modular hierarchy of the high school friendship Community-44 of the Add-Health dataset [27] was uncovered using several influence function calculation methods. All of these methods revealed four well-distinguishable main modules with a large amount of further sub-modules (Figure 4A and Figures S1.10, S1.11 and S1.12 in the Electronic Supplementary Material S1). Girls were less likely to form multiracial friendship communities (chi-square $p < 0.05$; Figure 4B), and boys were in the overlap of significantly more friendship communities than girls (chi-square $p < 0.0001$; Figure 4C). These differences are in agreement with the sociological observations indicating a larger cohesiveness of friendship circles of girls than that of boys [28,29].

## Efficient determination of central, key nodes of power-grid network

To test whether the ModuLand method family can identify key network nodes, we calculated the change of network integrity [30] during the disintegration of the USA Western Power Grid network [31]. Nodes were removed in the decreasing order of their degree, betweenness centrality and ModuLand bridgeness (measuring the bridge-like role of the nodes between the modules as defined in Section V.6.d. in the Electronic

Supplementary Material S1). Figure 5 shows that the impact of bridgeness-based node removal on network integrity was larger than that of the degree-based attacks and was well comparable to, or better than the result of betweenness centrality-based node removal. The equal-to-better performance of bridgeness-based disintegration compared to that using betweenness centrality is surprising all the more, since the global network integrity measure corresponds extremely well to the global betweenness centrality measure [30].

## Discrimination between date- and party-hubs

Discrimination of date- and party-hubs of protein interaction networks, i.e. proteins sequentially or simultaneously interacting with a large number of neighbors, is a rather difficult task [32-37]. We hypothesized, that among date-hubs and party-hubs of similar centrality, date-hubs may have a higher bridgeness (i.e. they are more overlapping between modules of the network). This assumption was substantiated by the inter-modular position of date-hubs [34,36] and by the similarly high efficiency of bridgeness-based and date-hub-based network disintegration (cf. Figure 5 with Figure 2 of [34] and [37]). The identification of the overlapping modules of a high-confidence yeast protein-protein interaction network [25] resulted in a number of modules with well-known functions (Figure 6A and Figure S1.14 in the Electronic Supplementary Material S1). We calculated the bridgeness and centrality measures of the individual proteins, and plotted these values on Figure 6B. The separation of date- and party-hubs represented by the line of Figure 6B classified 84 party-hubs correctly of the total of 201, and 307 date-hubs of the total of 318. This result becomes even more convincing, if we consider that 10 out of 11 incorrectly identified date-hubs (91%) and 89 out of 117 incorrectly identified party-hubs (76%) have been potentially misclassified, if comparing them to the consensus of classifications [32-36]. In conclusion, by the help of the novel measures of the ModuLand-based analysis, we were able to discriminate between date- and party-hubs, thus predicting the dynamic behavior of network nodes using only the topological information of their network.

## General characterization of the ModuLand method family

After the examples showing the utility of the ModuLand method family to determine overlapping modules of a variety of model and real world networks in this section we will summarize the characteristics of the ModuLand method family. In principle both the calculation of the influence functions and the determination of the community landscape hills are demanding problems, requiring specific solutions depending on the precise nature of the analyzed network. However, by constructing the community landscape, the small details of the influence functions get averaged out, therefore in practical cases fast and approximate solutions of the mentioned problems become possible and sufficient. This is the reason why rather simple influence function calculation methods (like the NodeLand algorithm) perform well on various kinds of real-world networks. On the other hand, the module membership value of any given node is obtained as the sum of the module membership value of the links of the given node, thus the small details of the hill determination step get also averaged out. The summation of the link module membership values provides an overlapping modularization of the nodes even in the absence of an overlapping modularization of the links themselves. (A similar situation is described in ref. [14].) To summarize, we divided the very challenging problem of module determination into two likewise hard subproblems, but fortunately in most cases a relatively fast, approximate treatment of these subproblems provided sufficiently fine modularizations in the end. However, depending on the precise nature of the application, it is possible, or even advised to devise a more elaborate treatment of the subproblems of influence function calculation and community landscape hill determination.

Several widely used efficient network modularization methods [2,7] can be interpreted as parts of the ModuLand method family either by 1.) identifying the underlying influence function calculation method or by 2.) identifying the community landscape directly (Section IV.4. in the Electronic Supplementary Material S1).

**Previous methods as potential influence function calculation methods of the ModuLand method family.** As an important example for the first case, Bagrow and Bollt [38] define local communities by the spreading of *l*-shells from the nodes of the network, which are suitable as influence functions for the ModuLand method family. The recent work of Roswall and Bergstrom [5] published during the course of the current study [39] uses the probability flow of random walks to construct a map of scientific communication yielding non-overlapping modules. Pons and Latapy also use [11] random walks in their algorithm called 'Walktrap' to define a similarity value for merging communities. The random walks used in these methods can be interpreted also as influence functions. The method of Lancichinetti et al. [23] iteratively finds local modules optimizing a local fitness function. However, instead of executing the local module finding for each node of the network, it is only executed for nodes not contained in any local module yet. This local module finding step can be inserted into the ModuLand method family as a (binary) influence function calculation method. (Note that executing the influence function calculation method only for a fraction of nodes is a possible valid approximation method inside the ModuLand method family, too.) The method of Lancichinetti et al. [23] does not yield fine information about the membership strength of the nodes to different modules as the ModuLand method family does, but yields binary containment information instead. The method and the ModuLand method family have different approaches for the determination of hierarchical levels (see Section VII. in the Electronic Supplementary Material S1).

**Previous methods as potential community landscape identification methods of the ModuLand method family.** As examples for the second case, namely, for the direct identification of the community landscape in previous methods, we briefly summarize the previously described network landscapes. Previous network landscape construction methods used clustering coefficients [33], edge number per visualized network unit area [40], loop-coefficients [41], or degrees [42] to define the landscape-height. The „*leading eigenvector method*" of [15] is able to divide the network in two (or if applied recursively, more) non-overlapping communities maximizing the modularity measure Q.

Both the ModuLand method family and modularity-based methods let their users adapt to the specificities of the analyzed network. However, in case of the ModuLand method family this adaptation is achieved by the choice of the sub-steps (like the community landscape construction method), while the Q modularity-based methods require a null-model to be chosen, which is reflecting the experimenter's expectations about the network. The "*community centrality*" introduced in [15], just as any centrality measure, is a valid basis of forming a ModuLand community landscape, therefore making it possible to include this modularity-based method into the ModuLand method family. Recently, a number of publications showed a 'hidden metric space' behind network topologies, which also links the network structure to a landscape-type representation [43]. Hinneburg and Keim [44] used the density function landscape to determine non-overlapping clusters for the traditional data clustering task, but did not calculate the overlaps based on the hill detection as defined in ModuLand method family. Actually none of the methods mentioned above and listed in Table S1.2 in the Electronic Supplementary Material S1 use the hills of the landscape to determine the modular structure. Evans and Lambiotte [45] show that meaningful modules can be found in networks by finding modules of links instead of nodes, so that nodes can trivially belong to multiple modules, if its links do. However, this method does not give the fine information about the membership strength of the nodes to different modules as can be uncovered with the ModuLand method family.

New modularization methods can easily be generated by taking an existing ModuLand modularization protocol, and changing any of its influence function calculation, community landscape generation, or hill determination methods. Additionally, former methods yielding non-overlapping modules (which can be interpreted as the application of the threshold-based hill determination method) can be upgraded to overlapping modularization methods using the local maxima-based module determination approach of the ModuLand method family (for details see Section IV.4. in the Electronic Supplementary Material S1).

Enriching the binary, yes/no module membership assignment of many previous methods, the ModuLand method family gives a continuous scale for the association of each link and node to all modules (Figure S1.7 in the Electronic Supplementary Material S1). To define the number of modules of a link or node the 'effective number' of modules was introduced (see Section V.6.b. in the Electronic Supplementary Material S1), which is a threshold-less, continuous measure based on the effective size of support of a probability distribution [46]. Additionally, the ModuLand method allowed the definition of further measures characterizing e.g. the centrality and bridgeness of network nodes and links (see Sections IV. and V.6. in the Electronic Supplementary Material S1).

## Selecting the appropriate method of the ModuLand method family

In the ModuLand approach we divided the very challenging problem of module determination into two likewise hard subproblems: the influence function determination (1); and the determination of hills of the resulted community landscape (2). Although in most cases a relatively fast, approximate treatment of these subproblems provides sufficiently fine modularizations in the end, in the following section we give a brief guide to select the optimal algorithms for these subproblems.

1) As we mentioned earlier, the determination of the influence functions requires a network-dependent perturbation-flow simulation on the network. However, we saw, that the details of the influence functions usually average out during the community landscape construction, which justifies the use of less specific, faster approximations. We prefer to use the LinkLand algorithm on undirected networks, which is found to provide an acceptable compromise between precision and speed. However, on directed networks we suggest to use the PerturLand algorithm (see Section IV.2. in the Electronic Supplementary Material S1) instead. We believe that our influence function calculation algorithms (NodeLand, LinkLand, and PerturLand) present only the first steps in the direction towards novel accurate and fast influence simulation techniques.
2) For the hill determination on the community landscape we presented two main approaches (for other possibilities see Section V. in the Electronic Supplementary Material S1): the centrality threshold-based hill determination and the local maxima based hill determination approaches. The centrality threshold-based hill determination approach is appropriate whenever the goal of the analysis is to find modules without overlapping regions. In order to determine the overlaps between the modules we suggest to use one of the local maxima based methods. While for practical purposes we suggest the use the ProportionalHill method, the most detailed module overlap information is acquirable with the computationally more expensive TotalHill method. If we need only the most important overlaps between the modules, we may use the GradientHill method, as described in Section V.2.b. in the Electronic Supplementary Material S1.

We note that although the local maxima-based approaches we described in this paper (including the ProportionalHill method suggested above) outperform the traditional threshold-based approach in terms of overcoming the giant-component problem and producing continuously overlapping modules, nevertheless they also have their own drawbacks. When applying the local maxima-based approach on a ‘noisy’ community landscape, each local maximum will result a new (and possibly highly overlapping) module. Therefore we routinely applied a simple, yet effective post-processing step for merging the groups of extremely overlapping modules (having a correlation higher than 90%) (see Section VI. in the Electronic Supplementary Material S1).

To summarize, the hill-finding approach, which is the second phase of the ModuLand methods, gives an additional layer of flexibility, where the relatively inaccurate results of simpler hill definitions, and the large computational costs of more accurate optimization processes can be tailored to the network and to the experimenter’s needs and possibilities.

## Discussion

The ModuLand method family we introduced in this paper and in part in an earlier patent application [39] is a novel, integrative approach, which includes also the usual partitioning techniques such as the threshold-based hill determination methods over an appropriate community landscape. However by using local maxima based hill determination methods, it yields overlapping modules over any community landscape. Thus, our approach is suitable to extend the previous partitioning techniques to find overlapping modules. We presented novel, special examples for influence function calculation (NodeLand, LinkLand and PerturLand algorithms), which form the basis of the community landscape construction. Moreover, the identification of modules as hills of the community landscape is a new approach, including traditional threshold-based algorithms and the novel local maxima based algorithms (as our ProportionaHill, GradientHill and TotalHill methods). The Moduland method family defines link-based modules and results in pervasively overlapping modules as some of the few most recent approaches [14],[45] published well after our initial studies [39]. Previous methods using local community detection or yielding overlapping modules (Table S1.2 in the Electronic Supplementary Material S1) [4,7] can be interpreted as special cases of our ModuLand method family, if appropriate hill determination techniques or community landscape construction methods are designed.

The extensive and rich overlaps, network hierarchy, as well as the novel centrality and bridgeness measures uncovered by the ModuLand method can be used for the identification of long-range, stabilizing weak links, for the determination of the recently described creative, trend-setting nodes governing network development and evolution [47], for prediction of missing links or nodes, for network classification and for the design of efficient information transfer to name only a few of the many possibilities. Module overlaps might play a key role in the disconnection and synchronization of modules of complex systems, and their re-assembly during and after crisis, respectively. We invite our colleagues to design novel versions of the framework we gave, and to explore the above and other examples.

## Materials and Methods

### Networks

**Network science co-authorship network.** The giant component of the undirected, un-weighted network science co-authorship network contained 379 nodes and 914 links [15].

**Karate club social network.** The weighted and undirected social network of a karate club has been reported by W. Zachary [19] containing 34 nodes and 78 links. As the members of the karate club have split into two factions later, the network became a gold-standard of module determination methods [1-5,7].

**Word association network.** The giant component of Appendix A of the University of South Florida word association network (http://www.usf.edu/FreeAssociation/) [20] with removed link directions contained 10,167 nodes and 63,788 weighted links, where weight refers to the association strength (see Section I.3. in the Electronic Supplementary Material S1).

**School friendship network.** The giant component of the high school friendship Community-44 of the Add-Health database (http://www.cpc.unc.edu/projects/addhealth) [27] with removed link directions contained 1,127 nodes and 5,096 weighted links, where weights represent the strengths of friendships (see Section I.4. in the Electronic Supplementary Material S1).

**Power-grid network.** The un-weighted and undirected network of the USA Western Power Grid [31] contained 4,941 nodes and 6,594 links (http://vlado.fmf.uni-lj.si/pub/networks/data/map/USpowerGrid.net).

**Yeast protein-protein interaction network.** The giant component of the un-weighted and undirected yeast protein-protein interaction network [25] contained 2,444 nodes and 6,271 links, covering approximately half of the yeast genome and the most reliable ('strongest') ~3% of the expected number of total links. All these network data are included in the ModuLand program package downloadable from our homepage: http://www.linkgroup.hu/modules.php.

# Supporting information

*Electronic Supplementary Material S1 (ESM1):* This supporting information contains a detailed description of the ModuLand method including the pseudo-codes of all specific algorithms and methods used, 14 Supplementary Figures, 5 Supplementary Tables (with 18 module definitions, 129 different modularization methods and 13 module comparison methods), a Supplementary Discussion and 396 references.

# Acknowledgments

We thank Gábor Szuromi and Balázs Zalányi for help in the analysis of networks, members of the LINK-group (www.linkgroup.hu) for discussions and Tamás Vicsek for his seminar in our lab on 16th June 2005 giving us the starting encouragement to work on the ideas of this paper and for his continuous suggestions. This research was supported by research grants from the EU (FP6-518230), Hungarian Science Foundation (OTKA K-69105) and by an unrestricted grant from Unilever Hungary to the Hungarian Student Research Foundation, which helped the research of the authors.

# Author contributions

I.A.K. conceived and designed most of the ModuLand method, performed part of the network analysis and wrote part of the manuscript of the paper, M.S.S. and R.P. helped to formulate details of the method, designed the final computer programs, performed part of the network analysis and wrote part of the manuscript of the paper, P.C. gave the basic idea, suggested the network examples, helped the interpretation of the data and wrote part of the manuscript of the paper. I.A.K., M.S.S. and R.P. started their research as members of the Hungarian Research Student Association (www.kutdiak.hu/en), which provides research opportunities for talented high school students since 1996.

## Competing financial interest

The authors declare that they have no competing financial interest.

**Box 1. Glossary.** Here we present a short guide to the algorithms and methods defined in this paper and in the Electronic Supplementary Material S1 in detail.

**GradientHill method**: a local maxima-based module determination approach, in which the module membership value of a link is determined by the module membership value(s) only of its neighboring link(s) having maximal centrality values (see Section V.2.b. in the Electronic Supplementary Material S1).

**LinkLand algorithm**: an influence function calculation method starting from a given link in undirected networks (see main text and Section IV.1.b. in the Electronic Supplementary Material S1).

**ModuLand method family**: the integrative name of our module determination approach based on the hills of the community landscape (see main text and Section II. in the Electronic Supplementary Material S1).

**NodeLand algorithm**: an influence function calculation method starting from a given node in undirected networks (see main text and Section IV.1.a. in the Electronic Supplementary Material S1).

**PerturLand algorithm**: an influence function calculation method starting from a given node in directed networks (see Section IV.2. in the Electronic Supplementary Material S1).

**ProportionalHill method**: a local maxima-based module determination approach, in which the module membership value of a link is determined by the module membership values only of its neighboring links having non lower centrality values (see main text and Section V.2.b. in the Electronic Supplementary Material S1).

**TotalHill method:** a local maxima-based module determination approach, in which the module membership value of a link is determined by all the module membership values of its neighboring links (see Sections V.2.c. and V.2.d. in the Electronic Supplementary Material S1).

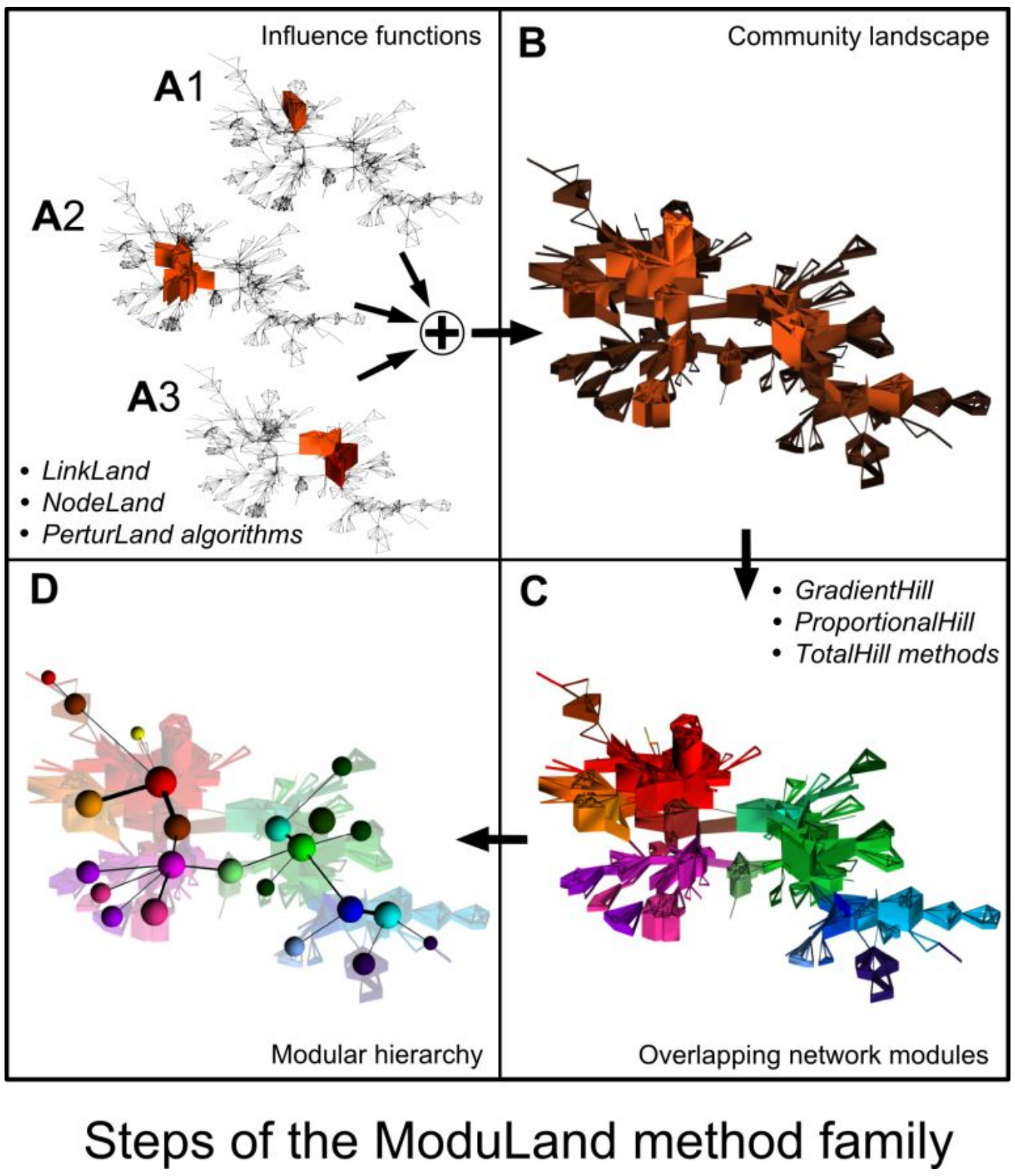

**Figure 1. Description of the ModuLand method family.** For this illustrative example we used the network science co-authorship network [15] without link weights using the LinkLand influence function calculation method with the TotalHill module membership assignment method. The network was laid out using the Kamada-Kawai algorithm and was visualized with a custom Blender script. On the vertical axes influence function values (panel A), or community landscape values (panels B, C and D) of the links are shown. Influence functions of panels A1, or A2 belong to the Barabási—Vicsek, or Girvan—Newman author-pairs, respectively. Panel A3 shows the merged influence function of the Arenas—Pastor-Satorras and Guimera—Amaral co-authorship links. Links and nodes of panels C and D are colored in proportion of the colors of the modules they belong. **Panel A:** influence function calculation. First, the influence function of each link (or node) of the network were identified. If a link is in the 'middle' of a module, it is affected by many influence functions (all the three widely collaborating author-pairs, whose influence functions are shown by the arrows, are from this category). On the contrary, links at module 'edges' are affected by few influence functions only. At the bottom of the panel the names of the three algorithms we described in details are shown. **Panel B:** community landscape construction. Next, the community landscape is constructed by summing up the influence function values for all nodes or links. The hills of the community landscape correspond to the modules of the network. **Panel C:** determination of overlapping modules. Last, modular centers are identified as the links at the local maxima of the community landscapes, and memberships of links in all network modules are determined. At the top of the panel the names of the three methods we described in details are shown. **Panel D:** determination of network hierarchy. Optionally, a higher level hierarchical representation of the network can be created, where nodes of the higher level correspond to modules of the original network, and links of the higher level correspond to overlaps between the respective modules. Sizes of higher level nodes correspond to the log size of the respective lower level modules, where the module size is the sum of the membership assignment strengths of all nodes to that module.

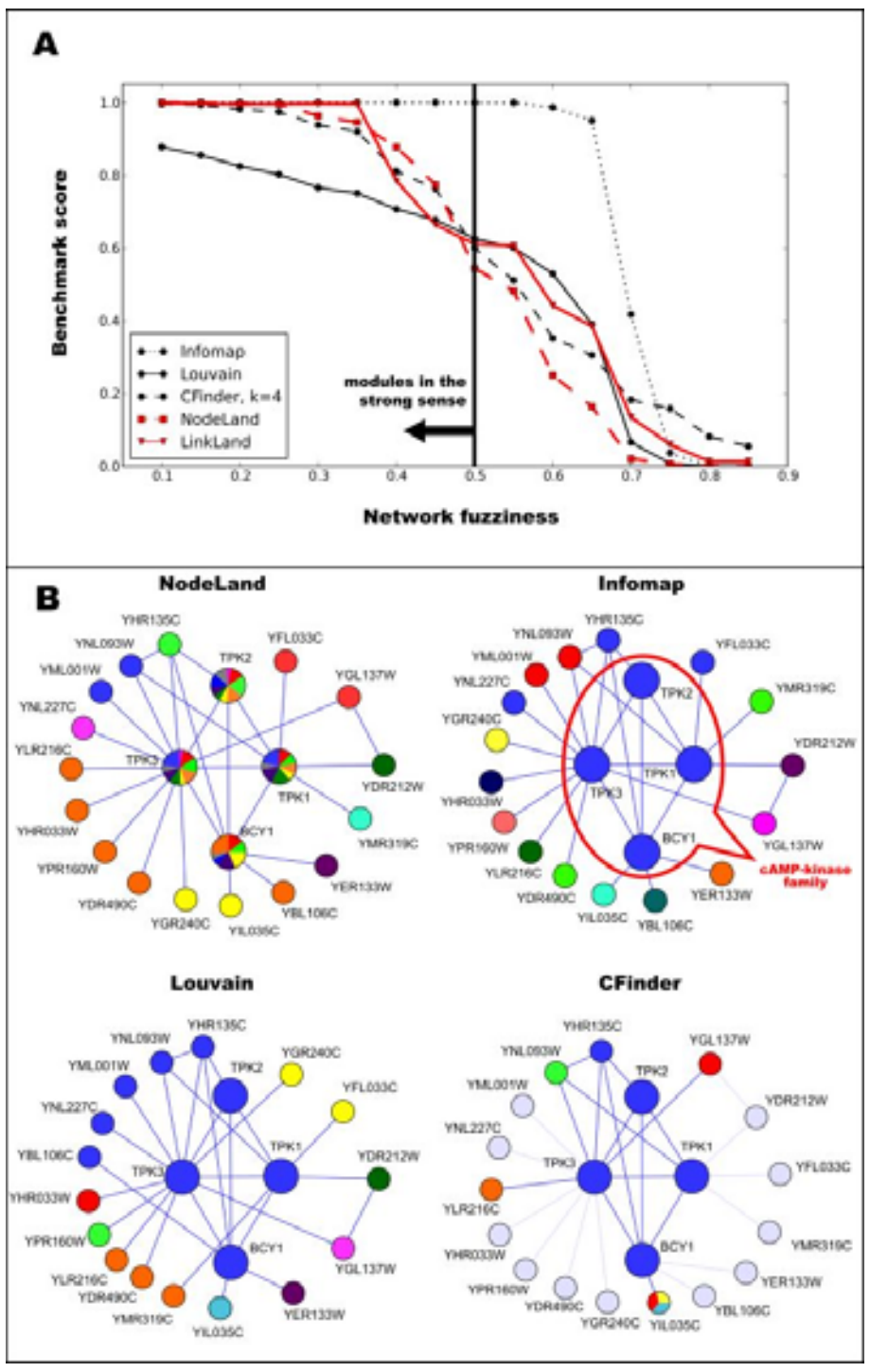

**Figure 2. Comparison of the ModuLand method with other modularization methods. Panel A:** Comparison of the identified modules with the modules of the benchmark graph of Lancichinetti et al. [21]. Modularization has been performed on benchmark graphs with degree and module size distribution exponents $\gamma=2$ and $\beta=1$ using the NodeLand or LinkLand influence function calculation algorithm with the ProportionalHill module membership assignment method with merging highly correlated modules using an arbitrary chosen correlation threshold of 0.9 (see Section VI.1. of ESM1; red squares with dashed line and red rectangles with solid line for NodeLand and LinkLand, respectively), the InfoMap method (black circles with dotted line, [5]), the Louvain method (black circles with solid line, [22]) and the CFinder method with k=4 cliques (black circles with dashed line, [7]). The number of nodes of the benchmark graphs was N=1000, the maximum degree was Kmax=50, the average degree was K=15 and the network fuzziness ($\mu$ of the x-axis of Panel A) was ranging from 0.1 to 0.85, where $\mu$>0.5 means that the modules are no longer defined in the strong sense. Higher normalized mutual information (shown on the y-axis) represents a better recovery of the original modules. The panel shows the averaged results of 50 representations. **Panel B:** comparison of module assignment of the cAMP-dependent protein kinase family in the yeast protein-protein interaction network. The panel shows the modular assignment of the 3 catalytic and the regulatory subunit of the yeast cAMP-dependent protein kinase together with that of their first neighbors in the high fidelity protein-protein interaction network of Ekman et al. [25]. For the sake of simplicity only intra-subnetwork contacts have been included. The top left, top right, bottom left and bottom right figures show the modular assignment using the NodeLand, InfoMap, Louvain and CFinder methods, respectively, determined as described in the legend of Panel A. Various colors correspond to different modules. Overlapping ModuLand modules of cAMP kinase members and CFinder modules of their casein kinase II neighbor are marked with pie-charts, where the area of color-codes is proportional to the module membership value of the given node in the given module. To simplify the figure, in case of the NodeLand method (top right figure of Panel B) all neighbors of cAMP-dependent protein kinase family members (which should all have similar pie-charts to the 4 central members due to their multiple modules) were assigned to their maximal modules only.

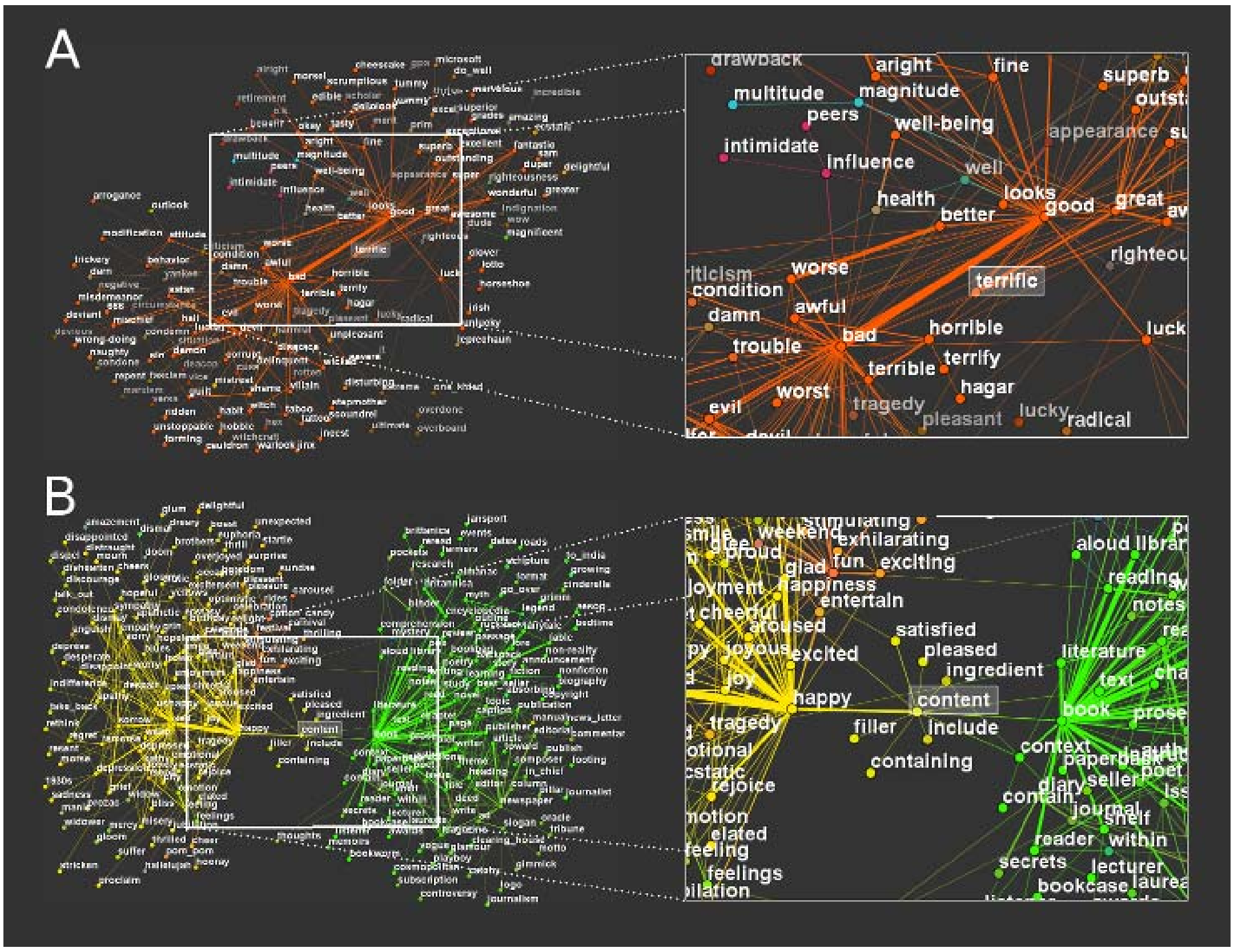

**Figure 3. Overlapping modules of a word-association network.** Modules of the University of South Florida word association network [20] were determined using the LinkLand influence function calculation method and the TotalHill module membership assignment method. During the post-processing of the module assignment, we merged the modules with ProportionalHill module membership assignment-based correlation higher than 0.9 (see Section VI. of ESM1, we received similar results without this merging process; data not shown). The network was laid out using the Kamada-Kawai algorithm of Graphviz [48] and visualized using a custom program written in Python language using OpenGL graphics. Links were colored in proportion to the colors of the modules they belong. **Panel A:** modules around the antagonym word, “terrific”. **Panel B:** modules around the heteronym word, “content”. In addition to the selected words “terrific” and “content” similar words above a similarity threshold of 10% are also shown with a contrast corresponding to their degree of similarity. The extent of similarity between two words was calculated as the sum of the two pair-wise minima of their unity-normalized module membership vector giving the membership assignment strength of the given word to all modules of the network (for more details see Section V.6.e. of ESM1).

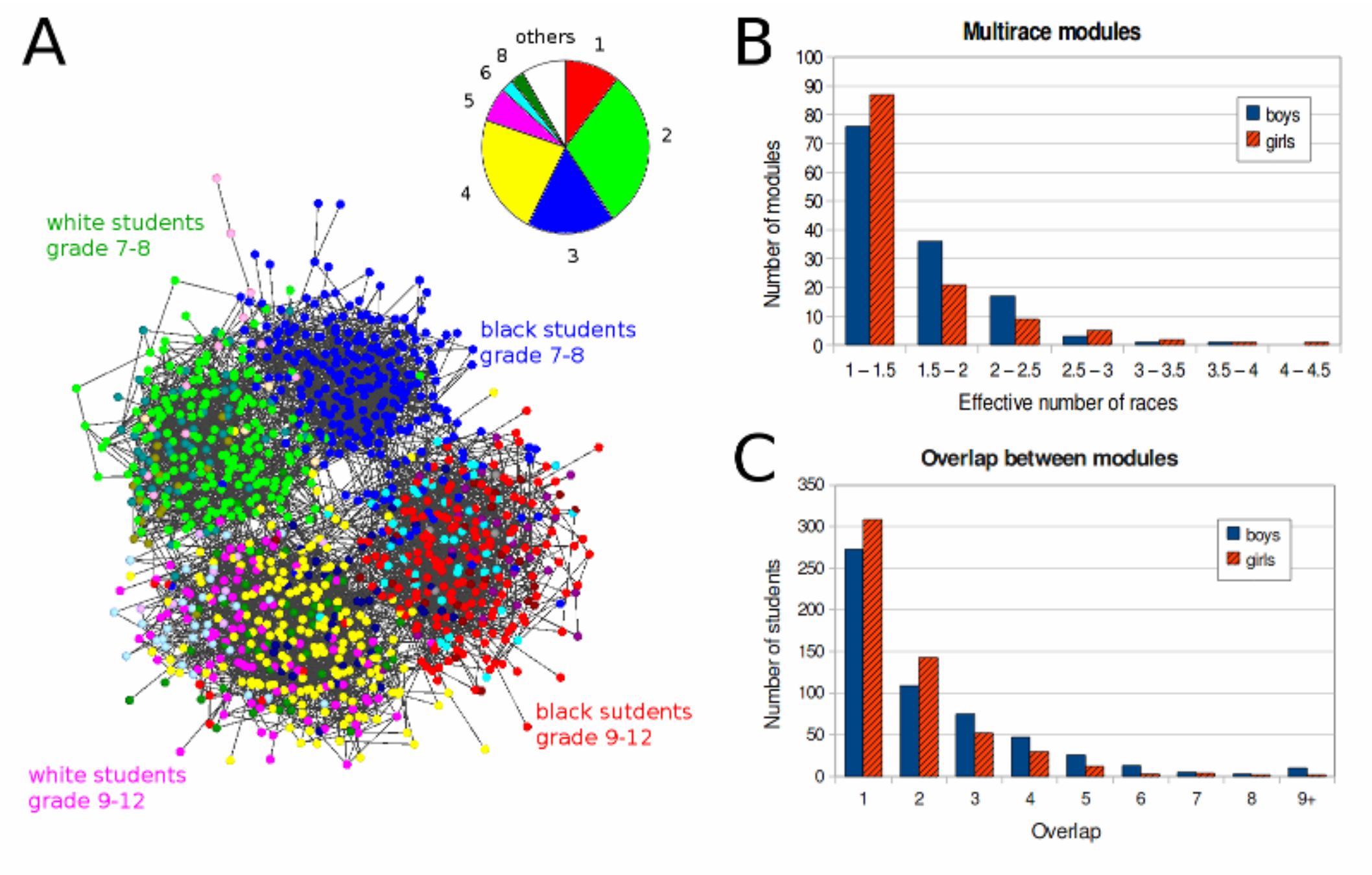

**Figure 4. Overlapping modules of a school-friendship network.** We have determined the modular structure of Community-44 of the Add Health survey [27] using the LinkLand influence function calculation method together with the ProportionalHill module membership assignment method. During the post-processing of the module assignment, we merged the modules with ProportionalHill module membership assignment-based correlation higher than 0.9 (see Section VI. of ESM1, we received similar results without this merging process; data not shown). **Panel A:** modules of Community-44. The school friendship network was laid out using the Kamada-Kawai algorithm. Nodes represent the individual students, and were colored according to the color of the friendship module they assigned the most. We show the modular structure of the first hierarchical level having 18 modules. The *inset* of Panel A shows color-codes of the modules with an area proportional to the size of the respective module. **Panel B:** the number of network modules in case of boys (blue, solid bars) and girls (red-black hatched bars) with mixed racial contents at the lowest hierarchical level (level 0). The extent of mixed racial content was monitored using the 'effective number of races' (Section V.6.b. of ESM1) with a bin-size of 0.5. **Panel C:** overlaps of boys and girls in friendship circles. The number of boys (blue, solid bars) and girls (red-black hatched bars) having different overlaps in friendship circles were determined in the first hierarchical level with a bin-size of 1. Overlap was measured as the 'effective number' (Section V.6.b. of ESM1) of modules of the given student.

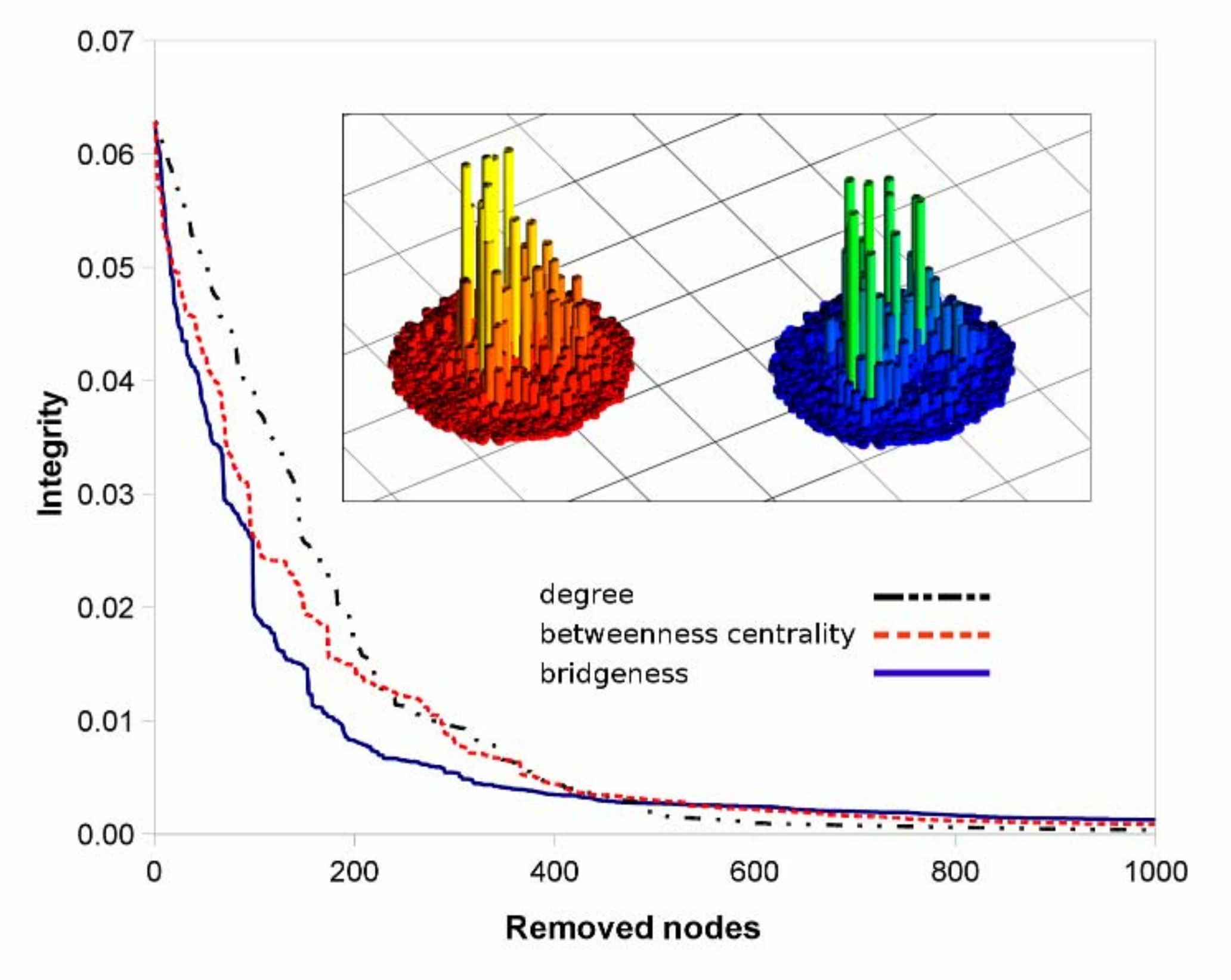

**Figure 5. Determination of key nodes of the USA Western Power Grid network.** The figure shows the decreasing integrity of the USA Western Power Grid network [31] as a function of the number of nodes removed. Nodes were removed in the order of their decreasing degree (black alternating dashes and dots) betweenness centrality [2] (red dashed lines) or 'bridgeness' (solid blue lines), where 'bridgeness' measures the overlap of the given node between different modules as described in detail in Section V.6.d. of ESM1. Network integrity has been calculated after Latora and Marchiori [30]. Bridgeness was calculated from the modular structure of the lowest hierarchical level as determined by the LinkLand influence function calculation method and the TotalHill module membership assignment method. During the post-processing of the module assignment, we merged the modules with ProportionalHill module membership assignment-based correlation higher than 0.9 (see Section VI. of ESM1, we received similar results without this merging process; data not shown). On the vertical axis of the *insets* the betweenness centrality (left, color-coded from red to yellow) and bridgeness (right, color-coded from blue to green) of the nodes of the USA Western Power Grid network are shown. Networks on the *insets* were laid out using the Kamada-Kawai algorithm and visualized with a custom Blender script.

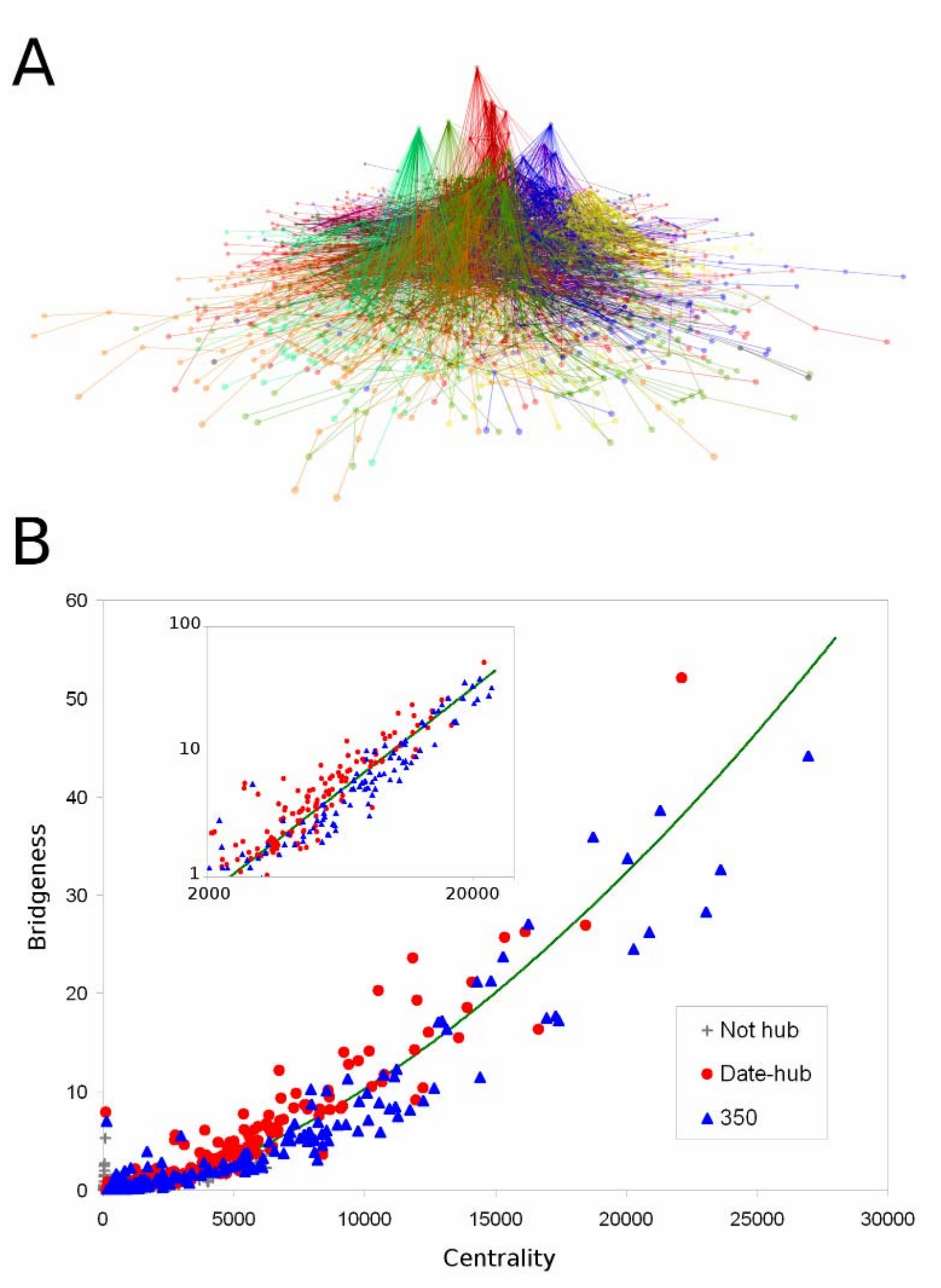

**Figure 6. Prediction of the dynamical behavior of network nodes: segregation of date- and party-hubs based on their modular overlaps.** Overlapping modules of the yeast protein-protein interaction network of Ekman et al. [25] were identified using the LinkLand influence function calculation method with the TotalHill module membership assignment method using the modular structure of the lowest level of hierarchy. During the post-processing of the module assignment, we merged the modules with ProportionalHill module membership assignment-based correlation higher than 0.9 (see Section VI. of ESM1, we received similar results without this merging process; data not shown). **Panel A**: 3D view of the yeast protein-protein interaction network. The underlying 2D network layout was set by the Kamada-Kawai algorithm. The vertical positions reflect the community landscape values of the nodes on a linear scale. Nodes were colored as the module of their maximum membership. **Panel B:** centrality and bridgeness of yeast date- and party-hubs. Hubs having more than 8 neighbors and non-hubs with less neighbors were positioned on the scattergram according to their ModuLand centrality (x-axis, the height of the community landscape) and ModuLand bridgeness (y-axis) as defined in Section V.6.d. of ESM1. Date- and party-hubs are marked with red circles and blue triangles, respectively, while non-hub proteins are represented by gray crosses. The *inset* shows a double logarithmic plot of hubs with large centrality.

Electronic Supplementary Material S1

# Community landscapes: an integrative approach to determine overlapping network module hierarchy, identify key nodes and predict network dynamics

István A. Kovács, Robin Palotai, Máté S. Szalay, Péter Csermely*

Department of Medical Chemistry, Semmelweis University, Tűzoltó str. 37-47, H-1094 Budapest, Hungary

*E-mail: csermely@eok.sote.hu

## Summary

In this Electronic Supplementary Material (S1) we give a detailed description of the ModuLand network module determination method family. This integrative method is based on the construction of community landscapes from influence functions. In Section IV. we describe three versions of the influence function calculation algorithms, the NodeLand, LinkLand and PerturLand algorithms in detail. As the next step. the combination of influence functions to a community landscape is shown. We demonstrate the wide applicability of the ModuLand method to accommodate previous community detection methods in the examples of the BetweennessCentralityLand (BCLand) and CliqueLand community landscape determination methods resulting in distinct and overlapping network modules, respectively. In Section V. we show the local maxima-based identification of modules as hills of the community landscape. The module membership of network nodes and links is calculated using one of the developed module membership assignment methods, such as the GradientHill, ProportionalHill or TotalHill methods yielding modules of minimal, fair or detailed overlaps, respectively. In Sections VII. And VIII. we also show that the ModuLand method family enables a hierarchical analysis of network topology and the construction of a zoom-in network visualization method. Besides the detailed description of the ModuLand method the Electronic Supplementary Material also contains 14 Supplementary Figures and their Supplementary Discussion, as well as a detailed summary of 18 module definitions, 129 different modularization methods, 13 module comparison methods as 5 Supplementary Tables and 396 references.

**The Linux-based computer programs of the ModuLand-related methods or a Windows-based application with a User Guide can be downloaded from here: www.linkgroup.hu/modules.php.**[1]

[1]A Linux/x86 compatible operating system with kernel version 2.6 is required for running the programs provided in the ModuLand program package. If you do not have access to such system, you may use a prebuilt VirtualBox image of a Linux system with all necessary programs as described in the package.

# Table of content

# Supplementary Figures

**Figure S1. Time-scale of the development of modularization methods.**
The figure shows the number of modularization methods listed in Table S2 as a function of publication years between 1956 and 2009. Books and reviews were omitted to help the direct assessment of the methodological enrichment of the field.

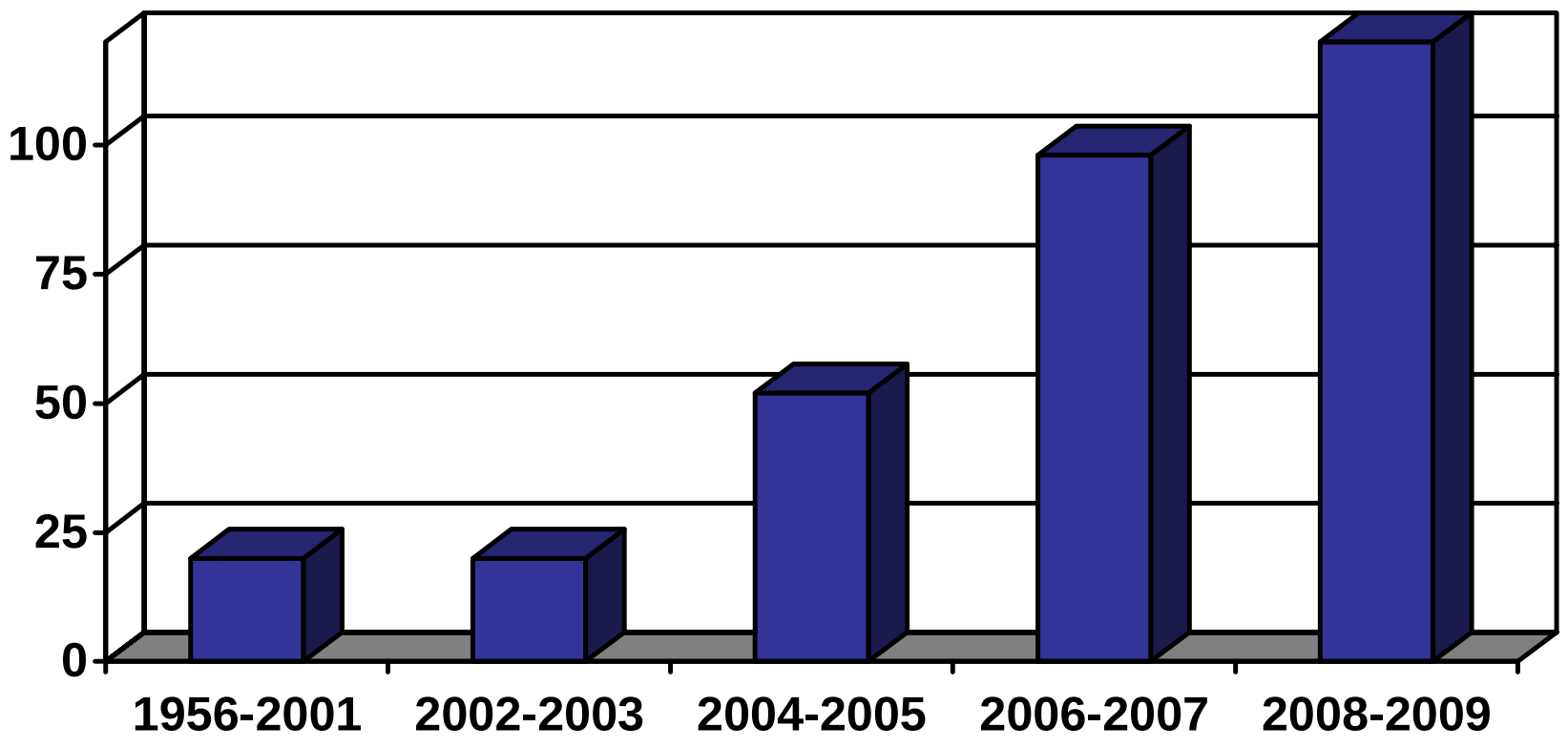

**Figure S2. Flowchart of the available ModuLand method algorithms.**
The figure shows the different phases and possible options of the ModuLand algorithm package. Cylinders represent data storage options, while boxes represent different operations. Box captions refer to the name of the operation, while the name in parentheses refers to the executable program name as found in the ModuLand program package downloadable from http://www.linkgroup.hu/modules.php. For a detailed explanation please see the User Guide included in the program package, Figure 1 and the main text as well as Sections IV., V. and VII.

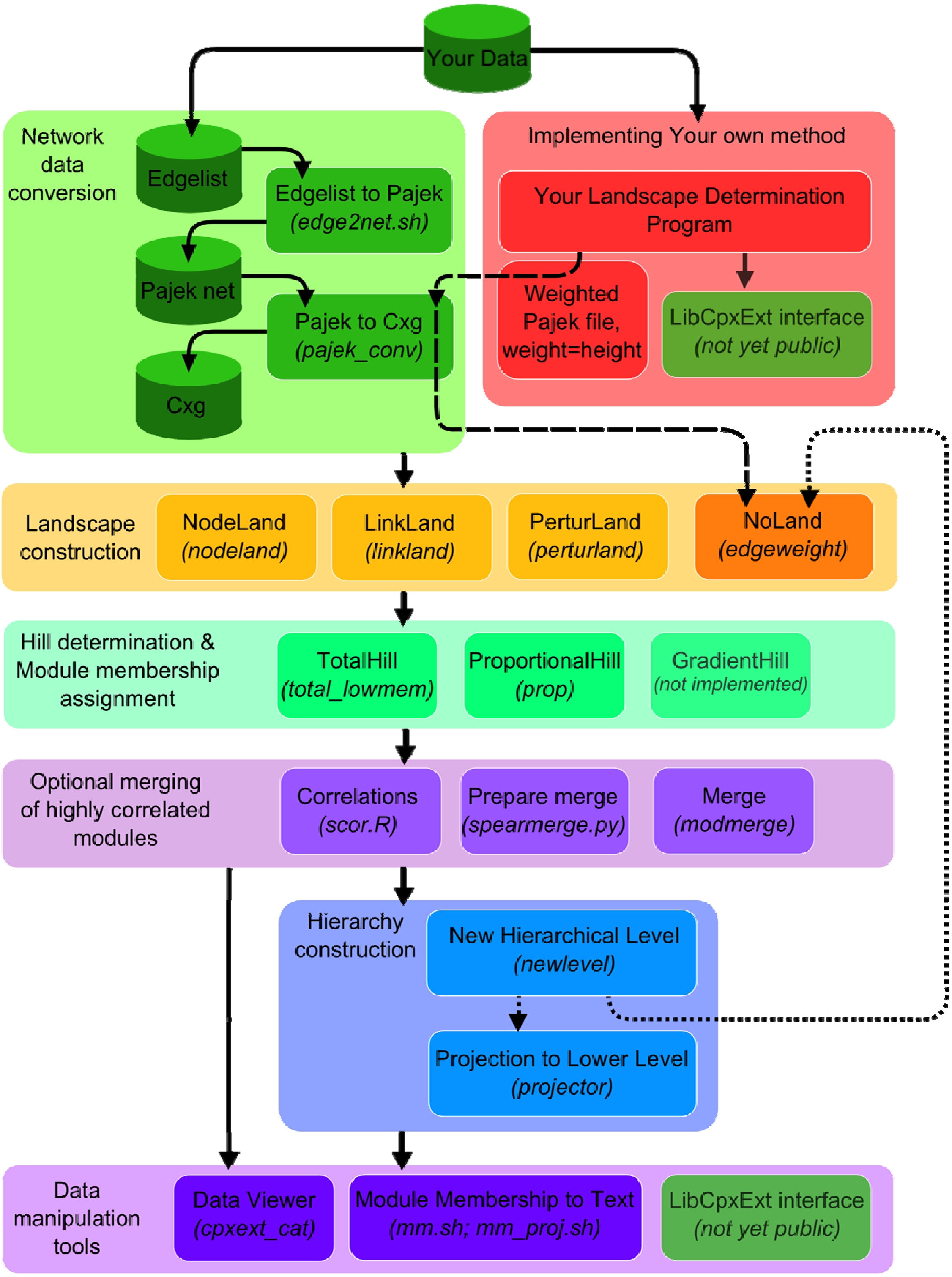

**Figure S3. Different approaches of community landscape construction.**

Panels **(a)** and **(b)**. Influence function-based community landscape construction. Panel **(a)** illustrates that an influence function (colored hill-like areas) may belong to each node or link of the network. Influence functions can be determined by any of the influence function calculation algorithms described (see Figure 1. of the main text and Suppl. Methods, Section IV). Panel **(b)** shows that a community landscape may be constructed by summing up the individual influence functions belonging to the nodes or links of the network (see Suppl. Methods, Section IV.1-3.). Panels **(c)** and **(d)**. Direct construction of a community landscape. It may be possible to construct the community landscape directly, omitting the influence function calculation step. As an example, data acquired by former community detection algorithms can be directly transformed into a community landscape. On Panel **(c)** the betweenness centrality values of a network are shown as heights of the links. The original algorithm by Girvan and Newman (2002) iteratively removes the links of highest betweenness, and defines modules as connected components (colored parts under the curve in the picture) of the remaining network. In this method the number of modules depends on the length of the iteration process. The horizontal line of Panel **(c)** represents the final round of the iterative process. As this algorithm removes links of high betweenness centrality from the network first, which are usually inter-modular links, the links removed last tend to be located at modular centers. Let the community landscape height $c_{ij}$ of the link$(i,j)$ be k $(k>0)$, if the given link was removed from the network in the $k$-th iteration. This way we get the 'BCLand community landscape' shown on Panel **(d)** (see Suppl. Methods, Section IV.4.b.), on which the original modularization can be gained by applying the height threshold-based hill determination method. A similar conversion of an other  partitioning technique to a community landscape yields overlapping modules as a rule. The horizontal lines below the panels represent links (or nodes) of the network with the appropriate colors of their modules.

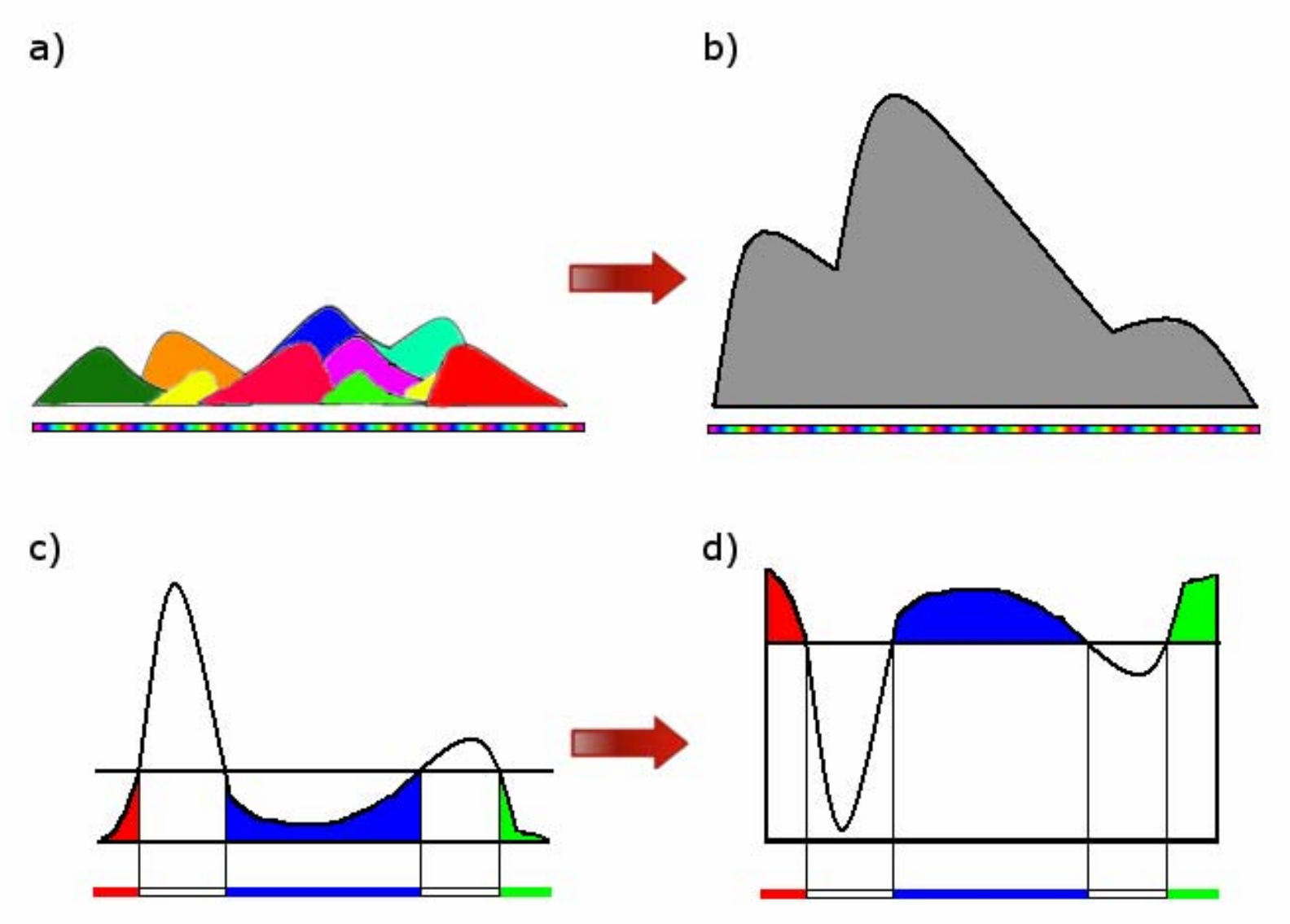

## Figure S4. Hill determination on community landscapes.

Once a community landscape had been constructed, modules can be identified in different ways. On this illustrative figure the horizontal line below the Panels represents nodes or links of the network with the color of their modules. The hill-like curve on the Panels represents the community landscape. The threshold-based hill determination method (see Suppl. Methods, Section V.1.) shown on Panel **(a)** identifies all connected components of the network above a given centrality threshold as modules. Another possibility (see SameHill method in Suppl. Methods, Section V.3.) shown on Panel **(b)** identifies connected components of height not differing more than a given percentage from the height of hill-top nodes or links as modules. We note that these approaches in principle do not assign all nodes or links into modules. The local maxima-based approach (see Suppl. Methods, Section V.2.) shown on Panel **(c)** identifies the hill-tops of the community landscape as module-cores, and uses one of the ProportionalHill or TotalHill etc. module membership assignment methods (see Suppl. Methods, Section V.2.b.-e. and Figure S5) to assign each node and link of the network into overlapping modules.

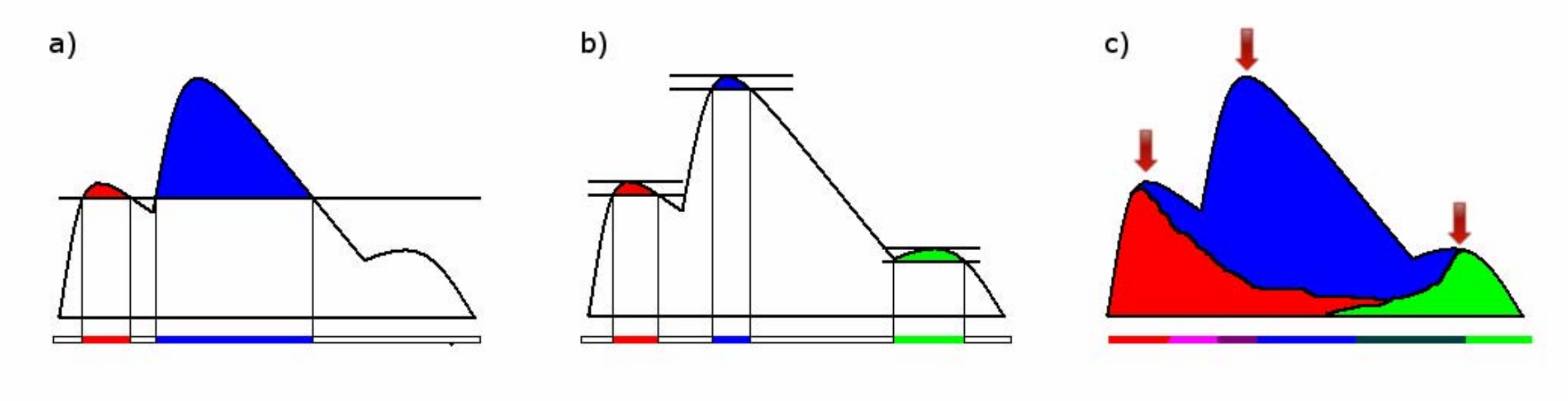

**Figure S5. Local maxima-based methods for module membership assignment.**

The figure illustrates three module-membership assignment methods resulting in increasing overlaps between the network modules. The horizontal colored bars illustrate links of the network, and the vertical lines connecting the links represent a node shared between the links. Link colors refer to the module membership assignment of the links. Vertical position of a given link indicates the centrality of that link (see Figure 1 of the main text). Links A and E are selected as module-cores, because they are local maxima (see Suppl. Methods, Section V.2.a.). Module-core links are always assigned to their own modules only. Panel **(a)**. The GradientHill module membership assignment method. In this method a non-core link is assigned only to the module of its highest neighboring link (or the module assignment is divided between modules of its highest neighbors, if more than one such neighbors exist). Based on this rule link C is assigned to the red module. As illustrated, the GradientHill method results in a minimal overlap between the modules. Panel **(b)**. The ProportionalHill module membership assignment method. In this method a non-core link is assigned to the modules of its higher neighbors proportional to the height of the respective neighbors. Based on this rule link C is assigned more to the red module and only minimally to the blue module. The ProportionalHill method results in a moderate overlap between the modules. Panel **(c)**. The TotalHill module membership assignment method. In this method a non-core link is assigned to the modules of its all neighbors proportional to the height of the respective neighbors. This type of assignment can be efficiently solved with a set of linear equations. Based on this rule not only link C is assigned both to the red and blue modules, but links B and D also become inter-modular links between the red module defined by the module-core, link A and the blue module defined by the module-core, link E. Additionally, the adjacent links of links A and E at the two links of the figure – being also non-core links – will be assigned to more than one modules illustrated by their mixed color. Thus, the TotalHill module membership assignment method results in an extensively high overlap between the modules.

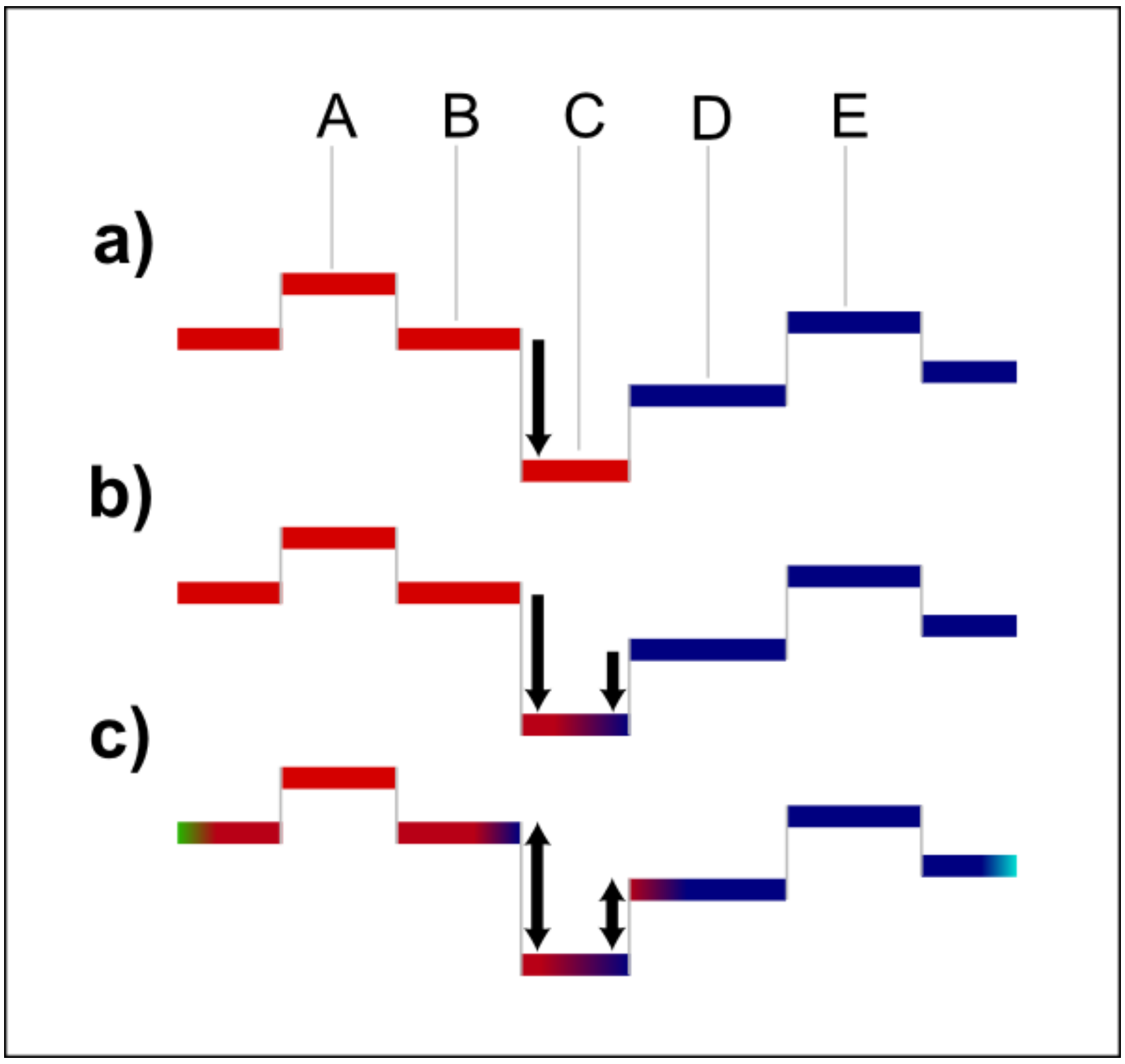

**Figure S6. Hierarchical levels of the network science co-authorship network.**

The first row of the figure shows the hierarchical levels of the network science collaboration network (Newman, 2006a) as uncovered by the LinkLand influence function calculation algorithm and the ProportionalHill module membership assignment method. Influence functions were determined only at the original network level. The network was visualized with the Kamada-Kawai algorithm. Modules of the original (zero-level) network shown on the left side of the figure became the nodes of the first hierarchical level of the network labeled as 'level 1'. Modules of 'level 1' became the nodes of 'level 2' until the network coalesced to a single node, which would constitute 'level 4' of the figure (not shown). It is also possible to project the module memberships of a higher level network back to the nodes of any intermediary level network (see Suppl. Methods, Section VII.3.). The result of this projection procedure is shown in the second row of the figure, where the module memberships of the respective level are projected back to the nodes of the original network and are represented by different colors.

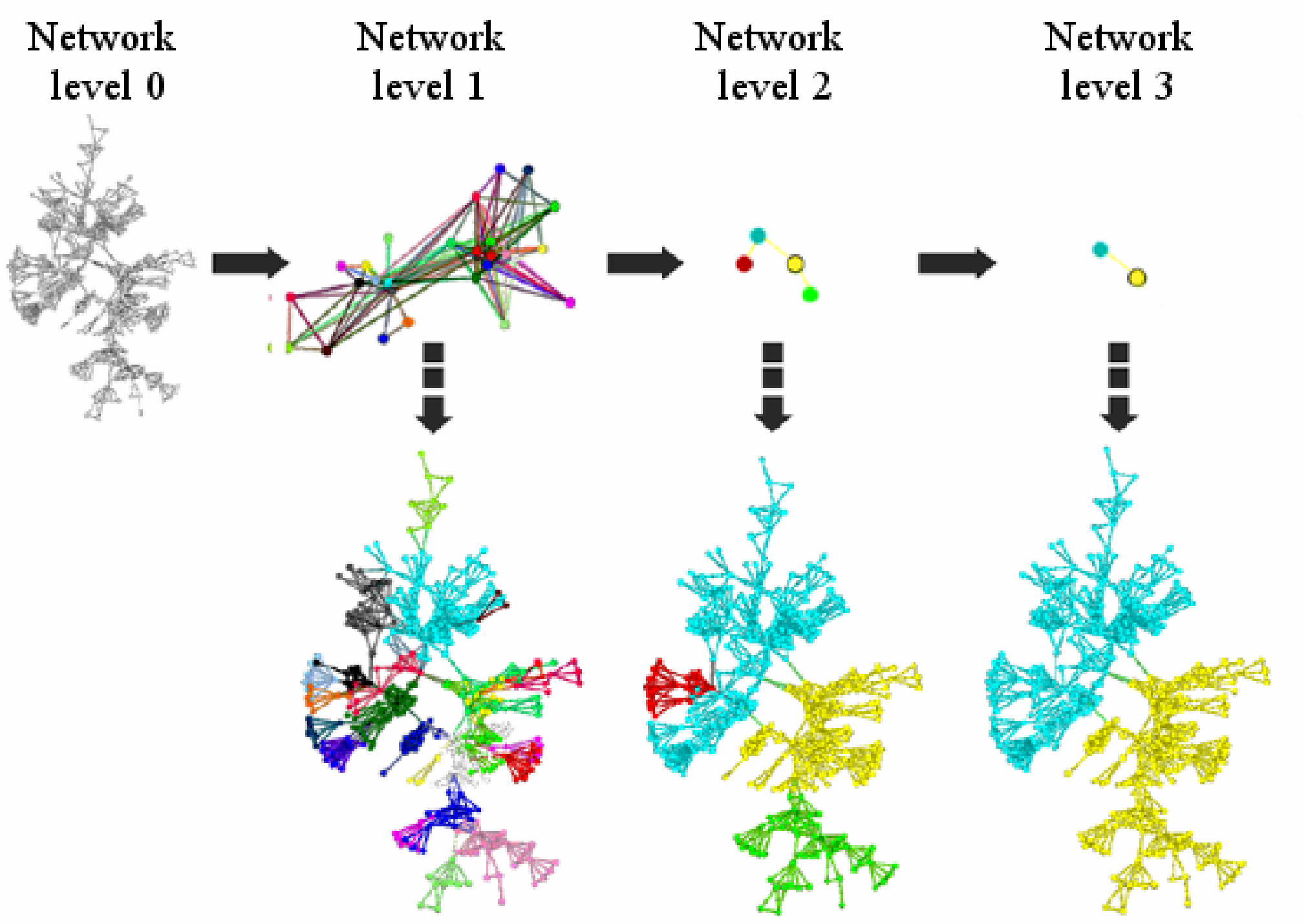

## Figure S7. Modularization of the Zachary network.

This figure shows the modularization results of one of the gold-standard social networks for modular analysis, the Zachary karate club (Zachary, 1977). Panels **(a)** and **(b)**. Overlapping modules of the Zachary karate club network. The three overlapping modules of the Zachary network, as identified by the NodeLand influence function calculation algorithm together with either the ProportionalHill or the TotalHill module membership assignment methods, are shown on Panel **(a)** or Panel **(b)**, respectively. The network was visualized with the Kamada-Kawai algorithm. Green squares, red circles and blue triangles refer to the three modules identified by the methods. Node colors are mingled to the extent of their overlapping module membership. Node shapes refer to the module, which is the maximal-strength module of the given node. Numbers refer to the number of the respective karate club member in the original study (Zachary, 1977). The method correctly identifies the observed split of the original network, while uncovering several club-members in modular overlaps and a third module also identified in previous studies (Zhang et al., 2007a; Leskovec et al., 2008; Shen et al., 2009). Numbers on the insets at the bottom of Panels **(a)** and **(b)** show the effective number of modules (see Suppl. Methods, Section V.6.b.), where the respective node belongs to. Applying the ProportionalHill method on Panel **(a)** results in a lower effective number of modules per node than the application of the TotalHill module membership assignment method on Panel **(b)** indicating a smaller modular overlap, which is in agreement with the illustrative model of Figure S5.

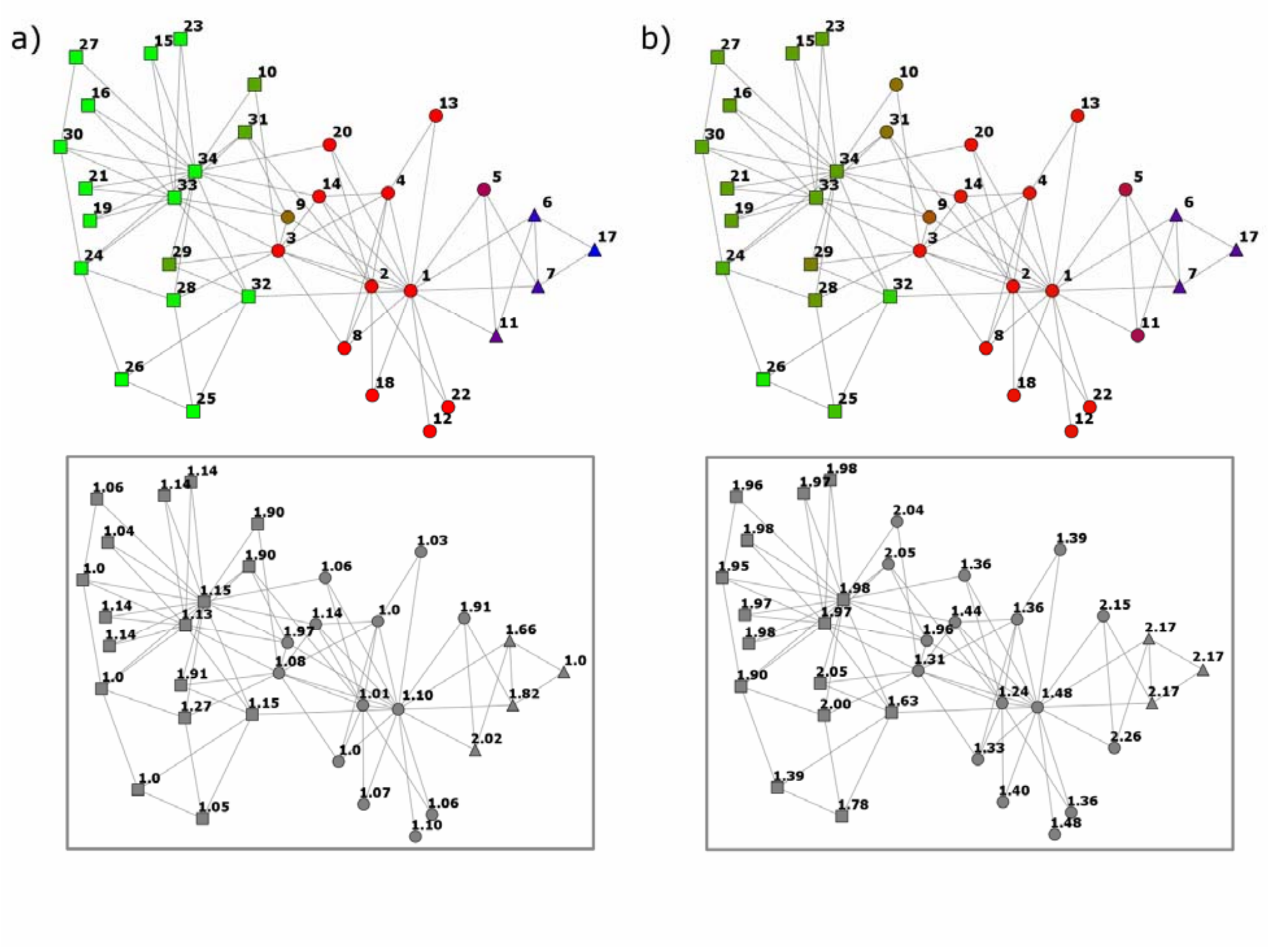

**Figure S8. Distribution of module sizes, module degrees, module overlap sizes and node membership numbers of the USF Word Association Network.**

Modules of the University of South Florida word association network (Nelson et al., 1998) were identified by applying the LinkLand influence function calculation algorithm and the TotalHill module membership assigment method. During the post-processing of the module assignment, we merged the modules with a ProportionalHill module membership assignment-based correlation higher than 0.9 (similar results were obtained without the merging step; data not shown). The modular structure was characterized by four metrics. Panel **(a)** shows the cumulative distribution of the size of modules, where the size of a given module is the summed membership assignment strength of each node to the given module. Panel **(b)** shows the cumulative distribution of the effective degree of modules, where weighted links are defined between modules based on their overlap as defined in Section VII.1., and the effective degree of a given module is the effective number (see Section V.6.b.) of such weighted links of the given module. Panel **(c)** shows the cumulative distribution of the overlaps between modules, where the overlap between two modules is the summed area-overlap between the respective modules (see Section V.6.d.) of each node of the network. Panel **(d)** shows the cumulative distribution of the effective number of modules per node (a similar measure is also known in the literature as 'node membership number'; González et al., 2007), where the effective number of modules of a node is given by the modular overlap measure of the given node (see Section V.6.c). All plots show the cumulative distributions on log-log scales. *P* on the vertical axes is defined as the fraction of modules (Panels **a** and **b**), the fraction of module pairs (Panel **c**) or the fraction of nodes (Panel **d**) for which the measured quantity equals or is greater than the value of the horizontal axes, for any given x.[2]

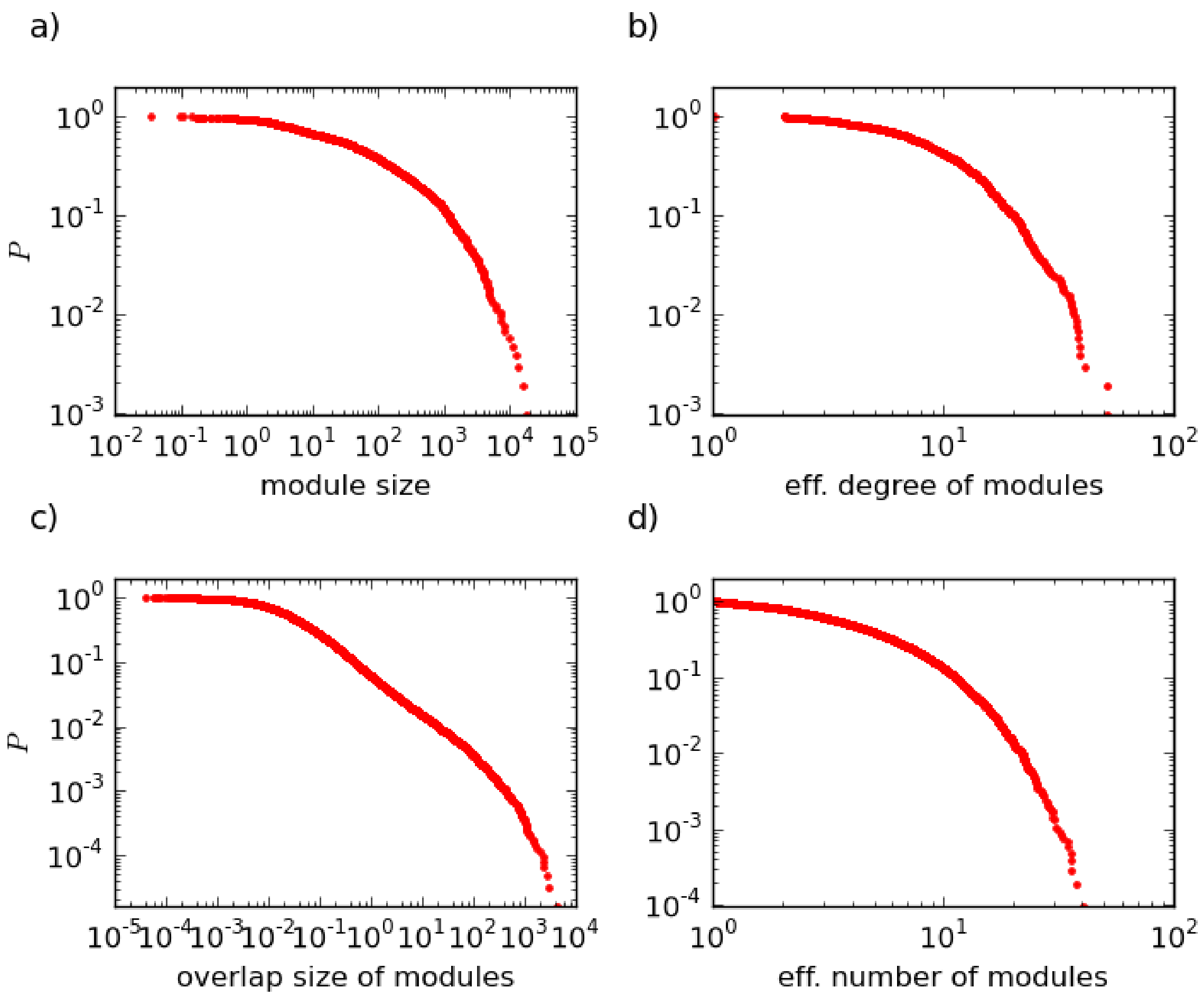

---

[2]For example, a point located at coordinates (x=10, y=0.42) on panel **(b)** means that 42% of the modules have an effective module degree of at least 10 (this corresponds to the situation that 42% of the modules are connected to at least 10 modules in the unweighted case).

**Figure S9. Modules of two homonym words in the USF Word Association Network.**

Modules of the University of South Florida word association network (Nelson et al., 1998) were determined using the LinkLand influence function calculation algorithm together with the TotalHill module membership assignment method. During the post-processing of the module assignment, we merged the modules with ProportionalHill module membership assignment-based correlation higher than 0.9 (similar results were obtained without the merging step; data not shown). The network was laid out using Graphviz (Gansner and North, 1999) and visualized using a custom program written in Python. Links were colored in proportion to the colors of their modules. In addition to the selected words "bright" (Panel **a**) or "focus" (Panel **b**), similar words above a similarity threshold of 13% or 20%, in case of "bright" or "focus", respectively, are also shown with a contrast corresponding to their degree of similarity. The calculation of similarity between any two words is described in section V.6.e. of Suppl. Methods.

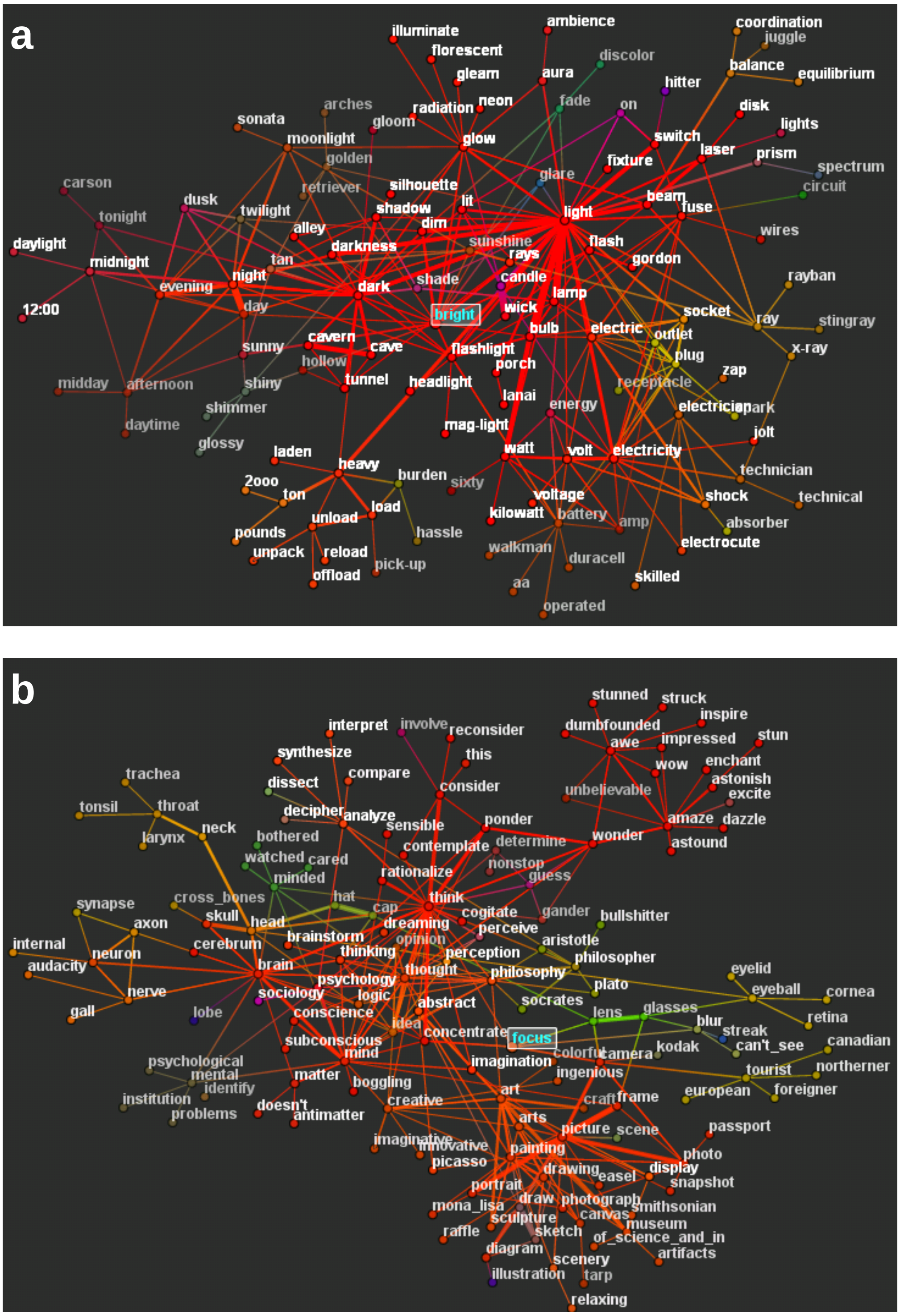

**Figure S10. Comparison of the effect of four influence function calculation algorithms on the modular structure of a school friendship social network.**

The modular structure of the school friendship social network of Community-44 of the Add Health dataset (see Suppl. Methods, Section I.4.; Moody, 2001; Newman, 2003) is shown at its first and second hierarchical levels (on the main images of the four panels and on their right-top insets, respectively), as uncovered by various influence function calculation algorithms of the ModuLand method family using the ProportionalHill module membership assignment method. Influence functions were determined only at the original network level. During the post-processing of the module assignment, we merged the modules with ProportionalHill module membership assignment-based correlation higher than 0.9 (similar results were obtained without the merging step; data not shown). The network was laid out with the Kamada-Kawai algorithm. Colors represent the color of the module, where the given node assigned most. The numbers in parentheses are the effective number of modules (see Suppl. Methods, Section V.6.b.). While the number of modules are rather consistent using the four influence function calculation algorithms, the least accurate NoLand method (which simply takes link weights as community landscape heights) fails to distinguish the known four main sections (Newman, 2003) of the network.

.

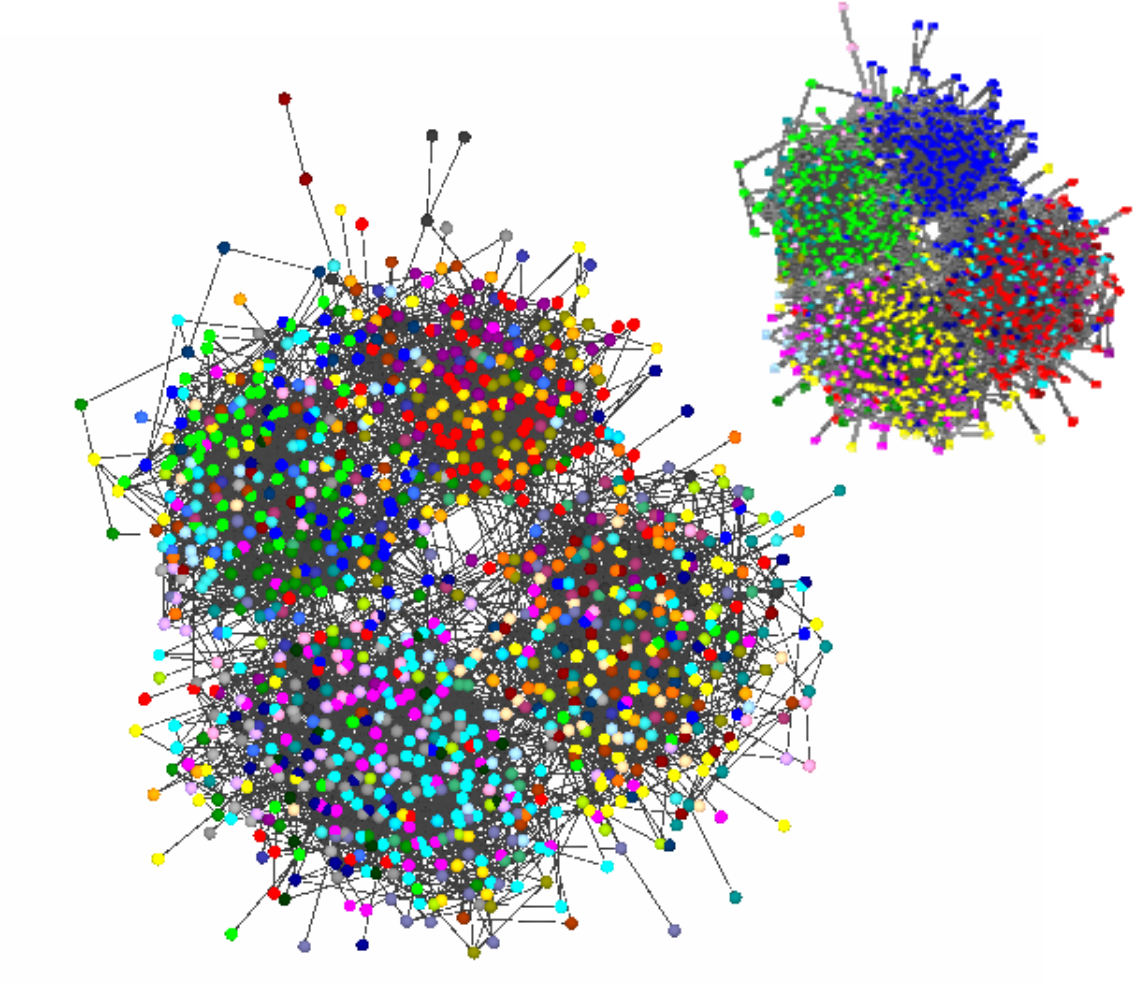

**LinkLand**

Level 0: 174 modules (27.55)
Level 1: 18 modules ( 7.18)

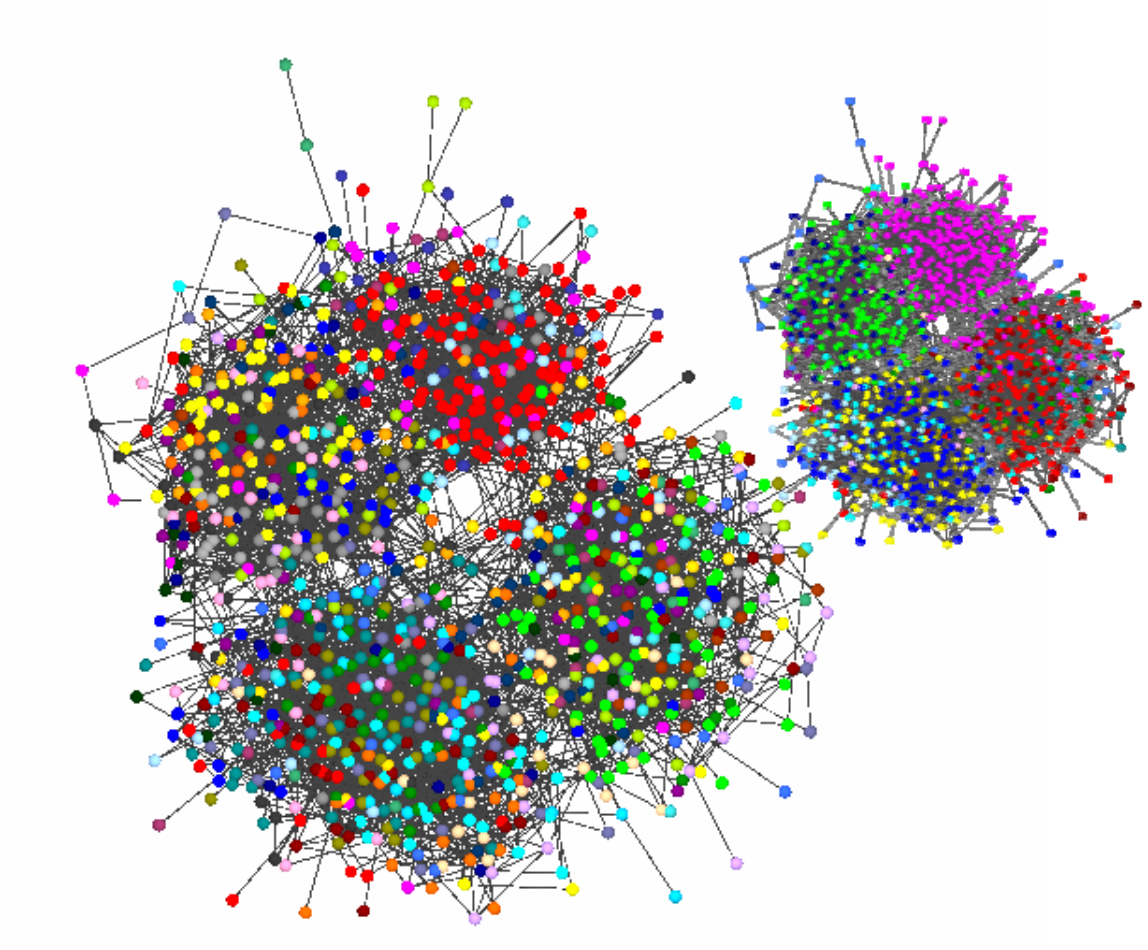

**NodeLand**

Level 0: 175 modules (31.48)
Level 1: 19 modules ( 8.22)

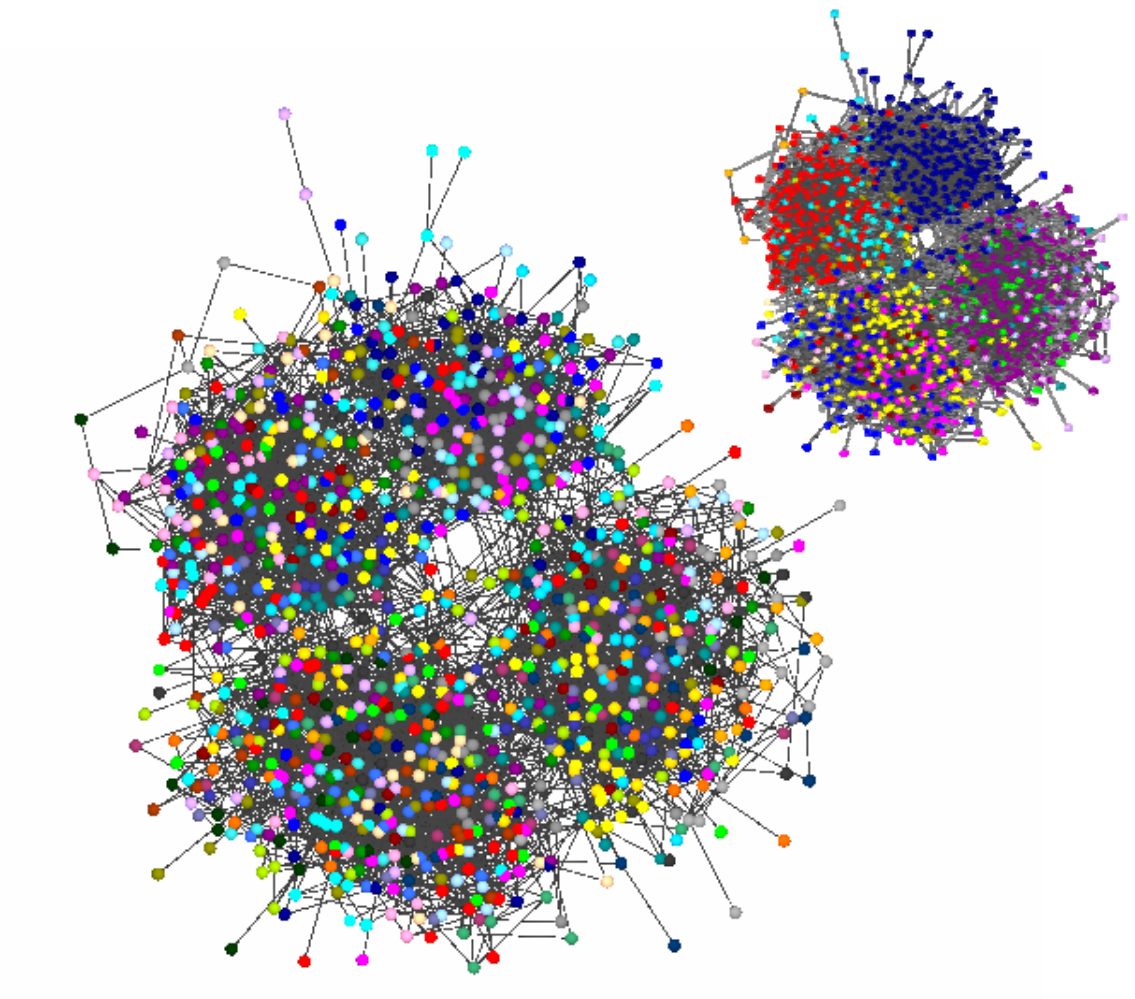

**NoLand**

Level 1: 229 modules (162.79)
Level 2: 24 modules (12.14)

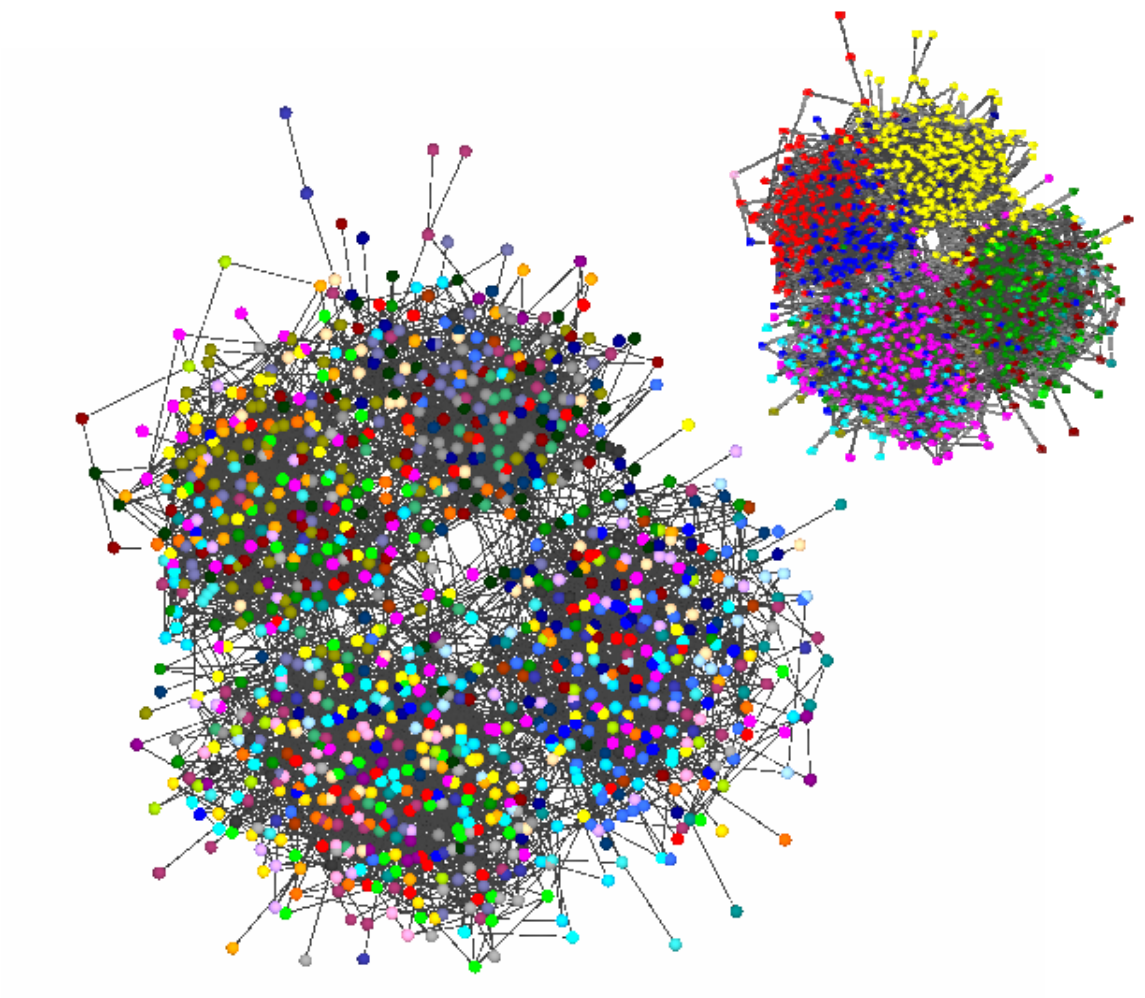

**PerturLand** (X=0.2)

Level 1: 206 modules (123.17)
Level 2: 18 modules ( 9.34)

## Figure S11. Modular hierarchy of a school friendship network obtained by the LinkLand algorithm.

The modular structure of the school friendship social network of Community-44 of the Add Health dataset (see Suppl. Methods, Section I.4.; Moody, 2001; Newman, 2003) is shown at various hierarchical levels, as uncovered by the LinkLand influence function calculation algorithm using the ProportionalHill module membership assignment method. Influence functions were determined only at the original network level. During the post-processing of the module assignment, we merged the modules with ProportionalHill module membership assignment-based correlation higher than 0.9 (similar results were obtained without the merging step; data not shown). The network was laid out with the Kamada-Kawai algorithm. Colors represent the color of the module, where the given node assigned most. The numbers in parentheses are the effective number of modules (see Suppl. Methods, Section V.6.b.). As shown, the higher hierarchical levels yield less, but more extended modules.

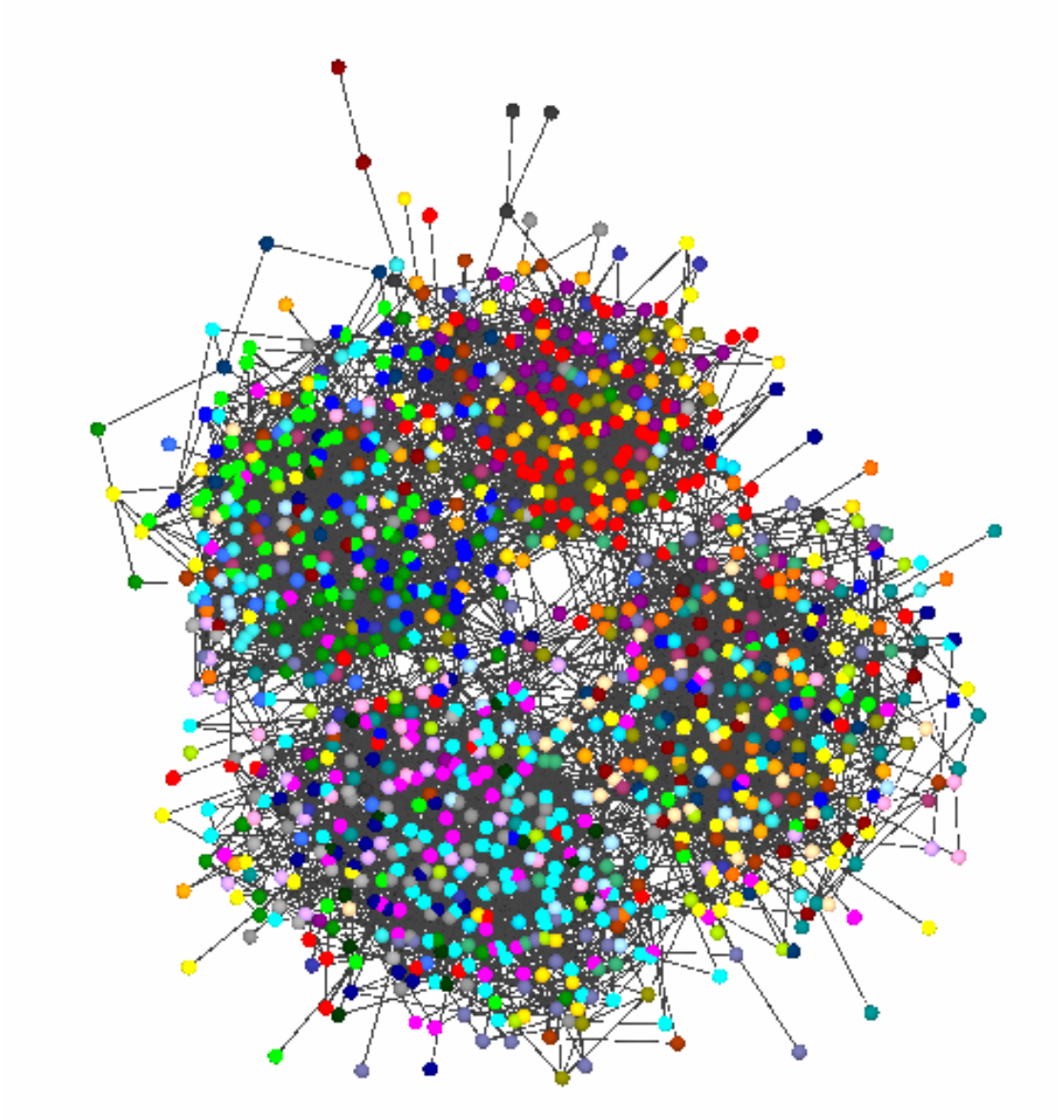

**LinkLand – level 0**
174 modules (27.55)

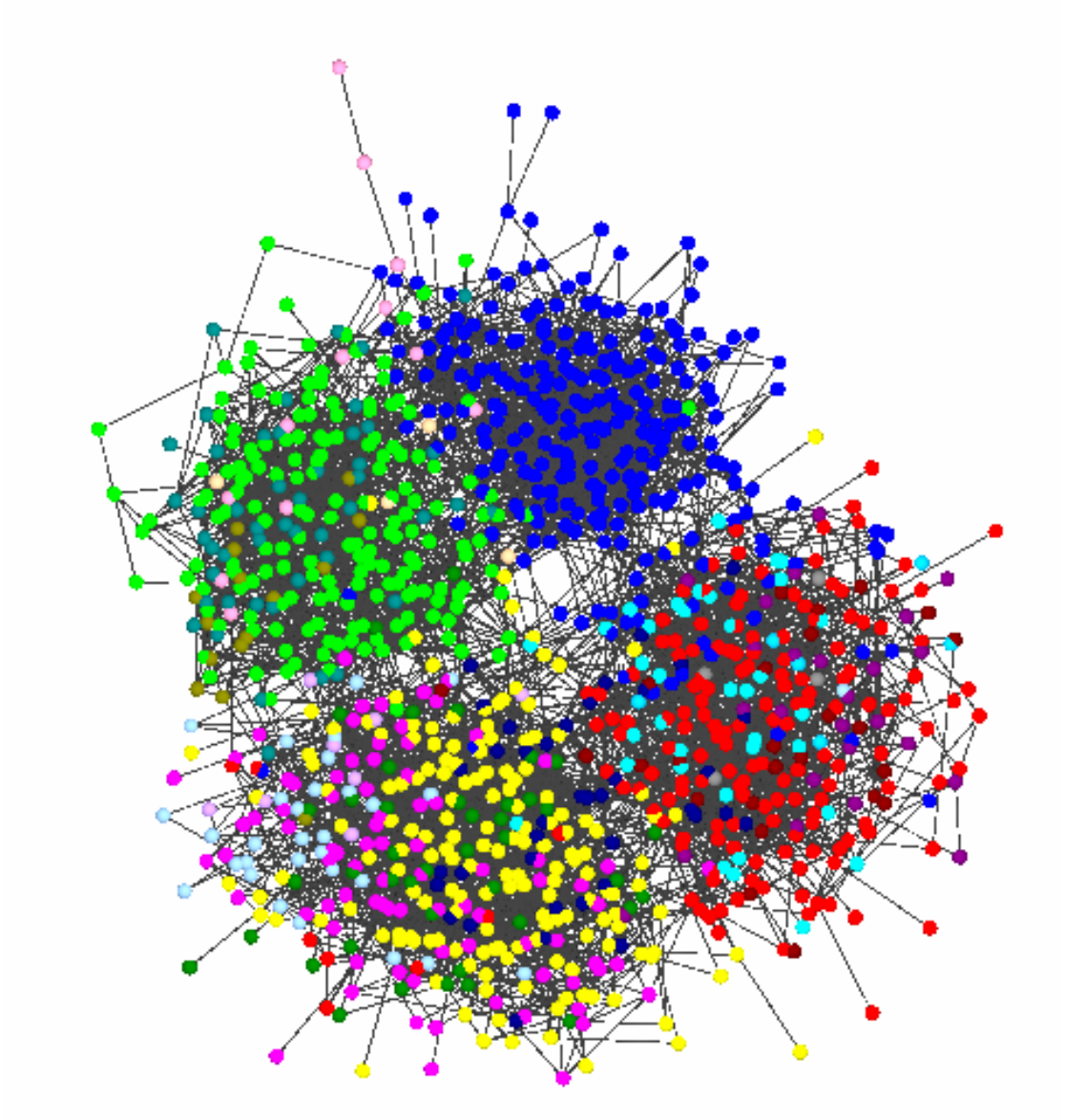

**LinkLand – level 1**
18 modules (7.18)

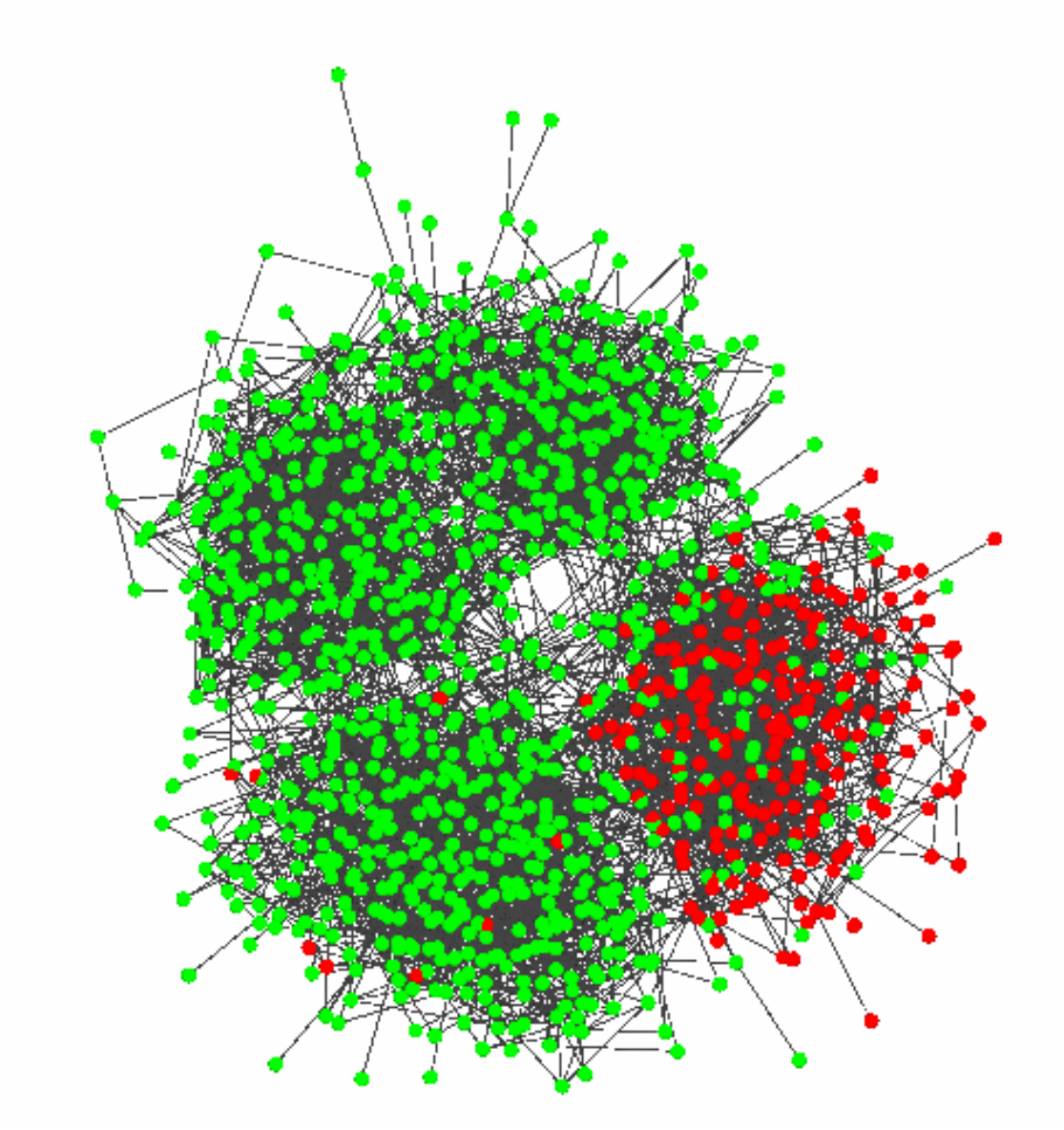

**LinkLand – level 2**
2 modules (1.62)

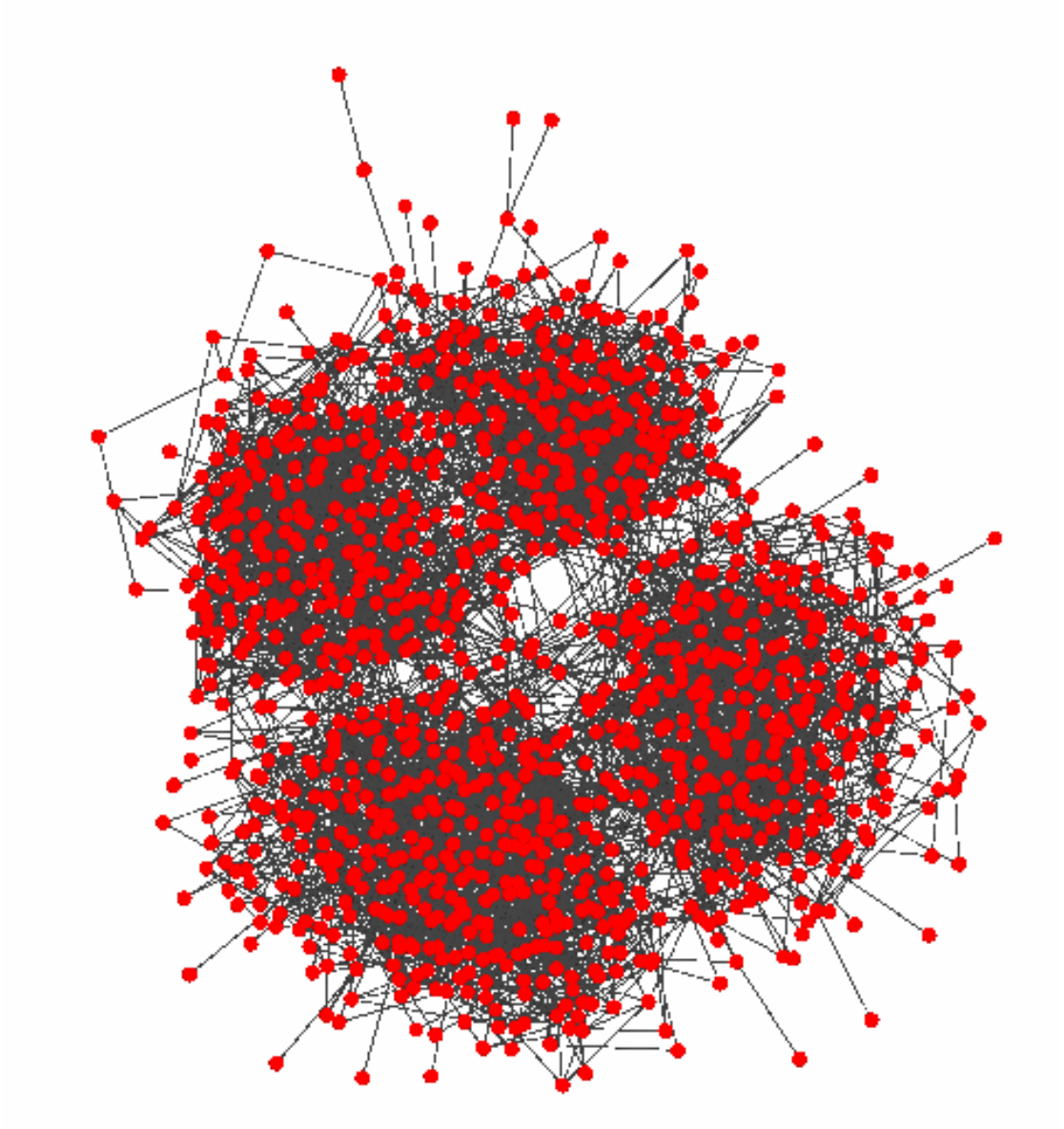

**LinkLand – level 3**
1 modules (1)

## Figure S12. Modular hierarchy of a school friendship network obtained by the PerturLand algorithm.

The modular structure of the school friendship social network of Community-44 of the Add Health dataset (see Suppl. Methods, Section I.4.; Moody, 2001; Newman, 2003) is shown at various hierarchical levels, as uncovered by the PerturLand influence function calculation algorithm with different *X* parameters, using the ProportionalHill module membership assignment method. Influence functions were determined only at the original network level. During the post-processing of the module assignment, we merged the modules with ProportionalHill module membership assignment-based correlation higher than 0.9 (similar results were obtained without the merging step; data not shown). The network was laid out with the Kamada-Kawai algorithm. Colors represent the color of the module, where the given node assigned most. The numbers in parentheses are the effective number of modules (see Suppl. Methods, Section V.6.b.). The *X* parameter of the PerturLand influence function calculation algorithm controls the decay of the simulated perturbation. *X* values near 0 cause a rapid decay and thus result in small, nuclear influence zones, while *X* values near 1 cause a minimal decay and result in extended influence zones. As seen in the figure, the larger *X* parameter of 0.5 (inducing a slower perturbation-decay and yielding more extended influence zones) merged many of the smaller and even two of the four major modules (Newman, 2003) of Community-44.

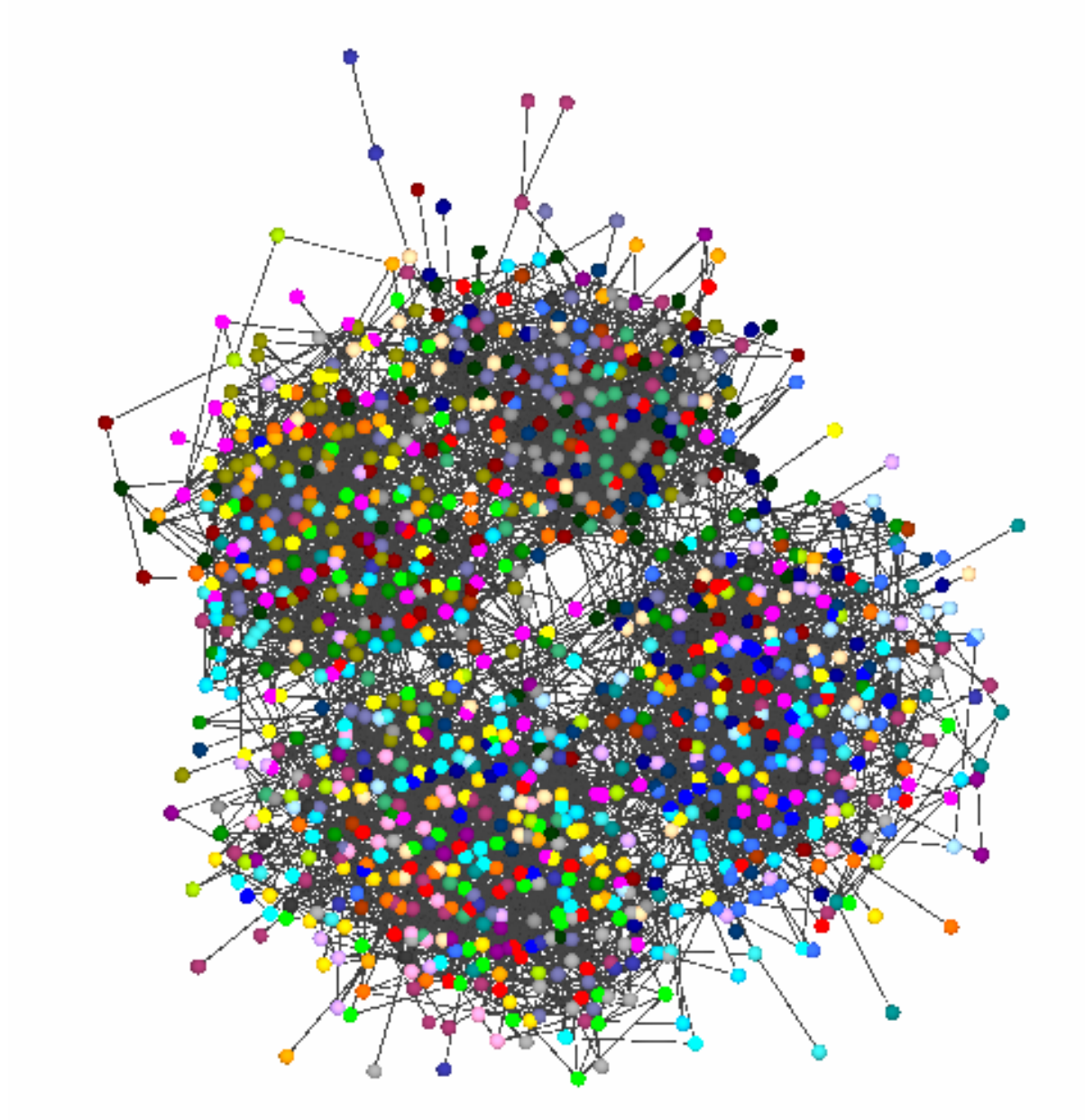

**PerturLand, X=0.2 – level 0**
206 modules (123.17)

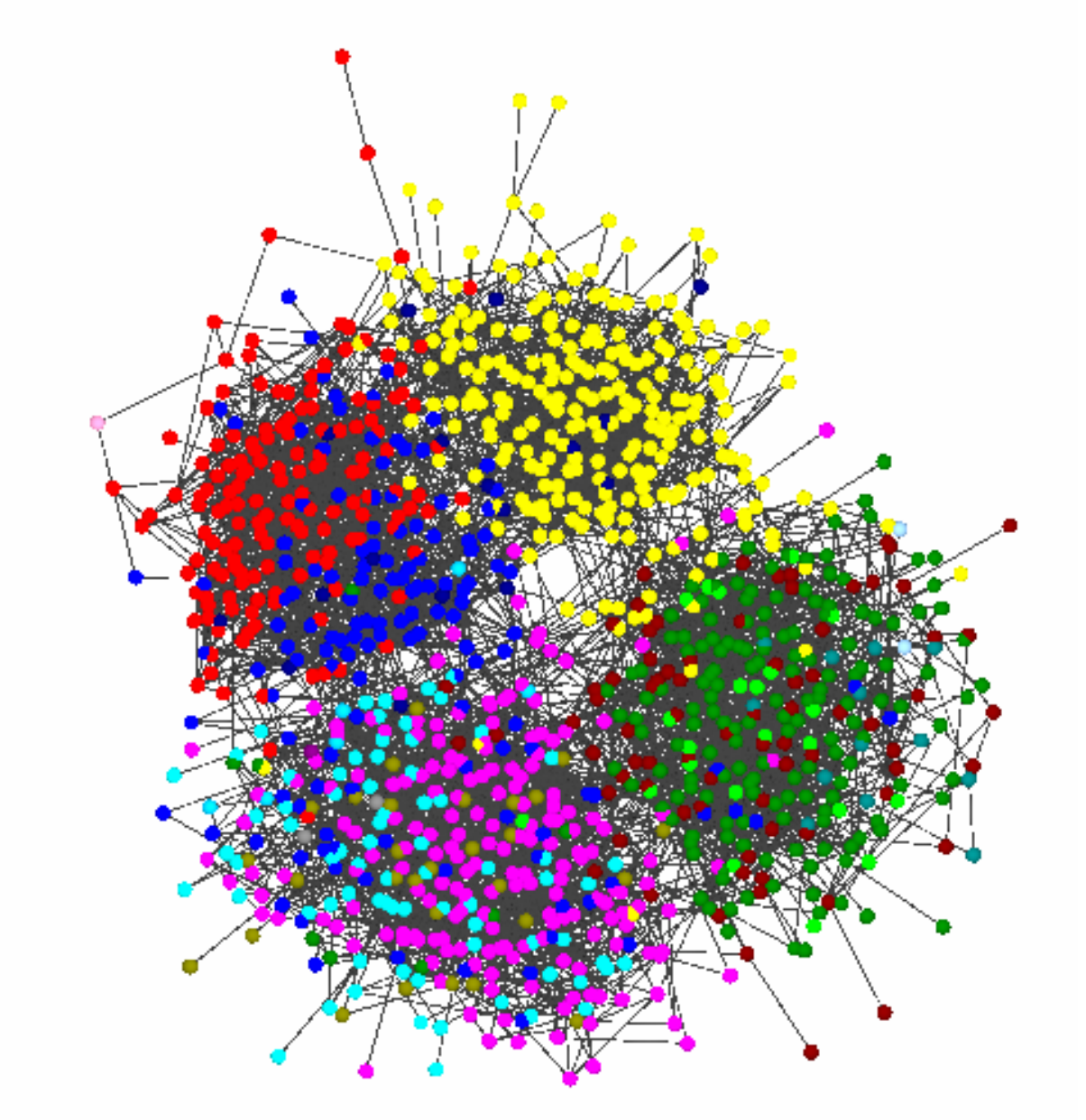

**PerturLand, X=0.2 – level 1**
18 modules (9.34)

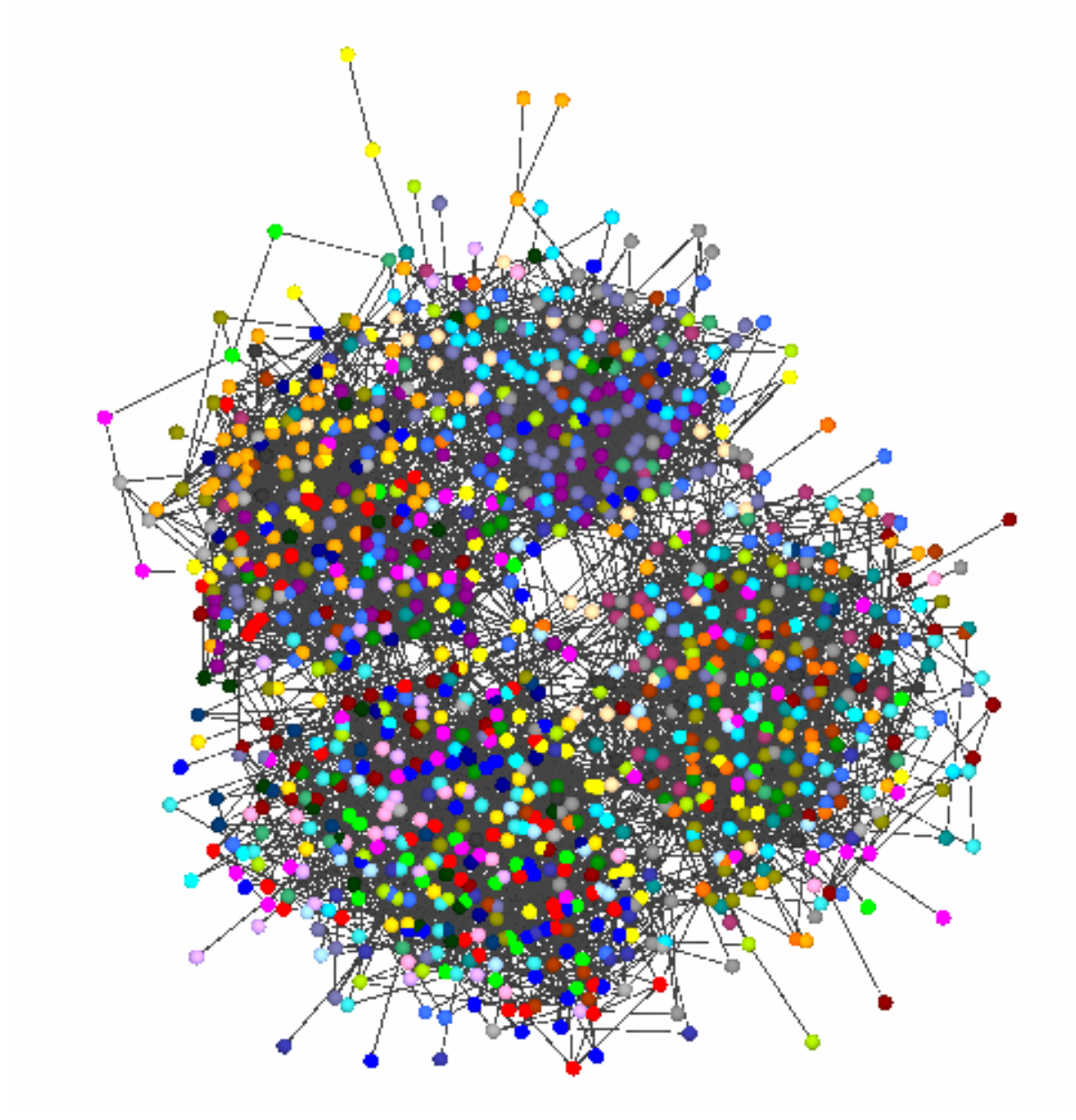

**PerturLand, X=0.5 – level 0**
170 modules (91.03)

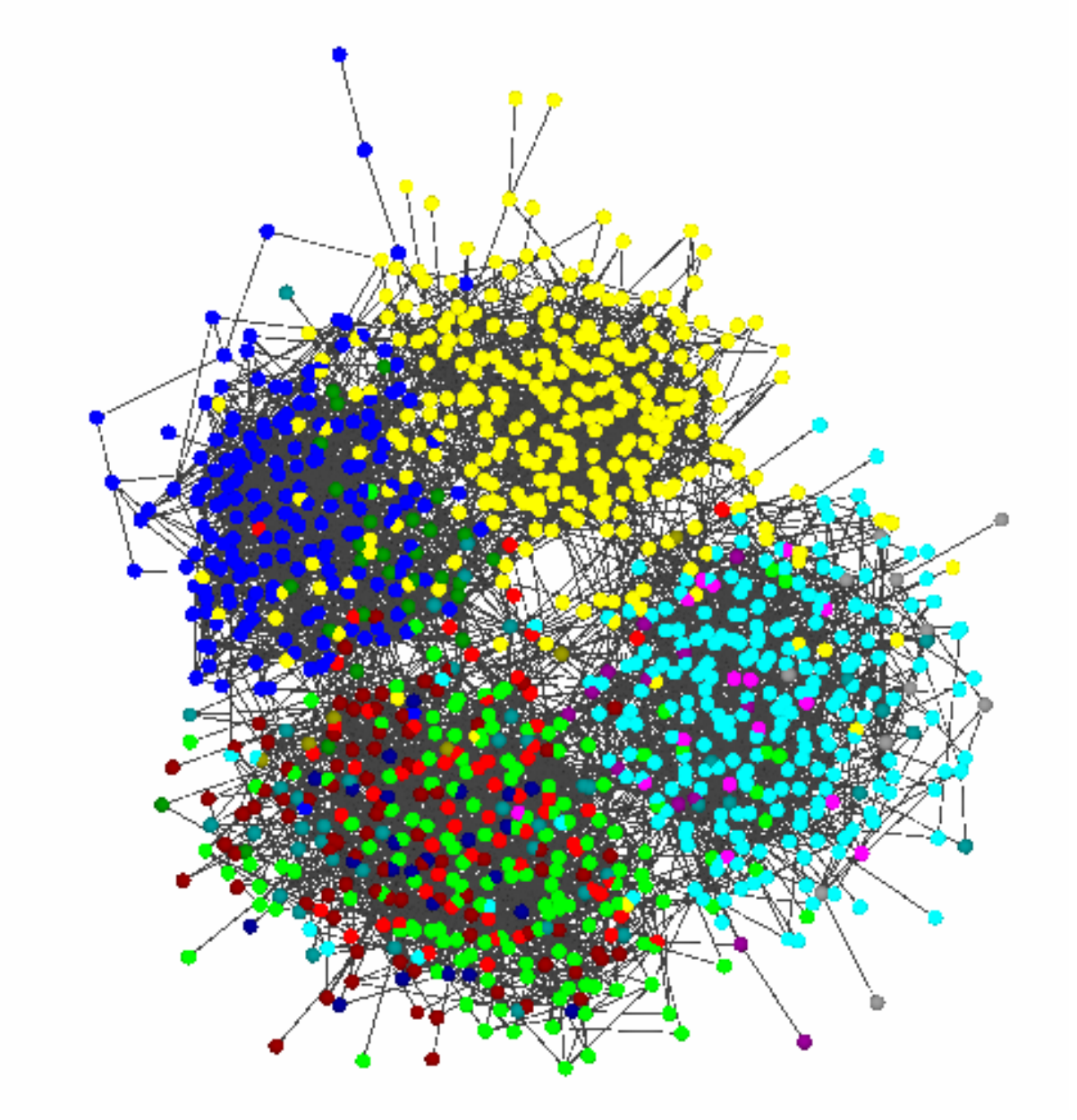

**PerturLand, X=0.5 – level 1**
14 modules (9.24)

## Figure S13. Robustness of the ModuLand method in recovering modules of benchmark graphs.

Panels **(a)** and **(b)** show the correspondence of the identified modules to the modules of the benchmark graph of Lancichinetti et al. (2008). It can be seen that modules of the benchmark graph are consistently identified until the point, while modules can be defined in the strong sense ($\mu<0.5$). Representations of the benchmark graph were generated with degree and module size distribution exponents $\gamma=2$ and $\beta=1$ (Panel **a**) or $\gamma=2$ and $\beta=2$ (Panel **b**). The number of nodes was N=1000, the maximum degree was $K_{max}=50$, the average degree varied as K=15 (blue rectangles), K=20 (yellow triangles), K=25 (red diamonds), and the network fuzziness ($\mu$ of the x-axis of Panels **a** and **b**) was ranging from 0.1 to 0.6 (x-axis), where $\mu>0.5$ means that the modules are no longer defined in the strong sense. Higher normalized mutual information (shown on the y-axis) represents a better recovery of the original modules. Modules were identified using the NodeLand influence function calculation algorithm and the ProportionalHill module membership assignment method with merging highly correlated modules using an arbitrary chosen correlation threshold of 0.9 (see Section VI.1.). The figure shows the averaged results of 100 representations. Panels **(c)** and **(d)** show the effect of choosing different correlation thresholds for merging correlated modules. Representations of the benchmark graph were generated with degree and module size distribution exponents $\gamma=2$ and $\beta=1$, respectively, the number of nodes was N=1000, the maximum degree was $K_{max}=50$, the average degree was K=20, and the network fuzziness was set as $\mu=0.5$. Data show the average of 100 representations. Panel **(c)** shows the relative frequency (y-axis) of the given module-module correlation values (binned x-axis, bin size=0.1). It can be seen that the majority of module-pairs are weakly correlated while higher correlation is less probable. However, the increased probability of extremely high correlations indicates the existence of nearly identical artifact modules, which artifacts can result from using a local maxima-based hill determination method on a noisy community landscape. Panel **(d)** shows the effect of removing artifacts by merging correlated groups of modules into single modules above a given correlation threshold (x-axis) on the module identification accuracy measured *via* the normalized mutual information (y-axis). It can be seen that a threshold too low results in the erroneous merger of distinct modules, while a threshold too high misses to merge some artifact modules. Generally, the correlation threshold for merging modules may be chosen by inspecting the frequency curve of module-pair correlations and setting the threshold to merge the modules of extremely high correlation.

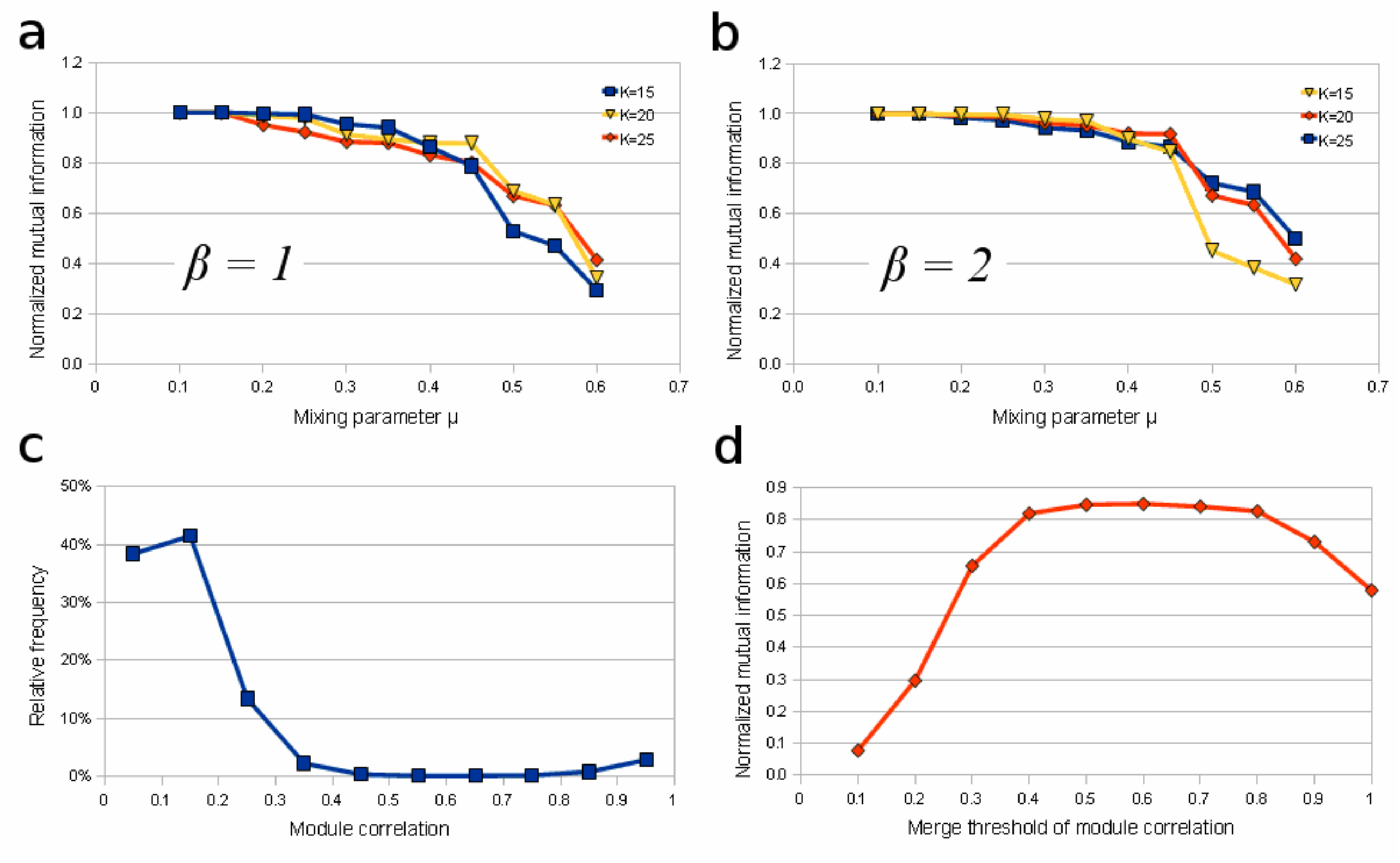

**Figure S14. Functional modules of the yeast protein-protein interaction network.**
Overlapping modules of the yeast protein-protein interaction network of Ekman et al. (2006) were identified using the LinkLand influence function calculation algorithm together with the TotalHill module membership assignment method. During the post-processing of the module assignment, we merged the modules with ProportionalHill module membership assignment-based correlation higher than 0.9 (similar results were obtained without the merging step; data not shown). The modular structure of the lowest level of hierarchy is shown. The underlying 2D network layout was set by the the Kamada-Kawai algorithm. The vertical positions reflect the centrality values of the nodes on a linear scale. Nodes were colored as the module of their maximum membership. The modular functions were assigned by the functions of the core modular proteins having at least 50% of the centrality of the local maximum of the module. In all cases these core-proteins showed a functional consensus. The functional labels and their arrows have the similar colors to their respective modules.

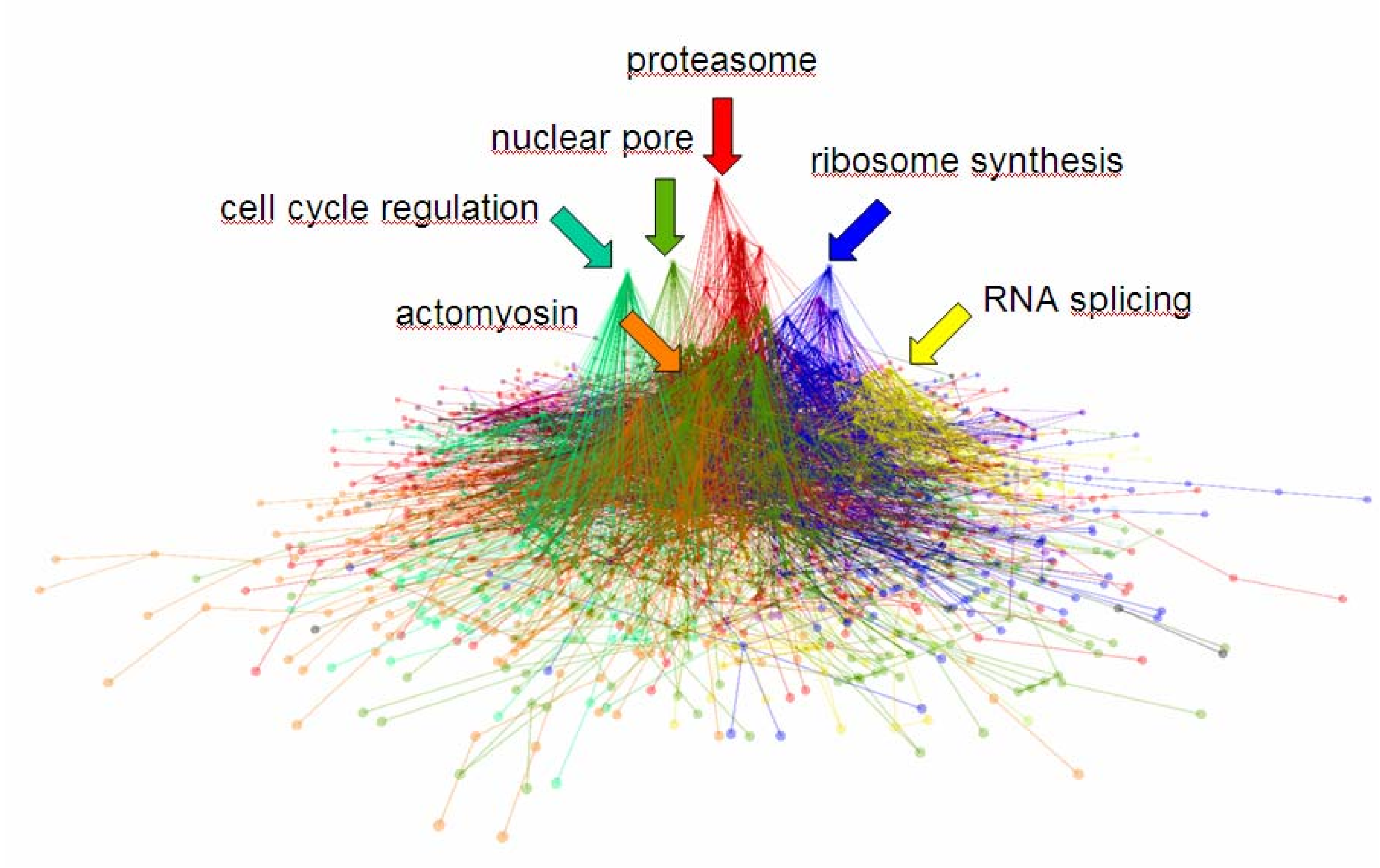

# Supplementary Methods

## I. Networks used for the analysis

### *1. Network science collaboration network*

This network is the giant component of the undirected network science collaboration network as compiled by Mark J. Newman (2006a) containing 379 nodes (network scientists) and weighted 914 links between them. Here a link represents a joint publication between two authors, and the weight is proportional with the number of these co-authorships. The original network data is available at the web-site: http://www-personal.umich.edu/~mejn/netdata/. The network data can also be downloaded from our web-site: <www.linkgroup.hu/modules.php>.

### *2. Zachary karate club social network*

We analyze the weighted and undirected social network of a karate club as recorded by W. Zachary from 1970 to 1972 (Zachary, 1977). This network has 34 nodes and 78 links, where nodes were (initially) members of a karate club, and links are weighted with their interaction strength. The karate club network is a favored test-case for community detection, because in the course of recording the network data, members of the karate club have split into two factions. These factions provide insight about the natural communities of the karate club network. The original network data is available at the web-site: http://vlado.fmf.uni-lj.si/pub/networks/data/Ucinet/UciData.htm#zachary. The network data can also be downloaded from our web-site: <www.linkgroup.hu/modules.php>.

### *3. Word association network*

We have used the University of South Florida word association network (http://www.usf.edu/FreeAssociation/), where 6,000 participants produced nearly three-quarters of a million responses to 5,019 stimulus words. This word association network gives a relative strength for each stimulus-response word pair, calculated by taking into consideration the count of associations to the response word given the count of stimuli by the stimulus word: The relative weight of an A → B link (called forward strength, FSG) is expressed as FSG = P/G, where G is the count of people who received the word A as the stimulus, and P is the count of people among them who responded with word B for that stimulus. Based on this data, a weighted and directed network can be built. While the direction of links provide insight to the complexity of human conceptual thinking, in the present study we considered the fact of association between words, and built an undirected network. Therefore, the parallel forward and backward links were collapsed into a single non-directed link, and weighted with the sum of the original weights. This process on the giant component of Appendix A of the University of South Florida word association network resulted in a weighted and undirected network. In this study we analyzed the largest connected component of this network consisting of 10,617 nodes (English words) and 63,788 links (associations) between them. The network data can be downloaded from our web-site: <www.linkgroup.hu/modules.php>.

The antagonym, heteronym and homonym words chosen for our investigation were selected from lists found on the following websites:

- http://www-personal.umich.edu/~cellis/heteronym.html
- http://www-personal.umich.edu/~cellis/antagonym.html
- http://sb058.k12.sd.us/multiple%20meanings/multiple_meaning_words.htm

### *4. School-friendship network*

We used the data of the high-scale Add-Health survey, which mapped social connections of high schools of the USA (Gonzaléz et al., 2007; Moody, 2001, Newman, 2003).[3] In the survey recorded between 1994 and 1995 social connections of 90,118 students in 84 schools were recorded. For each friend named, the student was asked to check off, whether he/she participated in any of five activities with the friend. These activities were:

1. you went to (his/her) house in the last seven days;
2. you met (him/her) after school to hang out or go somewhere in the last seven days;
3. you spent time with (him/her) last weekend;
4. you talked with (him/her) about a problem in the last seven days;
5. you talked with (him/her) on the telephone in the last seven days.

[3]This research uses data from Add Health, a program project designed by J. Richard Udry, Peter S. Bearman, and Kathleen Mullan Harris, and funded by a grant P01-HD31921 from the National Institute of Child Health and Human Development, with cooperative funding from 17 other agencies. Special acknowledgment is due Ronald R. Rindfuss and Barbara Entwisle for assistance in the original design. Persons interested in obtaining data files from Add Health should contact Add Health, Carolina Population Center, 123 W. Franklin Street, Chapel Hill, NC 27516-2524 (addhealth@unc.edu).

Based on these data, connections were assigned with weights from 1 to 6. A nomination as friend already resulted in a weight of one, and each checked category added one to that weight. In addition to the nomination data, these files include the gender, race, grade in school, school code, and total number of nominations made by each student.

In our study we analyzed one of the preferably examined school friendship network of the database, the Community-44 school network, because it contains a high number of students with a dense social network (Newman, 2003). This network has an approximately equal number of black and white students. The network contains 1,147 students with 6,189 directed links between them. In our current study directed parallel links were merged into a single undirected link with weight equal to the sum of the original weights and only the largest connected component of the network was used. This process resulted in a weighted undirected network consisting of 1,127 nodes and 5,096 links with weight between 1 and 12. The network data can be downloaded from our web-site: <www.linkgroup.hu/modules.php>.

## *5. Electrical power-grid of the USA*

In our studies we used the unweighted and undirected USA Western Power Grid network as an example from the field of engineered networks (Watts and Strogatz, 1998). The power grid network has 4,941 nodes and 6,594 links, and is a favored network for studying error propagation and the effect of malicious attacks. The original network data were downloaded from the website of Prof. Duncan Watts (University of Columbia, http://cdg.columbia.edu/cdg/datasets), which was not available at the submission of this paper. Data can also be found at the Pajek website (http://vlado.fmf.uni-lj.si/pub/networks/data/map/USpowerGrid.net). The network data can also be downloaded from our web-site: <www.linkgroup.hu/modules.php>.

## *6. Yeast protein-protein interaction network*

We used the unweighted and undirected yeast protein-protein interaction network assembled by Ekman et al. (2006) consisting of 2,633 nodes and 6,379 links covering approximately half the proteins of yeast genome. We analyzed the largest connected component of the network consisting of 2,444 nodes and 6,271 links. Besides the high confidence of its data, we have chosen this network, because it has been involved in the identification of party and date hubs, an interesting dynamic feature of protein-protein interaction networks (Ekman et al., 2006). The network data can be downloaded from our web-site: <www.linkgroup.hu/modules.php>.

## II. Overview of the ModuLand network module determination method family

The ModuLand network module determination method family is constructing a community landscape of networks, where the hills and highlands mark high community densities (corresponding to module-cores), and the valleys between them show the approximate positions of overlaps between network communities. In the closing step, the ModuLand method gives a network representation of the overlapping modules, and by a recursive process uncovers a hiearchical network enabling a zoom-in analysis and visualization of large networks. The ModuLand method family is an integrative method consisting of the following four major steps (see main text Figure 1).

1) Determination of influence functions belonging to each node or link, and calculation of a value (called influence function value) for every links and/or nodes showing how much the given link is affected by the influence function of the starting node or link.
2) Construction of community landscape, where the community landscape height of a link of the network is a value calculated by taking into consideration all influence function values, which were assigned to the given link in step 1.
3) Determination of overlapping network modules by finding the hills and highlands on the community landscape and assigning all links and nodes to them.
4) Constructing the community hierarchy of the network using the original or higher level modules as nodes of the second or higher layers of the community hierarchy, respectively.

All the four steps contain variable parameters. The appropriate way of calculating the influence functions may vary from network to network, depending on our information about the complex system and on the interaction, which we wish to analyze. In step 1) the influence function values may give a simple representation of the network communities setting the influence function value of those nodes or links as a constant larger than zero, which are affected by an influence function and zero to all others. However, more sophisticated influence function value structures can also be designed, and there is a wide variety of possible assignment criteria for the determination of the boundaries of effect (or gradual decay) of influence functions. It is also possible that we calculate the influence functions for an assembly of starting nodes or links (e.g. cliques or motifs) and not for individual network nodes or links what the detailed procedures of this paper will describe. From now on we assume that the influence function values are non-negative. (Numerous specific cases can be extended to other influence function values, including negative values, or even complex numbers, too.)

In a simple and straightforward version of step 2) the influence function values may be summed for each node or link. However, other community landscape construction methods may also give satisfying results. For instance, if the influence function value of the given node or link represents the local deviation of a measure, where the deviation is caused by the starting node of the influence zone,[4] the community landscape height may be calculated as the resulting local deviation of all influence functions based on the correlation matrix of their starting nodes. In the case when the impacts of the influence functions are uncorrelated, the result is the square root of the sum of the influence function value squares. Thus, the summation we use in the implementations shown in this paper is only one of the many alternative influence function integration possibilities. From now on we assume that the community landscape is constructed by summing up the influence function values for each link.

In step 3) we suggest to use one of the local maxima-based algorithms, which start the determination of community landscape hills by finding their hill-tops (or highlands). The rules of node or link assignment to the hill-tops/highlands may vary as we will describe in detail.

Importantly, several versions of the ModuLand methods give a continuous scale for the modular 'distribution' of all nodes and links in the network marking a fraction of the given node or link belonging to various modules. This gives a much more detailed representation of the network structure than the usual yes/no answers, which unequivocally assign a node or link to one module (see Table S2). The ModuLand methods, which include local maxima-based hill determination, do not require any previous knowledge on the possible number of network modules. Moreover, the ModuLand method family gives a series of hierarchical modular representations, where the hierarchical topology of the network is gradually uncovered from bottom to top. Top hierarchical levels give a bird's eye approach of the network, and thus provide better overview, while lower levels contain more and more detailed views of the network. Several ModuLand methods – such as the NodeLand, LinkLand and PerturLand algorithms described in Sections IV.1. and IV.2., respectively – do not use strictly local or global information of the network for the determination of network communites, but explore many scales of network topology giving less and less weight of network segments being further and further away from the actually examined node or link of the community landscape. It is important to remark, that the main steps of the ModuLand method, which perform well for weighted undirected networks, are also applicable for weighted directed networks (see e.g. the directed PerturLand algorithm in Section IV.2.).

[4]The influence zone is a subgraph of the network, which contains the starting node (or link, or network segment), and all other nodes (or links), which have an influence function value of the starting node (or link) higher than zero.

In Section IV.3. we will show two extreme cases of community landscapes describing one of the most and the least stringent method of community landscape construction. We will outline three specific representations of the ModuLand method family, the NodeLand, LinkLand and PerturLand algorithms, as well as the ModuLand adaptations of the well known community detection methods of Girvan and Newman (2002; 2004) and Palla et al. (2005) named as BetweennesCentralityLand (BCLand for short) and CliqueLand methods, respectively. Before starting all these, in the following sub-sections of this Section we list a few basic definitions and considerations we followed during our work, and will describe the 4 major steps of the ModuLand method in detail.

## III. Starting considerations and definitions

### *1. Terminology*

- ***Using all nodes/links.*** We have considered all (known) nodes/links of the networks, and did not introduce any threshold values for excluding any segment of the network, except the elimination of self-loops. The reason for eliminating self-loops is that we are interested in the true interactions between nodes.
- ***Non-existing links.*** Non-existing links (or non-identified, missing, hidden, 'secret' links) could be regarded as links having a zero weight.
- ***Influence function.*** Based on the direct interactions in the network we calculate the effective, indirect impact of the starting node to the rest of the network. In our work we represent this new indirect interaction as a property of the original links of the network, so we calculate a new weight, the so-called influence function value $f_i(j,k) \geq 0$ for every link (*j*, *k*) from a given starting node *i*. Let the influence function $f_i$ of node *i* affect a link (j,k) if $f_i(j,k) \neq 0$, and let the influence function affect a node *n* if any link of node *n* is affected. Let the influenced links be the links affected by the influence function, and let the influenced nodes be the nodes affected by the influence function. Let the influence zone denote the set of influenced links and nodes.[5] The influence zone (in our current, restricted interpretation) is a connected subgraph, surrounding the starting node in which the influence function values are all larger than zero. The influence function may also belong to a pair of nodes, or from a link (like in the LinkLand algorithm), or more generally from any selected motifs or subgraphs. Since the precise form of the effective interaction depends on the meaning of the links in the network, in principle each network may have its own, unique optimal influence function calculation algorithm. The optimal influence function calculation algorithm may differ from network to network, but for most networks general versions of both fast and accurate methods can be designed. Different versions of these generally applicable influence function calculation algorithms of the ModuLand method family will be described later in detail, and adequate hints for their optimal use will also be given.
- ***Influence function value.*** The influence function value is a non-negative number rendered to every link of the original network showing how much the given link is affected by the influence function of the starting node or link. The influence function values depend on the influence function calculation algorithm.
- ***Community landscape.*** Integrating all influence function values, which were assigned to a given link, we get a centrality-type value called as the community landscape height of that link. The community landscape height shows how much the given link is affected by the integrated indirect impact of all the starting nodes of the network. From now on in this paper the community landscape is simply defined as the sum of the influence function values. We usually represent the community landscape as a 3 dimensional image of the original network, where the horizontal plane is a 2 dimensional, 'usual' representation of the network, while on the vertical axis the community landscape values of network links are plotted.

### *2. Input data of the ModuLand method family*

The ModuLand method family requires a list of at least unweighted and undirected links between the nodes of the network. The ModuLand method family can be applied for weighted and directed links as well without increased resource requirements. The present versions of the ModuLand method family accommodate only a single type of links with non-negative weights, where a weight is greater, if the connection is stronger between the endpoint nodes. The available network data do not always fulfill these conditions, therefore a data conversion may be required. This is the case, if multiple types of links (colored graphs), links with negative weights, links with weights meaning distances between the nodes or multi-node interactions (hypergraphs) are present. There is no universal recipe for converting these systems into a suitable weighted network, but basically weights representing endpoint similarity are advised.

For the present implementation of the ModuLand method family, linear link weights are necessary, as we will see at the introduction of the PerturLand algorithm (Section IV.2.). In this case, parallel links of the same direction between the given nodes can, and in practice should be merged, where the resulting weight will be the sum of the original weights. We note, that loop links are discarded, as we are only interested in links between nodes.

In the usual case, the implementation of the ModuLand method family requires a detailed information on the links between the network nodes. However, we note that in case we do not know the links between the nodes, but do know certain features of each node, the ModuLand method family can also be applied by defining a (non-negative, linear) similarity metric between the network nodes. Treating these similarity values as link weights, a network representation can be derived.

---

[5]The 'influence zone' defined here can also be named as a community heap. When in the algorithms, or in the User Guide we refer to a 'community heap' or 'heap' we refer to the influence zone defined here.

The principle of the ModuLand method family is the possibly most precise determination of the influence functions, therefore any further information determining the effective interactions is preferred. As an example of such an information, the *activity* of a node ($p_i$) can be taken into account by the ModuLand method family, if available. The activity of a node represents the relative strength of the influence function of the given node. If we know the activity, the influence function values of the respective node are multiplied by the activity of that node. This option has high relevance, if we wish to investigate the behavior of a complex system under different circumstances, defining different activity of their nodes. For example, news or an epidemic may not spread with equal probabilities from all nodes, and this presumption can be modelled by assigning different activities to the nodes.

When applying the ModuLand method family, nodes in different connected components are always assigned into componentwise distinct sets of modules, therefore it is practical to analyze different connected components separately. This procedure is also useful by sparing computational resources.

## IV. Determination of the community landscape

In a typical version of the ModuLand method family the community landscape is constructed by a specific integration of the influence functions of the network. However, there are situations, where the community landscape of the network is directly known instead of knowing the link-structure of the network. In these cases the influence function calculation step can be omitted, because we may start directly from the community landscape to determine the modules.

We define the effective, indirect interactions as influence functions. In principle, an optimal method of influence function calculation is a unique method for each given network, which may differ from network to network. However, different, generally applicable versions of influence function calculation algorithms can be designed, and will be described here in detail. In all these methods the influence function value of a node for a given influence function means the indirect effect on that given node.

In the current applications it is not our aim to model the network dynamics, but to give static prediction of modules based on a static, or time interval-averaged description of the network. Therefore we only consider influence function calculation algorithms, which result in time-independent, static indirect interactions, but we note that many nodes of network dynamics may also be included to the framework we describe here.

### *1. The NodeLand and LinkLand influence function calculation algorithms for weighted and undirected networks*

The NodeLand and LinkLand algorithms are fast, but approximating methods for the determination of the influence functions in weighted, undirected networks.

In the generalized version of the NodeLand and LinkLand algorithms the influence function belonging to the starting node or link is determined by a network walk. During this walk the neighboring nodes and links of the starting node or link are explored, and this procedure is continued and the influence zone is extended until an appropriately selected criteria is fulfilled. Examples for the these criteria are given below, and will be listed in Section IV.1.a. for the NodeLand algorithm, Section IV.1.b. for the LinkLand algorithm, and Section IV.4. for the BetweennessCentralityLand (BCLand) and CliqueLand methods.

In the case of NodeLand and LinkLand algorithms, the calculation of the influence function is governed by the so-called density, which is defined as ([the sum of the weights of the influenced links] / [number of influenced nodes]).

As an example for the above criteria, the algorithms of the NodeLand and LinkLand algorithms follow the principle of ‘the density of the growing set of influenced nodes and links is not allowed to decrease’. This principle is extended further in the NodeLand algorithm by the conjuncture that ‘maximal growth is preferred’. These principles result in the following illustrative behavior of influence zone growth.

- If the starting node of the influence zone was at a connection-poor part of the original network, the influence zone will extend towards more and more dense parts of the network. Having found a connection-rich region, the growth of the influence zone stops.
- Conversely, if the starting node of the influence zone was at a connection-rich part of the original network, the extension of the influence zone will stop very soon, since there will be only such nodes in the further neighborhood of the initial influence zone, with which the density could only be decreased.

As a result of the above behavior illustrated on Figure 1A of the main text, nodes and links belonging to a local connection rich region will be affected by many influence functions, while nodes and links of connection-poor segments will not be ‘attractors’ of influence zone growth, and will only be affected by very few influence functions.

In the end, after creating the community landscape by combining (in the methods detailed in the present paper simply summing up) influence functions, connection-poor segments of the network will have lower community landscape height, and thus will be valleys of the community landscape, while connection-rich segments will have high community landscape height, and thus will form hills of the community landscape.

#### a. The NodeLand influence function calculation algorithm

In the NodeLand influence function calculation algorithm the starting point of each influence zone is a node of the original network. The starting node and later, its growing influence zone are extended by only that neigboring node and its links linking it to the existing influence zone, which will increase the density (as defined above: [the sum of the weights of the influenced links] / [number of influenced nodes]) of the existing influence zone, and this increase will be maximal among all the possible increases supplied by any of the neigboring nodes. If more than one nodes exist, which fulfill the above criteria, all of them are added to the influence zone, including the links connecting these nodes both with each other and with the existing influence zone. If additional nodes are added to the influence zone of the starting node, the density of the influence zone generally increases, making the chances of node additions to the influence zone from any further rounds of neighboring nodes even more difficult. If there are no neighboring nodes fulfilling the above, rather stringent requirements, the influence function of the starting node is considered to be ready, and the method continues with the determination of the influence function belonging to the next node of the original network. **Algorithm 1** below describes the influence function calculation of the NodeLand algorithm in detail.

**Algorithm 1: Algorithm of the NodeLand influence function calculation algorithm**

```
/*
Important variables used in the algorithm:

        startNode: the starting node.

        infNodeList: influenced nodes (initially empty).

        infLinkList: influenced links (initially empty).

        tempList: nodes to be added to influenced nodes in the next round.

        actualDensity: sum of the weight of all links in infLinkList divided by the number of nodes in
        infNodeList.
*/
tempList := [ startNode ]
while tempList is not empty {
        add all nodes of tempList to infNodeList
        for each link e connected to any nodes of tempList {
                if endpoints of e are already in infNodeList { add e to infLinkList }
        }
        clear tempList
        recalculate actualDensity
        maxNewDensity := actualDensity

        for each node n not in infNodeList but having non-zero links lks with an endpoint in infNodeList {
                newDensity := sum weight of links in infLinkList + sum weight of link in lks
                newDensity := newDensity / (number of nodes in infNodeList + 1)
                if newDensity > maxNewDensity {
                        clear tempList
                        maxNewDensity := newDensity
                }
                if newDensity = maxNewDensity { add n to tempList }
        }
        if maxNewDensity = actualDensity {
                // cannot increase the threshold any more, stop algorithm
                clear tempList
        }
}
```

In the end of **Algorithm 1**, we find the nodes and links of the influence zone in the **infNodeList** and **infLinkList** lists of **Algorithm 1**, respectively. Identifying the influence function of one node in the NodeLand algorithm is structurally similar to a breadth-first search, therefore the runtime complexity of the algorithm is O(n(n+e)), where *n* is the number of nodes and *e* is the number of links in the network. However, in practice the algorithm is extremely fast as an influence zone of any given node rarely covers the whole network.

For downloading the ModuLand program package including the NodeLand influence function calculation algorithm of **Algorithm 1** as the **nodeland** program see our homepage <http://www.linkgroup.hu/modules.php>.

In the NodeLand algorithm the influence function value of a link in the influence zone is set as the original weight of this link, while the influence function value of a link not belonging to the influence zone is set as zero. Furthermore, if the activity of the nodes is known, then all influence function values are multiplied by the activity of the starting node of the influence zone, as described earlier in Section III.2.

The community landscape value of each nodes or links are constructed by summing up the influence function values of that node or link for all influence functions. We note that in case of the NodeLand influence function calculation algorithm, the community landscape height value of non-zero weighted, but locally weak links may become zero, if that given link does not belong to any influence functions.

The NodeLand algorithm resembles to the *l*-shell method of Bagrow and Bollt (2005) with the important difference that in the NodeLand algorithm the influence zones resembling the *l*-shells are later summarized in order to create the community landscape. The NodeLand algorithm can be easily applied for directed networks, too. In this case we take into consideration only the outgoing links from the existing influence zone by determining the next nodes of the influence zone.

**b. The LinkLand influence function calculation algorithm**

The LinkLand influence function calculation algorithm gives a less stringent method for the determination of network modules than the NodeLand algorithm described in the previous section. In agreement with this property, the NodeLand algorithm usually gives smaller modules and faster results than the LinkLand algorithm.

In the LinkLand influence function calculation algorithm the same density is used as in the NodeLand algorithm. Thus, the density is defined as ([the sum of the weights of the links belonging to the original influence zone] / [number of nodes in the influence zone]). The starting point of each influence zone is a link of the original network with its two end-nodes. Since the starting link of the influence zone connects its two end-nodes, we define the starting density as half of the original weight of the starting link. The starting link and later its growing influence zone are extended by those neigboring nodes and their links linking them to the existing influence zone and to each other, which will at least not decrease the density of the existing influence zone. Please note, that the LinkLand algorithm neither requires a direct increase in the density, nor asks for the maximal increase among all the possible increases, which were the characteristic features of the NodeLand algorithm.

To see the meaning and scope of the LinkLand algorithm from an other point of view, we can define the influence function connection strength of a given node in the network as the sum of the weights of those links, which are connecting the node to the existing influence zone. With this definition, we can express the LinkLand algorithm as follows. The growing influence zone is extended by those neigboring nodes, which have an influence function connection strength to the existing influence zone *at least equal to* the density of the influence zone.

After all eligible neighboring nodes and their linking links have been already added to the influence zone of the starting node, the density of the influence zone is re-calculated. If there are no further neighboring nodes having an equal or higher influence function connection strength than the density of the influence zone, the influence function of the starting node is considered to be ready, and the method continues with the determination of the influence function belonging to the next link of the original network.

**Algorithm 2** below describes the LinkLand influence function calculation algorithm in detail.

**Algorithm 2: Algorithm of the LinkLand influence function calculation algorithm**[6]

*/**
*Important variables used in the algorithm:*

***startLink****: the starting link of the actual influence zone.*

***heapNodeList****: nodes of the influence zone (initially empty).*

***heapLinkList****: links of the influence zone (initially empty).*

***tempList****: nodes to be added to the influence zone in the next round.*

***actualHeapThreshold****: sum of the weight of all links in* **heapLinkList** */ number of nodes in* ***heapNodeList****.*

****/***

[6]The 'heap' of the variables in this description refers to the influence.

```
clear templist
add the two end-nodes of startLink to tempListwhile tempList is not empty {
        add all nodes of tempList to heapNodeList.
        for each link e connected to any nodes of tempList {
                if endpoints of e are already in heapNodeList { add e to heapLinkList }
        }
        clear tempList
        recalculate actualHeapThreshold
        maxNewHeapThreshold := actualHeapThreshold

        for each node n not in heapNodeList but having non-zero links lks with an endpoint in  heapNodeList {
                newHeapThreshold := sum of the weight of all links in heapLinkList + sum weight of link in lks
                newHeapThreshold := newHeapThreshold / (number of nodes in heapNodeList + 1)
                if newHeapThreshold > maxNewHeapThreshold {
                        clear tempList
                        maxNewHeapThreshold := newHeapThreshold
                }
                if newHeapThreshold = maxNewHeapThreshold { add n to tempList }
        }
}
```

In the end of the **Algorithm 2**, we find the links and nodes of the influence zone in the **heapLinkList** and **heapNodeList** lists of **Algorithm 2**, respectively. Identifying the influence function of one link in the LinkLand algorithm is structurally similar to a breadth-first search, therefore the runtime complexity of the algorithm is O(e(n+e)), where $n$ is the number of nodes and $e$ is the number of links in the network. However in practice the algorithm is very fast as an influence zone of any given link rarely covers the whole network.

For downloading the ModuLand program package including the LinkLand influence function calculation algorithm of **Algorithm 2** as the **linkland** program see our homepage <http://www.linkgroup.hu/modules.php>.

While the influence function value of a link outside the influence zone is set to zero, the influence function value of the link in the influence zone is calculated by multiplying the weight of the actual link by the weight of the starting link of the influence zone. Thus influence functions with a starting link of significant weight gain higher influence on the community landscape, and communities of non-existing links (or links of zero weight) are automatically excluded from influencing the community landscape.The LinkLand influence function calculation algorithm could be regarded as taking pairs of nodes instead of links as starting point of the influence zones. From this it follows that the LinkLand algorithm will assign a zero influence function value to all links of the graph, which belong to the influence zones of a pair of starting nodes having no link between them.

In case the activity of the nodes is known, then in the LinkLand algorithm the influence function values are multiplied by the average activity of the endpoint nodes of the starting link of the respective influence zone.

The community landscape value of each nodes or links are constructed by summing up the influence function values of that node or link for all influence zones. We note that, unlike the case of the NodeLand construction method (where the community landscape height value of non-zero weighted, but locally weak links may become zero, if that given link does not belong to any influence zones), in the LinkLand algorithm the community landscape height value of a link with non-zero, positive weight will always be a non-zero, positive value.The LinkLand influence function calculation algorithm can be applied for directed networks as well by only considering the outgoing links for determining the neighboring nodes of the growing influence zone in the influence function calculation process.

## *2. The PerturLand influence function calculation algorithm for weighted and directed networks*

In the PerturLand influence function calculation algorithm the indirect effect of a given starting node is modelled as the amount of information reaching other nodes and links of the network spreading out from the given starting node. The piece of information spreading from the starting node remains ‘identical’ throughout the spreading process, which can be envisioned as the very same piece of information reaching the other nodes, but with decreasing reliability or confidence. Therefore, the maximum incoming and outgoing information quantity of a given node (and not the sum information quantity) is the measure of how much information the given node receives from the network and sends to the network, respectively. The initial link weights (or more precisely, these values multiplied by the value of the $X$ parameter, as seen later) give the ratio of the information, which is spreading from the starting node to the ending node. The initial information content of the nodes is uniformly one. If we know and use the activity (defined in Section III. 2.) of the nodes, it corresponds to the initial condition, where the initial information content of a node is the activity of that node. The PerturLand algorithm assumes the link weights to be in the range of [0;1]. If the link weights are not in the range of 0 to 1, the link weights have to be normalized. In the followings $s_{ij}$ will denote the link weight, if the link$(i,j)$ exists, and $s_{ij}=0$, if no such link exists.

The influence functions of different starting nodes are independent from each other. From their independence it follows that the influence functions can be determined in any order. The information-spread determining the influence function generally corresponds to a perturbation spreading process. In this process a perturbation (disturbance, impulse, information) spreads from the starting node through directed links, while its effect gradually weakens. If a unit of the perturbation is propagating from node $a$ through the $(a,b)=e$ link towards node $b$, then the quantity of perturbation arriving to node $b$ from node a is $s_{ab}X$, where $X$ is a parameter describing the attenuation of the perturbation during the propagation process. $X$ can be in the range of [0;1]. The higher the value of $X$, the smaller the attenuation of the perturbation is spreading in the network. The value of $X$ should be chosen by taking into account the properties of the analyzed network. $X$ can be determined by estimating the information transfer efficiency between the network nodes. The formal definitions and rules of the perturbation spreading process are given below. In the example detailed by these formalism the starting point is a node. In principle the starting point may also be a link, or a set of nodes and/or links.

We define the vector $P[i]$ to store the maximum quantity of perturbation that reached node $i$. Initially

$$P[i] = \begin{cases} 0, & i \neq a, \\ p_i, & i = a, \end{cases}$$

where $p_i$ ($0<p_i\leq 1$) is the activity of node $a$, where $a$ is the starting node of the perturbation. During the perturbation spreading process the values of the vector $P[i]$ are iteratively modified. The alteration of the vector $P[i]$ in one round of iteration can be described as

$$dP[b] := \begin{cases} \max_i(P[i]s_{ib}X), & P[b] = 0 \\ 0, & P[b] \neq 0 \end{cases}$$

where the $X$ parameter is the attenuation coefficient of the perturbation as described before. Component $b$ of the vector $dP$ is the quantity of the perturbation that would reach node $b$, which was perturbation-free in the previous round of the iteration process, if the perturbation would flow towards node $b$. Knowing these values, perturbation only spreads towards the nodes, which would be maximally effected by the perturbation. Matrix node $M[j][m]$ denotes perturbation flowing from node $j$ to node $m$ through the link$(j,m)$, if there is such link, otherwise $M[j][m]$ is equal to zero. Thus matrix $M$ keeps an account of the perturbations flowing in the network. Initially, matrix $M$ is filled with zeros.

$$M[j][m] := P[j]s_{jm}X,\ \forall m : dP[m] = \max_i dP[i], P[j] > 0$$

After determining the quantity of perturbation flowing from the influence zone to the newly reached nodes, the quantity of perturbation flowing into these newly reached nodes can be calculated:

$$P[m] := dP[m],\ \forall m : dP[m] = \max_i dP[i]$$

After merging these new nodes $m$ to the influence zone, the perturbation spreads on the $(m,n)$ links sourcing from these new nodes to other nodes $n$ in the influence zone (including other newly merged nodes), and the following update reflects this spreading:

$$M[m][n] := P[m]s_{mn}X,\ \forall m : dP[m] = \max_i dP[i], P[m] > 0$$

Please note that the normalization of link weights, the range of the $X$ attenuation parameter and the max-greedy nature of the perturbation process described here together yield that the $P[i]$ perturbation quantity of node $i$ never increases once it was set to a non-zero value. This statement can be explained by the procedure first transforming the initial perturbation quantity and the ($s_{mn}X$) perturbation affinities of links by taking their negated logarithmic values, applying Dijksta's shortest path algorithm (Dijkstra, 1959) and finally transforming back the resulting perturbation quantities, which yields the same result as the perturbation process described here.[7] It is known that Dijkstra's algorithm will never update the value of a node after updating it for the first time (Dijkstra, 1959). Similarly, the *M[m][n]* perturbation flow from node *i* to *j* will never increase once it has been set to a non-zero value.

The above steps describe one round of the iterative process for determining the perturbation quantity of each link and node of the network. The iteration stops once all components of the vector *dP* become zero, meaning that the non-zero (but maybe infinitesimally small) perturbation quantity reaching each link and node is already determined. At the end of the iteration process component *i* of vector *P* is the maximum perturbation quantity flowing into node *i*. The influence function value, $f_a(i,j)$ of any link *(i,j)* to the influence of node *a* (the starting node of the perturbation spreading process) is defined as the perturbation quantity flowing through link *(i,j)*, that is

$$f_a(i,j) = M[i][j].$$

Let the influence function value of node *i* be $f_a(i,j)=max_j\{M[i][j]\}$, i.e. the maximum outflowing perturbation quantity of node *i*. With this defintion, the determination of the influence function is completed. This process of determining influence functions of a directed network describes the PerturLand influence function calculation algorithm. Once all influence functions have been determined, the community landscape height of a given link is calculated by summing the influence function values of the given link for all influence zones. Note that the *X* parameter is not required for constructing the influence functions of higher hierarchical level networks, because these networks already have link weights proportional to the information flow rate on these links (see Section VII. for details).

While the PerturLand algorithm is defined in the above desription for directed networks only, it can easily be applied in case of undirected networks by substituting undirected links with two directed links of opposing direction having the same weights as the original link. In this version of the PerturLand algorithm the community landscape height of a given undirected link is defined as the maximum of the community landscape height of the substituted directed links.

**Algorithm 3** below describes the PerturLand influence function calculation algorithm in detail.

**Algorithm 3: Algorithm of PerturLand influence function construction method**

*/**
*Input variables:*

> ***startNode:*** *the starting node of the actual influence zone.*
>
> ***startPerturbation:*** *the starting perturbation (activation) of startNode*
>
> ***X:*** *the intensity of the perturbation flow (free parameter, chosen with regards to the dynamical properties of the given network, 0<X≤1)*
>
> ***normWeight[i,j]:*** *the normalized weight of the directed i→j links*

*Important variables used in the algorithm:*

> ***maxInPert[i]:*** *the maximal inflow perturbation of node i*
>
> ***maxOutPert[i]:*** *the maximal outflow perturbation of node i*
>
> ***tempMaxInPert[i]:*** *the maximal inflow perturbation reaching node i, if it would be added to the influence zone in the next round*
>
> ***newNodes:*** *the list of the nodes to be added to the influence zone*
>
> ***linkPert[i,j]:*** *the perturbation spreading trough the directed i→j link*

*Output variables:*

> *In the end of the algorithm, the influence function values of the nodes will be stored in the* ***maxOutPert*** *array, and the influence function values of the directed links in the* ***linkPert*** *array*

[7]Although the behavior of the PerturLand algorithm for determining the centrality of nodes can be described by calculating shortest paths via the said transformations, the centrality measure provided by the PerturLand algorithm greatly differs from the betweenness centrality, because PerturLand assigns centrality to nodes based on properties of shortest paths, while the betweenness centrality characterizes a centrality based on the number rather than the properties of shortest paths.

```
*/
clear maxInPert, maxOutPert, tempMaxInPert, linkPert variables (set all nodes of the arrays to zero)
tempMaxInPert[startNode] := startPerturbation

do {
        clear list newNodes
        newNodes := the positions of the max values in tempMaxInPert

        for each node i of the newNodes array {
                // setting the influence zone -> newNode links
                For each each j->i link where j not in newNodes array but
                maxInPert[j]>0, so node j already belongs to the influence zone {
                        linkPert[j,i] := X * maxInPert[j] * normWeight[j,i]
                }
                // setting the newNodes
                maxInPert[i] := tempMaxInPert[i]
        }

        for each each node i of the newNodes array {
                // setting the newNode->influenced links (including the cross-links between
                // the newNodes)
                for each i->k link where i is the given newNode and
                maxInPert[k]>0, so node k already belongs to the influence zone; and
                linkPert[i,k] is zero so the given link does not belong to the influence zone {
                        linkPert[i,k] := X * maxInPert[i] * normWeight[i,k]
                }
        }
        clear array tempMaxInPert (set all nodes to zero)
        // check the non-influenced neighbors of the influenced nodes
        for each i->j link where maxInPert[i]>0 and maxInPert[j] is zero {
                tmp := X * maxInPert[i] * normWeight[i,j]
                if tmp > tempMaxInPert[j] then { tempMaxInPert[j] := tmp }
        }
} while there are any non-zero nodes of the tempMaxInPert array

// setting the maxOutPert array
for each node i of the network {
        for each i->j link {
                if maxOutPert[i] < linkPert[i,j] then {
                        maxOutPert[i] := linkPert[i,j]
                }
        }
}
```

The runtime complexity of the PerturLand community landscape determination method as given by **Algorithm 3** is O(ne) per influence function, where *n* is the number of nodes and *e* is the number of links in the network. However, there is space for more optimal algorithms, since the result of perturbation flow process can either be calculated with a modified Dijkstra algorithm (Dijkstra, 1959) effectively enabling a runtime complexity of O(e + n log n) per influence function, or by the Floyd–Warshall algorithm (Floyd, 1962) enabling a runtime complexity of $O(n^3)$ for the whole community landscape construction process.

For downloading the ModuLand program package including the PerturLand influence function calculation algorithm of **Algorithm 3** as the **perturland** program see our homepage <http://www.linkgroup.hu/modules.php>.

## *3. Two extreme cases of centrality landscapes for weighted and directed networks*

We continue the explanation of the ModuLand method family by showing two extreme examples having one of the most, and one of the least stringent assignment criteria (see Figure S3 for an illustration).

One of the most stringent methods of the determination of the influence functions is, if only the links starting from the node $A$ will form the influence zone of node $A$ and their influence function values will be equal to the weight of the given link. With this definition, in the directed case the summation of the influence function values will give the original weights of all links as their community landscape height (in the undirected case the community landscape height will be twice as much for each link).

One of the least stringent methods for the determination of influence functions, if we take all nodes or links as parts of the influence zone, which are in the same connected subgraph than our starting node or link, and we set their influence function values all equal, e.g. 1. Summing up the influence function values at the end of the assignment procedure we will get all connected subgraphs as plateaus of the community landscape with a community landscape height equal to the number of nodes or links they contain.

These two extreme examples do not seem to be suitable for community detection, but at least show the general applicability of the ModuLand method for the analysis of network topology. However, in specific original networks (depending on the meaning of the given links in the network) even these extreme methods may be appropriate for the construction of the community landscape.

## *4. Transformation of widely used, former modularization methods to the centrality landscape framework of the ModuLand method*

Numerous former module detection methods can be implemented in the ModuLand method family framework. In this process the former module detection gains the ability to detect modular overlaps even in the case, when this property was not among its original features. Furthermore, as a result of the implementation, the number of modules can be obtained automatically, instead of a previous choice or a pre-set parameter tuning. The implementation process may often result in the community landscape directly omitting the influence function calculation step. The following CliqueLand and BetweennessCentralityLand (BCLand) community landscape construction methods are such examples.

### a, The CliqueLand community landscape determination method for unweighted and undirected networks

The CliqueLand method is an adaptation of the widely used, efficient clique percolation method (named as CFinder by the authors) of Tamás Vicsek and his collegaues (Adamcsek et al., 2006; Farkas et al., 2007; Palla et al., 2005; 2007a) for the determination of overlapping network modules using the terminology of the ModuLand method family. In the original version of the clique percolation method (Palla et al., 2005) links below a suitably selected threshold are left out from the analysis, and the residual links have a Boolean representation giving a weight of 1 to the link, if it exists, and zero otherwise.[8] The nodes and links of the modules determined by this method form $k$-clique communities meaning a union of all $k$-cliques that can be reached from each other through a series of adjacent $k$-cliques (two $k$-cliques are adjacent, if they share $k$-1 nodes). The clique percolation method determines the $k$-clique communities of the network for all $k$ values, and selecting the optimal $k$-clique distribution treats these interconnected $k$-cliques as network communities. Two interconnected $k$-clique sets may contain $k$-2 common nodes, which makes the clique percolation method very suitable for the detection of overlapping network communities.

We may define a community landscape showing the modules of the clique percolation method in frame of the ModuLand method family. Let the community landscape height of a given link be equal to the maximal value of $k$, for which there exists a $k$-clique-community containing this link. If there is no $k$-clique containing the link, than the community landscape height is set to 1.

[8]Although we here only consider the original version of the clique percolation method for unweighted and undirected networks (Palla et al., 2005), it is important to note that the authors of the clique percolation method have later introduced important and efficient extensions of the original method, which are able to identify overlapping modules in directed (Palla et al., 2007a) and weighted (Farkas et al., 2007) networks as well.

**b, The BetweennessCentralityLand (BCLand) community landscape determination method for weighted and directed networks**

The BetweennessCentralityLand (BCLand) method is an adaptation of the 'gold-standard' network module determination method of Girvan and Newman (2002; 2004) using the terminology of the ModuLand method family. The Girvan and Newman (2002; 2004) method first calculates the betweenness centrality of the network links (the number of shortest paths containing the given link), then iteratively deletes the link of highest betweenness centrality, and recalculates the betweenness centralities of the residual network. This way the original network is decomposed into non-overlapping components.

A community landscape over the links can be assigned to the Girvan-Newman method (Girvan and Newman, 2002; 2004), which is iteratively removing links in the order of their decreasing betweenness centrality. Let the community landscape height $c_{ij}$ of the link*(i,j)* be k *(k>0)*, if the given link was removed from the network in the *k*-th iteration. The community landscape of this betweenness-based method is illustrated on Figure S3.

**c, Methods yielding partitions of the network**

In addition to the previously described community landscape implementation of the Girvan and Newman (2002; 2004) method, traditional agglomerative or divisive modularization methods, or any methods yielding partitions can be implemented as community landscape-based methods, which determine modules by applying a threshold on the community landscape.

For example, in the case of agglomerative methods, the later a node or link is merged to the already existing clusters, the lower their community landscape height can be set. Figure S4a gives an illustration of the community landscape derived from an agglomerative method. In this interpretation of Figure S4a the positions of the horizontal lines correspond to cuts on the dendrogram of a hierarchical modularization method. The modules resulting from these horizontal cuts are the connected components of the original network corresponding to the community landscape segments above the horizontal lines (see the centrality threshold-based hill determination method in Section V.1.). While the number of modules depends on the position of the horizontal lines (the cuts on the dendrogram), the same method implemented in the frame of the ModuLand method family with one of the local maxima-based hill determination methods (see Section V.2.) determines the number of modules automatically, and may yield overlapping modules, if needed.

**d, Stochastic module detection methods**

Methods yielding multiple possible partitions or yield partitions by heuristic optimalization of some measure (see Table S2) can easily result community landscapes by defining the community landscape height of the link*(i,j)* to be the number of cases, in which the endpoints *i* and *j* belong to the same module, taking into account the possible module structures or repeated stochastic runs.[9] Note that the different module structures do not need to be equally important, for example structures may be weighted by a fitness measure or probability function of the given structure.

## *5. Summary of the community landscape determination methods*

In the examples above we showed that two of the widely applied previous methods, the method of Girvan and Newman (2002; 2004) finding distinct network communities as well as the method of Palla et al. (2005) finding overlapping network communities may be implemented as a wide sub-group of methods within the ModuLand network module determination method family. We also showed how several classes of other, previously described modularization methods of Table S2 may also be transcribed to the frame of the ModuLand method. Furthermore, we showed three illustrative examples, that the NodeLand, LinkLand and PerturLand algorithms use adaptively broader and broader scales of topological information to result in a community structure. We note, that our new methods (the NodeLand, LinkLand and PerturLand) experience no problem to detect the communities of *both* very low and very high density networks. Moreover, these methods can be efficiently applied to networks having any type of hierarchical topology including the simultaneous presence of very dense and very sparse segments.

Obviously, many other influence function calculation algorithms of Table S2 such as those based on the *k*-clan, *k*-club, *k*-plex and other community definitions of Table S1 can be used in other variants of the ModuLand method. It is also possible that we construct the influence functions for an assembly of starting nodes or links (e.g. cliques or motifs) and not for individual nodes or links what we described here. Moreover, different influence function summation protocols and community landscape exploration methods can also be designed. The borders of the influence zone must be defined by different algorithms for different types of networks depending on the meaning of the links. We believe that many more methods of the ModuLand family exist beyond the shown methods, which are optimal for one or another type of networks and/or community structures. We invite our colleagues to the exciting journey to explore and try them including – potentially – such methods, which are better for one or another type of networks than those we described in our current study.

[9]This idea can also be found in the work of Sales-Pardo et al. (2007), altough the authors do not describe their method as a community landscape.

## V. Determining modules based on the community landscape

In this step we have to find the hills of the community landscape representing the overlapping communities of the original network. In the case of local maxima-based hill determination methods described in this section, we first identify the module-cores, i.e. the links or connected plateau of links (hills or highlands), which have a local maximum on the community landscape (for illustration, see Figure S4). For the assignment of links (and nodes) to module-cores we may use any module membership assignment methods described in Section V.2., such as the GradientHill, the ProportionalHill or the TotalHill methods.

As it can be seen on Figure S4, it is also possible to define modules by simple horizontal cuts on the community landscape. First, we will describe this method in detail, then we will continue with the identification of local maxima, and finally we will introduce three module membership assignment methods (the GradientHill, the ProportionalHill and the TotalHill methods) yielding overlapping modules.

### *1. Horizontal cut of the centrality landscape as a detection threshold of previous community detection methods*

To represent the modularization methods yielding partitions of the network in the ModuLand method family we discuss the centrality threshold-based hill determination method. Most previous clustering or divisive methods of network community detection (see Table S2) can be represented as a horizontal cut of the community landscape. Cutting the community landscape at an appropriately selected height, these methods treat the connected subgraphs above (or occasionally: below) this horizontal plane as network modules (see Figure S4). If we treat the nodes having no links above the horizontal cutting height as separate modules, we have assigned all nodes to the community structure. These methods result in distinct network modules (having no overlaps) like in case of the widely used Girvan and Newman method (Girvan and Newman, 2002; 2004).

The horizontal cut model (making a visually conceivable meaning for the detection limit of network communities) helps us to explain why conventional methods are successful only seldom in finding big and small communities at the same time. These methods either raise the detection limit too high, and find only the largest communities, or set the detection limit too low, where most of the overlapping, large communities are already merged. This is called as the giant-component problem of the community detection methods, since by setting the detection limit too low, the modules collapse to a single giant-component (Berry et al., 2009; Fortunato, 2007; Fortunato and Barthélemy, 2007; Kumpula et al., 2007). The effect of the gradual shift in the detection limit on the development of more and more details of the modular network structure can be nicely followed in the hierarchical clustering methods. Another solution to find both small and large communities, if we simplify the network by leaving out the links below an appropriately selected arbitrary link weight threshold (Palla et al., 2005). This network simplification makes the communities more isolated, and enables to lower the detection limit to see smaller communities but leaving larger communities still more-less separated showing only a reasonably minor overlap. As we will describe in Section V.2., local maxima-based hill determination methods solve the giant-component problem, since they find the smaller and larger modules simultaneously.

If we modify the community landscape height values by a strictly increasing or decreasing transformation, the network modules determined by a horizontal cut of the community landscape will not change. However, in the case of local maxima-based hill determination methods the same strictly increasing or decreasing transformations can significantly modify the overlaps between the network modules on the community landscape. This property of the community landscape gives an another example, how a whole family of ModuLand methods can be designed around a previously applied, horizontal cut-type method, such as that of Girvan and Newman (2002; 2004) as well as Palla et al. (2005).[10]

### *2. Local maxima-based methods*

In this, closing step of the ModuLand method family we will first describe procedures to find the hill-tops/highlands of the networks as module-cores, and then we will assign all remaining links of the network to these hill-tops/highlands constructing a complete procedure, where the community landscape height of the link will be fully distributed between the module-cores (now: modules) it belongs. We will describe three different methods for the assignment of links, the proportional, the gradient and the total distribution methods naming them ProportionalHill, GradientHill and TotalHill methods, respectively.

[10]In order to get the *k*-clique-communities as hills based on the CliqueLand method, we have to modify the hill definition of the centrality threshold based hill determination method. In the CliqueLand method the connected subgraph above the selected community landscape height threshold may contain more than one hills. Therefore, we define the *k*-connected subgraph of the community landscape above the threshold as the 'hill', i.e. as the module of the CliqueLand method. (In *k*-connected subgraphs all nodes are reachable through a series of adjacent *k*-cliques; see Section IV.4.a.) Hills defined by this definition may have overlaps.

In the case of local maxima-based hill determination methods, the number of hill-tops/highlands will be equal to the number of network modules. Therefore these methods need no previous assumption on the number of network modules to give a final result.

Let us naïvely illustrate the ProportionalHill, GradientHill, TotalHill and other module membership assignment methods as a paint-flow process on the community landscape. Let us allow paints flow down from hill-tops pouring differently colored paints on each different hill-top. With this process the whole community landscape will be painted at the end. With appropriate flow rules certain links may be painted by more than one color. These links will represent the overlapping links between the modules.

**a. Finding the hill-tops and highlands of the community landscape**

We give the definiton of the hill-tops (or highlands) of the community landscape as follows.

*Directed networks:* The centers (or cores) of modules are defined as the hill-tops of the community landscape. A hill-top is either a link, whose outbound links have a smaller community landscape height [by the outbound links of a link*(i,j)* we mean the outbound links of node *j*, the end-point of link*(i,j)*], or a strongly connected component (this means that every node is reachable from each other node by a directed walk on this component) consisting of links with equal community landscape height, whose outbound links have a smaller community landscape height.

*Undirected networks:* The hill-top of the community landscape contains all connected links, which have the same community landscape heights, and whose neighboring links all have lower community landscape heights than theirs.

**Algorithm 4** describes the method for finding of community landscape hill-tops or highlands in detail.

**Algorithm 4: Algorithm to find the hill-tops or highlands of community landscapes**

/*
***As a result of the algorithm, network links will be assigned to the four different categories below:***

> ***'c':*** *This link is a module-core link (a hill-top of the community landscape).*
> *Please note that in the current implementation a module core consisting of multiple links still has just one link marked as the module-core link. This link the representative link for that core.*
>
> ***'h':*** *This link is part of a highland (a module-core consisting of several links).*
>
> ***'s'****: This link is a small link having at least one neighboring link, which has a higher community landscape height.*
>
> ***'e':*** *This link is "equal", having no neighbors with higher community landscape height, but at least one neighboring link with an equal community landscape height. Such a link can be a part of a highland forming a module-core, or part of a plateau. This is a temporal type, since the OldBoyWalk procedure will determine, whether such links are of type 's' or 'h'.*

*Important variables used in the algorithm:*

> ***maxNeighbor:*** *the community landscape height of the highest neighboring link (zero, if there is no neighboring link).*
>
> ***height(e):*** *the community landscape height of link e.*
>
> ***type(e):*** *the type of link e as described above.*

*/

```
for each link l in the network {
        if l has no neighbor { maxNeighbor := 0 }
        else { maxNeighbor := maximum of height(l') for l' in neighbors of link l }

        if height(l) = maxNeighbor{ type(l) := 'e' }
        else if height(l) < maxNeighbor{ type(l) := 's' }
        else if height(l) > maxNeighbor{ type(l) := 'c' }
}

// let OldBoyWalk resolve the type of "equal" links
for each link l in the network {
        if type(l) = 'e' { Call the OldBoyWalk procedure on link l }
}
```

**Description of the OldBoyWalk procedure:** the OldBoyWalk procedure behaves like a fitt, quick old man with a cradle. OldBoyWalk can walk everywhere on a field containing links with equal community landscape heights, but cannot step higher or lower than the field. After exploring the whole plateu, OldBoyWalk checks every neighbor of the plateau. If OldBoyWalk finds at least one neighboring link having a higher community landscape height than that of the plateau, sets every link of the plateau to type **'s'**, otherwise sets every link of the plateau to type **'h'**.

This hill-top detection method is incorporated into the downloadable implementation of the ProportionalHill and TotalHill membership assignment methods. For downloads, examples and usage instructions please see our homepage <http://www.linkgroup.hu/modules.php>.

The directed version of the local maxima-based hill determination method differs from the undirected version in two points. First, for a given directed link$(i,j)$ only the links outbound from node $j$ are regarded in the assignment process, because these links can be regarded as neighbors in the directed case. Second, the module cores identified by the OldBoyWalk procedure are strongly connected, that is each link of a module core can be reached from any other link of the same module core by a directed walk.

**b. The ProportionalHill and the GradientHill module membership assignment methods (undirected and directed versions)**

In the ProportionalHill module membership assignment method network links are assigned to modules of their non-lower neighboring links in the proportion of the absolute community landscape height of the respective neighboring links. The already mentioned module-core sets of links are assigned with full weight to the respective modules defined by themselves. In the GradientHill module membership assignment method a link is assigned to modules of the highest neighboring links of the given link.

In the start of the ProportionalHill module membership assignment method all links are marked as unassigned. After this, multiple rounds of link-assignments are performed: in all rounds, links are assigned to modules based on the assignment of previously assigned links. In each round, we descend to next slice of links, starting from the top community landscape slice. In this procedure a community landscape slice is formed by all links having the same community landscape height.

Here we describe the steps of a single round of the ProprotionalHill module membership assignment method:

- The first step: the hill-tops/highlands of the community landscape are marked, with the understanding that each of them becomes a new module-core. Each link of all these connected components are assigned to their own modules with an assignment-strength of their full community landscape height.
- In consecutive steps, unassigned links of the next descending community landscape slice, having at least one neighboring link already assigned to the growing modules, are assigned to modules proportional to the assignment-strength of their neighbors already assigned to existing modules. In such a step, links assigned in the current step are not considered as 'assigned neighbors' during the whole length of the respective step. The step described here is repeated until there are any unassigned links remained at the actual community landscape slice. Once all links of the actual community landscape slice have been already assigned to modules, the round is over, and the next round begins, unless there are no more (lower) community landscape slices left, where the whole assignment procedure ends.

As an outcome of the ProportionalHill module membership assignment process, for each link the sum of the assignment-strength values of the given link is equal to the community landscape height of the link.[11]

The GradientHill module membership assignment method is structurewise very similar to the ProportionalHill module membership assignment method described above.[12] However, in the GradientHill module membership assignment method we do not assign the links of the given round proportionally to the community landscape heights of all their neighbors assigned previously, but only to those neighbors, which have the maximal community landscape height among all neighbors. Obviously, this assignment procedure may also give an assignment of a new link to multiple network modules (if the new link has more than one neighbors having an equally maximal community landscape height), but produces much smaller modular overlaps than the ProportionalHill module membership assignment method. We suggest to apply the GradientHill module membership assignment method, if large overlaps between various modules are not interesting, or are not feasible.

**Algorithm 5** below describes the ProportionalHill and GradientHill module membership assignment methods for link assignment in detail.

**Algorithm 5: Algorithm of the ProportionalHill and GradientHill module membership assignment methods**

*/**
*Important variables used in the algorithm:*

> ***tmpEqualList:*** *Contains all links of the actual slice. (A slice of the community landscape is a set of links, where all links have the same community landscape height.)*
>
> ***actualLinkList:*** *Contains links set to be assigned to a module.*
>
> ***markedList:*** *Links from which the module membership of the current link is calculated.*
>
> ***height**:**(e):*** *Community landscape height of the link e.*
>
> ***neighLinks(e):*** *List of neighboring links of link e.*
>
> ***moduleMembership[e][m]=x**: link **e** belongs to module **m** with a strength of **x**.*
> *In the beginning all **moduleMembership[… ][… ]** values are set = 0, except module core links, whose module membership for the respective module is set to the height of the link.*

**/*

***cycle** on the slices in the descending order of the community landscape height of slices **{***

```
        put every link of the given slice to tmpEqualList
        while length of tmpEqualList > 0 {
                clear actualLinkList
                for each link e of tmpEqualList {
                        if e is not yet assigned to any modules and e has at least one assigned neighbor {
                                add e to actualLinkList
                        }
                }
```

[11]We may set the summation of assignment-strengths equal the weight of the starting, module-core link, by multiplying the assignment-strengths by the weight of the starting link and divide by the community landscape height. Henceforth we use only the assignment-strengths fulfilling the requirement that the sum of these assignment-strengths equals the community landscape height of a link. (Please note that in the current implementation a module core consisting of multiple links still has just one link marked as the module-core link – this link the representative link for that core.)
[12]Therefore, the GradientHill method is not found as a separate method in the algorithm package.

```
            // module membership assignment
            for each link e of actualLinkList {
                    // module membership assignment
                    clear markedList
                    if using the Proportional method {
                            for e' in neighLinks(e) {
                                    if height(e') >= height(e) { add e' to markedList }
                            }
                    }
                    else if using the Gradient method {
                            maxh := maximum of height(e') for e' in neighLinks(e)
                            for e' in neighLinks(e) {
                                    if height(e') = maxh { add e' to markedList }
                            }
                    }

                    sum_nHeight := sum of height(e') for e' in markedList
                    for m :=1..number of modules {
                            for e' in markedList {
                                    delta := moduleMembership[e'][m] / sum_nHeight * height(e)
                                    increment(moduleMembership[e][m], delta)
                            }
                    }
            }
            remove all links of the actualLinkList from tmpEqualList
        }
}
```

The runtime complexity of the ProportionalHill and GradientHill module membership assignment methods is O(edm), where $e$ is the number of links of the network, $d$ is the average degree of nodes and $m$ is the number of modules. Assuming practically that the average degree is bounded by a constant and that the number of modules is not more than the number of nodes, the runtime complexity is $O(n^3)$.

For downloading the ModuLand program package including the ProportionalHill module membership assignment method see our homepage: <http://www.linkgroup.hu/modules.php>.

*The ProportionalHill and GradientHill module membership assignment methods for directed networks.* The basic idea of the ProportionalHill and GradientHill module membership assignment methods was to assign a link to the modules of the neighboring (and in terms of community landscape height, non-lower) links of the given link. The directed and undirected cases have the single difference that in case of a given directed link*(i,j)* only the links outbound from node $j$ are regarded in the assignment process, because these links can be regarded as neighbors in the directed case.

**c. The TotalHill module membership assignment method for undirected networks**

In the TotalHill module membership assignment method the assignment of module-cores is performed as described previously for the ProportionalHill module membership assignment method, but when assigning a non-core link to modules of the neighboring links in proportion of the community landscape height of the neighboring links, the neighboring links of both non-lower and lower community landscape height are regarded. This module membership assignment method is especially important, since it yields the most detailed information of the network module structure.

From now on we use the following notations:

$s_{ij}$: the original weight of the link*(i,j)*

$m_{ij}$: community landscape height of the given link*(i,j)*

$b_{ij}[k]$: is defined only for not module-core links. $b_{ij}[k]$ is the module assignment strength of link$(i,j)$ to the $k$-th module giving us the value, that how strongly link$(i,j)$ belongs to the $k$-th module.
$a_{ij}[k]$: is the module assignment strength of link$(i,j)$ to the $k$-th module, if the link$(i,j)$ is the core of the $k$-th module. $a_{ij}[k]$ is the community landscape height of the given link$(i,j)$, if the link$(i,j)$ is the core of the $k$-th module, otherwise $a_{ij}[k]$ is the zero vector. This vector does not change during the procedure. For module-core links $b_{ij}$ does not exist.
$d_{ij}[k]$: the module assignment strength of link$(i,j)$ to the $k$-th module: for a module-core link$(i,j)$, $d_{ij}$ equals $a_{ij}$, and for non-core links, $d_{ij}$ equals $b_{ij}$.

In the following steps we will get an equation system with the number of variables equal to the non-core links of the network. Note that the $b_{ij}$ variables are not scalar, but vector values with the vector length equal to the number of modules. However, the components are independent of each other, so the equation system can be decomposed into equation systems, which have scalar-variables, and can be solved independently. Therefore for the sake of simplicity we write simply $b_{ij}$ instead of $b_{ij}[k]$ in the equation.

Unlike the directed case, in the undirected case each quantity is symmetric for index swapping, so $b_{ij}=b_{ji}$. For non-core links:

$$b_{ij} = m_{ij}\frac{\sum_{k,k\neq j} d_{ki} + \sum_{l,l\neq i} d_{lj}}{\sum_{k,k\neq j} m_{ki} + \sum_{l,l\neq i} m_{lj}}$$

Module-core links are exceptions, these links are assigned to the respective modules with their full height. If the link$(i,j)$ is the core of module $k$, then $a_{ij}[k] = m_{ij}$, while other components of $a_{ij}$ are zero. For non-core links $a_{ij}$ is the zero vector. In the current implementation, the community landscape height of a link is distributed between modules, and a node is assigned to modules with a quantity equal to the sum of the assignment quantity (now community landscape height) of the links of the given node. Let $A$ be the set of module-core links, $B$ the set of non-core links of non-zero community landscape height. The main point here is, that altough the equation system has the same form for all modules, the $A$ and $B$ sets depend on the choosen module.

Now we define the module assignment strength vectors of nodes based on the module assignment strength of links. Let we use the

$$a_i \equiv \sum_{k,(i,k)\in A} a_{ik}$$
$$b_i \equiv \sum_{k,(i,k)\in B} b_{ik}$$
$$d_i \equiv a_i + b_i = \sum_k d_{ik}.$$

notations.

Now we give an alternative form of the described equation system, where variables are assigned to nodes instead of links. After solving the equation system, the links can be assigned to modules based on the assignment of the nodes.

$$b_{ij} = m_{ij}\frac{\sum_{k,k\neq j} d_{ki} + \sum_{l,l\neq i} d_{lj}}{\sum_{k,k\neq j} m_{ki} + \sum_{l,l\neq i} m_{lj}} = \\ = \frac{m_{ij}}{m_i + m_j - 2m_{ij}}(d_i + d_j - 2b_{ij})$$

The module membership assignment equations can only be written for non-core links. We can rewrite the equation so that the module membership assignment of links can be easily calculated, given that the module membership assignment of nodes is known:

$$b_{ij} = m_{ij}\frac{d_i + d_j}{m_i + m_j} = r_{ij}\left(d_i + d_j\right)$$

where we introduced $r_{ij} = \frac{m_{ij}}{m_i + m_j}$. With

$$b_i = \sum_{j,(i,j)\in B} b_{ij} = \sum_{j,(i,j)\in B} r_{ij}\left(d_i + d_j\right)$$

we give a form of the equation system, where variables are assigned to nodes. Now $b_i$ can be written as

$$b_i = \eta_i + \sum_{j,(i,j)\in B} \rho_{ij} b_j,$$

$$b = (I-\rho)^{-1}\eta,$$

$$\eta_i = \frac{r_i a_i}{1-r_i} + \frac{1}{1-r_i} \sum_{j,(i,j)\in B} \frac{m_{ij} a_j}{m_i + m_j}$$

$$\rho_{ij} = \frac{m_{ij}}{(m_i+m_j)(1-r_i)}$$

.

**d. The TotalHill module membership assignment method for directed networks**

In the directed case, besides assigning a node *i* to modules, that is, determining how strongly node *i* is connected to the modules (measured by the module assignment strength of node *i* noted by the symbol, $d_{i*}$) another quantity (the 'module influence strength' of node *i* noted by the symbol, $d_{*i}$) can also be determined, namely that how strongly the modules connect to node *i*.

If a link*(i,j)* is a module-core link of *k*-th module, then the module assignment strength of link*(i,j)* to the *k*-th module is $a_{ij}[k] = m_{ij}$ (i.e. the community landscape height), while other components of $a_{ij}$ are zero. For non-core links, $a_{ij}$ is the zero vector. For a non-core link*(i,j)* the module assignment strength of the link, b*(i,j)* is

$$b_{ij} = m_{ij} \frac{\sum_k d_{jk}}{\sum_k m_{jk}},$$

where $d_{jk}$ is the module assignment strength of link*(i,j)* to the modules: for a module-core link*(i,j)*, $d_{ij}$ equals $a_{ij}$, and for non-core links, $d_{ij}$ equals $b_{ij}$. That is, the given link is assigned to modules based on the module membership assignment of the neighboring outbound links of the given link, in proportion of the community landscape height of the respective links. The module membership assignment strength of node *i*, $d_{i*}$ is calculated by summing the module membership assignment-strengths of the outbound links of node *i*: $d_{i*} \equiv b_{i*} + a_{i*} \equiv \sum_j d_{ij}$ . The module influence strength of node *i*, $d_{*i}$ is calculated by summing the module membership assignment-strengths of the inbound links of node *i:* $d_{*i} \equiv b_{*i} + a_{*i} \equiv \sum_j d_{ji}$.

This way we get an equation system with the number of variables equal to the non-core links of the network. Note that the variables are not scalar, but vector values with the vector length of the number of modules. However, the components are independent of each other, so the equation system can be decomposed into equation systems, which have scalar-variables, and can be solved independently.

Based on the module membership assignment rule of nodes and the equation system assigning links to modules, we give an alternative form of the equation system, where variables are assigned to nodes instead of links. Once the module membership assignment of nodes is calculated, links can also be assigned to modules. If $(i,j) \in B$, then

$$b_{ij} = \frac{m_{ij}}{m_{j*}} (b_{j*} + a_{j*})$$

Given the module membership assignment of nodes, the module membership assignment of links is easily calculated. Introducing

$$m_{i*} = \sum_{j,(i,j)\in B} m_{ij}$$

$$m_{*i} = \sum_{j,(j,i)\in B} m_{ji}$$

we can write, that

$$b_{i*} = \sum_{j,(i,j)\in B} b_{ij} = \sum_j \frac{m_{ij}}{m_{j*}} (b_{j*} + a_{j*})$$

$$b_{*i} = \sum_{j,(j,i)\in B} b_{ji} = \frac{m_{*i}}{m_{i*}} (b_{i*} + a_{i*}) .$$

The sought module membership assignment-strengths are

$$d_{i*} = a_{i*} + \sum_{j,(j,i)\in B} \frac{m_{ij}}{m_{j*}} (b_{j*} + a_{j*})$$

$$d_{*i} = a_{*i} + \frac{m_{*i}}{m_{i*}} (b_{i*} + a_{i*})$$

This way we rewrote the equation system in an alternative form assigning variables to nodes.

**e. The optimized version of the undirected TotalHill module membership assignment method**

**Algorithm 6** of the optimized version consists of two procedures: an equation system solving function and the TotalHill membership assignment method itself. We have chosen the Gauss-Seidel iterative equation solver algorithm, which does not have a run-time guarantee but in practice finds a solution with a low error extremely fast.

**Algorithm 6: Optimized algorithm for the undirected TotalHill module membership assignment method**

```
Procedure GaussSeidel ( A, b, x, maxError ) {
    x:=b
    error := maxError
    while error >= maxError {
        error := 0
        for i:=1..N {
            sum := b[i]
            for j:=1..N {
                if i != j { sum :=sum-A[i,j] * x[j] }
            }
            sum:=sum /A[i,j]
            error := error + |x[i] - sum|
            x[i] := sum
        }
    }
}

Procedure TotalHill ( G(V,E), maxError ) {
    // get the sum height of all nodes of the network
    SumHeight : Array[N]
    for i:=1..N { SumHeight[i] := 0 }
    for every e_ij link { SumHeight[i] := SumHeight[i]+ height(e) }

    // make the NxN matrix, representing the linear equation system
    // (in practice it could be more optimal to use a sparse matrix representation)
    A : Matrix [NxN]
    for i:=1..N {
        lineSum := 0
        for j:=1..N { A[i,j] := 0 }
        for every e_ij link {
            If notModuleCenter(e) {
                prop := (SumHeight(i) + SumHeight(j)) / height(e)
                lineSum := lineSum + prop
                A[i,j] := prop
            }
        }
        A[i,i] := lineSum-1
    }

    LinkBelong : Matrix[NumberOfModules, NumberOfLinks]

    // For every module-core link we set and solve the equation system (Ax=b)
    // Please note that in the current implementation a module core consisting of multiple links still
    // has just one link marked as the module-core link making this link is the representative link for that
```

```
        // core. Thereby there are as many module-core links as many modules.

        for every e_ij where e_ij is a module-core link {
                m:=the index of module, which has e_ij as module-core

                // set the right column vector of of the equation system
                b : Array[N]
                for k:=1..N { b[k]:=0 }
                b[i] := - height(e)

                // solve the system
                GaussSeidel( A, b, x, maxError)

                // by the result (x vector) – the module membership of nodes – we calculate the
                // module membership of links

                for every l_ab, where l_ab is a non-module-core link {
                        sumHeights := SumHeight[a] + SumHeight[b]
                        if sumHeights > 0 {
                                LinkBelong[m,l] := (x[a] + x[b]) * height(l) / sumHeights
                        } else {
                                LinkBelong[m,l] := 0
                        }
                }

                // only the e module-core link belongs to the m module,
                // the other module-core links not
                for every l_ab, where l_ab is a module-core link { LinkBelong[m,l]:=0 }
                LinkBelong[m,e] := height(e)
        }
        return LinkBelong
}
```

The runtime complexity of the TotalHill module membership assignment methods is $O(n^2 + mg)$, where $n$ is the number of nodes, $m$ is the number of modules and $g$ is the runtime complexity of solving a linear equation system of $n$ variables and $n$ equations. Using a trivial Gauss-elimination, $O(g)$ equals $O(n^3)$, and assuming practically that the number of modules is not more than the number of nodes, the theoretical runtime complexity of the TotalHill method is $O(n^4)$. However, we would like to note that the Gauss-Seidel linear equation system solver is very fast in practice, and that the computations can be easily parallelized regarding the number of modules. The linear equations could also be solved by a sparse LU decomposition-based methods (Duff et al., 1986).

For downloading the ModuLand program package including the TotalHill module membership assignment method see our homepage: <http://www.linkgroup.hu/modules.php>.

## *3. Other definitions of hills on a centrality landscape (the SameHill and NumberHill methods)*

In this section we introduce two hill definitions, which use a different perspective than those introduced previously.

*SameHill:* Inspired by the Potts model and spectral modularization methods (see Table S2), we introduce the SameHill hill definition. In this method values are assigned to nodes or links, and modules are defined as the parts of connected components with ‘similar’ values. For example, in the simplest spectral case nodes with negative values constitute one module, and positive nodes form the other. Hills of the community landscape of these methods are the links or nodes of connected components in the neighborhood of a hill-top, whose community landscape height is not lower than $p$% of the community landscape height of the respective hill-top. If hills of the community landscape are flat enough, with this method we get a fast, non-overlapping modularization for most of the nodes.

*NumberHill:* Let us chose a hill-top not-yet assigned (on the community landscape over the nodes or links), and assign it to a new module totally. Then let us continue adding the neighboring nodes or links of the nodes or links already in this new module with the highest community landscape height, until the module reaches a predetermined size (for example a given percent of the *N* number of nodes). This method can result overlapping modules by adding a node to more than one hill-tops during the process. In general there will be unassigned nodes too, because of the fixed size of the modules.

Both of the SameHill and NumberHill methods are fast, and may yield overlapping links or nodes. Naturally, several other hill definitions are possible, and their applicability is determined by the properties of the community landscape.

## *4. Finding the nodes of the overlapping communities*

After assigning all links of the network to various modules by any of the procedures described in Sections V.1-3. the assignment of the nodes may use a relatively simple method, called detailed module membership assignment. A node is assigned to a given module to the extent its links belong to that module. Nodes having no links are obviously not cores of any conventional modules. However, for other purposes, such as the visualization of the network, it may be useful to assign them as one-node modules of the network. From the node module membership assignment procedures we tried, the above, detailed module membership assignment procedure explores and utilizes the richest information on the network community structure. However, for some applications the detailed node module membership assignment might be too complex. In these cases we may assign the given node to that module, where its links belong the most.

The local maxima-based hill determination methods described in Section V.2. greatly differ from the horizontal cuts resembling to most of the previously designed network module determination methods of Table S2 described in Section V.1., since

1.) it assigns all links and nodes of the network to network modules;
2.) automatically determines the total number of modules;
3.) avoids the giant-component problem giving a detail-rich picture on the community structure of the network to the required extent.

## *5. Summary of the hill finder methods*

Obviously, besides the module membership assignment methods we described above, a large number of assignment procedures can also be designed. Such a module membership assignment method can be a variant of the ProportionalHill method, where the module membership assignment is performed proportionally to the community landscape height-difference between the actual link and its already assigned neighbors, not to the absolute height of the neighbors described above. The ProportionalHill or TotalHill module membership assignment methods can use non-linear (e.g. exponential, or other) functions for the assignment of links to the modules of their previously assigned neighbors. We are inviting our colleagues to explore this rich field of possibilities.

## *6. Metrics based on the ModuLand modularization methods*

### a. Effective size of support

The effective size of support (ESS) measure of a random variable was defined by Grendar (2006). For a better understanding of what the effective size of support means, let us see first the usual definiton of the size of the support of a random variable. Let *X* be a discrete random variable with probability mass function *p* consisting of *m* nodes. The definiton for the support of *X* is the set of those $x \in X$ whose weight (probability) is non-zero, that is

$$A(p(X)) = \{x \in X : p_x > 0\}$$

and $|A(p(X))|$, the size of the support equals the number of the members of this set.

While the size of support for random variables of the probability mass function *p = [0.5, 0.5]* and *q = [0.99, 0.01]* both equal two, the properties of the two random variables differ significantly. While the former random variable takes both possible values with equal probability, the latter takes only one of the two possible values most of the time. To distinguish between such differences, it would be useful to devise a measure $S \in [1;m]$, where the value of this measure describes the random variable from the introduced point of view, i.e. that the number of the members is not the discrete number of members occurring with a non-zero probability, but it is expressed as a continuous measure taking into account their probability. This new measure is the effective size of support *S(p(X))* or *S(X)*.

ESS should have certain properties, dictated by common sense (Grendar, 2006):

P1) $S(p)$ should be continuous, symmetric function (i.e., invariant under exchange of $p_i$ and $p_j$, $i,j = 1... m$).

P2) $S(\delta_m) = 1 \leq S(p_m) \leq S(u_m) = m$, where $u_m$ denotes the uniform probability mass function on m-node support, $\delta_m$ denotes an m-node probability mass function with probability concentrated at one point, $p_m$ denotes a probability mass function with $|A(p)| = m$.

P3) $S([p_m, 0]) = S(p_m)$.

P4*) $S(p(X, Y)) \leq S(p(X))S(p(Y))$, and equal if and only if $X$ and $Y$ are independent random variables. This means that the effective number of states of the joint system $(X,Y)$ cannot be greater than the number resulting from multiplying the effective number of states for the single systems $X$ and $Y$.

Grendar concludes that an $S(p(X))$ conforming to these properties can be written as $exp(H(p(X)))$, where $H(p(X)) = -\sum_{i=1...m}(p_i \ln p_i)$ is the Shannon-entropy.

So we can define the effective size of support as $S(p(X)) = exp(H(p(X)))$. Earlier the term 'perplexity' has also been defined (Jelinek et al., 1977; Bahl et al., 1983) for this quantity.

**b. Effective number**

The effective size of support measure described above can be used for the measurement of the effective number of members belonging to any group (e.g. network community or module). The 'effective number of members' from now on will be called shortly as *effective number*. Here 'effective' means that the number of the members is not counted in a discrete way, but it is expressed as a continuous measure taking into account the weight of the members, by giving larger weight to the important nodes. With other words, the effective number shows us the number of important or frequent nodes in the set. As an example a member with a relatively small weight does not count as one full member, just as a small fraction of a full member. We introduce the effective number as follows.

$$n_i\{V[i]\} = \exp\left(-\sum_i p_i \log p_i\right)$$

The equation above defines the effective number of the members of set $V$, where $V[i]$ is the weight of member $i$, and

$$p_i = \frac{V[i]}{\sum_j V[j]}.$$

While counting the number of significant members in the traditional, discrete way introduces a threshold, and discards members of relatively insignificant weight, counting the effective number of members does not require any threshold, and summarizes the total information available. Moreover, the effective number is a continuous function of the weights of the members. See Lee et al. (2010) for an example of the application of the effective number for network degrees.

**c. Modular overlap**

Let us assume that the module membership assignment-strength of node $i$ to module $j$ ( $j = 1..m$ ) is $d_i[j]$. The modular overlap of node $i$ is the effective number of modules that node $i$ is assigned to:

$$O[i] := n_a\{d_i[a]\}.$$

**d. Bridgeness**

Graph theory traditionally refers to a link as a *bridge*, whose removal would increase the number of components of the graph. In sociology, a *bridge node* is a node connecting two otherwise distinct groups (Burt, 1995; Nepusz et al., 2008; see Suppl. Discussion). We introduce the *bridgeness* measure of a node or link as the overlap of the given node or link between two or more modules relative to the overlap of the other nodes or links.

$$B[a][b][i] := \frac{T[a][b][i]}{\sum_j T[a][b][j]},$$

is the bridgeness of node $i$ between modules $a$ and $b$, where

$$T[a][b][i] := \min\{d_i[a], d_i[b]\}$$

is the area-overlap, or common area of node between modules $a$ and $b$.

The total bridgeness of node *i* describes the bridgeness of that node between all modules:

$$B[i] := \sum_{a=1}^{m} \sum_{b \neq a, b=1}^{m} B[a][b][i].$$

**e. Similarity of the nodes**

The similarity of the nodes *i* and *j* is based on their module membership vectors, $d_i$ and $d_j$:

$$Sim(d_i, d_j) = \sum_k \min\left(\frac{d_i[k]}{\sum_l d_i[l]}; \frac{d_j[k]}{\sum_l d_j[l]}\right)$$

The value of $Sim(d_i, d_j)$ is in the $[0; 1]$ interval, and it is maximal, if the $d_i$ and $d_j$ module membership vectors are exactly the same.

## VI. Further opportunities and analysis of the ModuLand method

### *1. Merging the overlapping communities*

The local maxima-based hill determination methods (as opposed to many other network module determination methods) does not require a pre-set value for the maximal number of modules expected. This may result in a rather large number of modules in networks with a complex community structure yielding a 'rough' community landscape with many local maxima. However, modules of such networks with rough community landscapes can be merged. This also allows the tailoring of the final module-structure to the needs of the investigator. Large overlaps, small differences in both the distance and community landscape height of modular hill-tops/highlands are all appropriate reasons for merging of two neighboring modules. In many practical applications merging the adjacent modules can be very useful (such as the reorganization of overlapping firm departments, or computer program subroutines, to name but a few). While in Section VII. we will describe a hierarchical representation of the community structure, which merges the appropriate modules based on the detailed topological information, here we introduce a simple yet effective post-processing step for merging artificial modules.

Our post-processing step relies on the fact that the nodes of the network can be ordered based on their module membership assignment strength to any given module. Therefore we define the correlation of two modules as the Spearman's rank-order correlation of the corresponding orderings of the nodes of the network. After calculating the correlation of all module pairs, groups of highly correlated modules, where the correlation is above a chosen threshold, are merged into a single module. The module membership strength of the nodes of the network to a given resulting module equals the sum module membership strength to the corresponding merged modules. In Figures S13c and S13d we show the effectiveness of this post-processing step and also investigate the effect of choice of threshold. Generally the correlation threshold for merging modules may be chosen by inspecting the histogram of module-pair correlations and setting the threshold to merge the modules of extremely high correlation, but throughout this paper we used an arbitrary chosen correlation threshold of 0.9 for the sake of simplicity.

We note that applying the TotalHill module membership assignment method yields massively overlapping modules unsuitable for merging modules based on this threshold, therefore in this case we merge the modules based on the ProportionalHill module membership assigment method-based module information.

### *2. Robustness of the ModuLand method for the measurement, sampling and computational errors*

Measurement and sampling errors obviously compromise any community detection method (Massen and Doye, 2006; Karrer et al., 2008; Yip and Horvath, 2007; Lusseau et al., 2008). If these errors are beyond the tolerance level set by the purpose of the study and/or the habits of the investigator, specially designed versions of the ModuLand method family may take this problem into consideration. If we have an estimate of the typical measurement error we may check the effect of this error by discretizing the link-weight values rounding them to the next available value of a scale in the range of the expected error and measure the differences in the outcome of the modular structure. Alternatively, both the measurement errors and the influence function-finding errors can be minimized by discretization of the community landscape heights to a scale comparable with the expected errors and examine the differences it caused in the modular structure. If none of these solutions works properly, it may happen that due to the sensitivity of certain influence function calculation methods the cumulative errors will occasionally show little, artificial hill-tops on a side of a hill of the community landscape. This yields to a number of unreal modules in the case of the local maxima-based hill determination methods. If experiencing this trouble, the merging procedures of Section VI.1. or specific ways of error reduction can be very helpful. However, in most cases we can neglect these artificial hill-tops, since at a higher level of the hierarchical community structure, these new, artificial communities will be often merged to real communities, and will not disturb the community structure any longer.

In our work we applied the benchmark graph generation method published by Lancichinetti et al. (2008), which produces non-overlapping modules, in order to check the correspondence of our highly overlapping modules with the surely known partitions of the benchmarks. The comparison of our results with the recently published updated benchmark graph generation method by Lancichinetti and Fortunato (2009a), which also supports overlapping modules, is an interesting and challenging problem of its own, and would require a study which is beyond the scope of our current paper.

In Figures S13a and S13b we show that the identified modules correspond consistently to the modules of the benchmark graph of Lancichinetti et al. (2008) over a range of parameter settings, where modules can be defined in the strong sense. Strong sense means here, at least the half of the neighboring nodes are assigned to the same module as the given node, see Lancichinetti et al. (2008).

## VII. Construction of the higher hierarchical level representations of the network

The ModuLand method family offers the way to create higher levels of hierarchical representation of the original network. Here the nodes of the higher level correspond to the modules of the original network, and the links of the higher level correspond to the overlaps between the respective modules (Figure 1D and Figure S6). This hierarchical representation can be established recursively in several steps, until the whole original network becomes represented by one or more single nodes without interaction between them. The higher levels of hierarchical representation are very useful, since several versions of the ModuLand method family, like the PerturLand algorithm may yield a rather detailed community structure at the first hierarchical level of modules, which gives an extremely rich information on the communities causing an 'overflow' of normal human cognitive capacities. In such cases (which often occurs at larger networks having more than 10,000 nodes) the complexity of the community structure can be easily reduced to the comprehension limit of a few modules examining a higher hierarchical level. Moreover, the occasional artificial hill-tops and modules, mentioned in Section VI.1., will also be eliminated at higher hierarchical levels. In the following Sections we describe a detailed assignment procedure of these hierarchical representation layers allowing a fast, zoom-in type analysis of large networks.

### *1. The strength of the link between two modules*

Let $s^{(2)}_{ab}$ denote the strength of the directed or undirected link between module *a* and module *b* (thus the strength of a link at the second hierarchical level of the modular representation noted by the upper index), which will be defined for both cases in this section. We will see, that $s^{(2)}_{ab}$ has usually a nonzero value even in case if $a=b$, meaning that loop links may occur on higher network levels. These loop links will be discarded, since they do not give information about inter-node connections on the higher level.

In the followings $d_i[j]$ ($d_{i*}[j]$ in directed networks) will denote the strength of node *i* belonging to module *j*, and $m_i$ will denote the centrality of node *i*.

In directed networks the activity (see Section III.2.) of a module (where the module is regarded as a node of the second – or any higher – hierarchical level), $p_a^{(2)}$ is given by the sum activity of the nodes belonging to the given module, weighted by the module assignment-strength of those nodes. This can be expressed by the formula $p_a^{(2)} = \sum_i p_i d_{i*}[a]$. If we set the activity of the nodes all equal to 1 (as it is in case of the test-networks used in this paper) having the assumption of $p_i \equiv 1 \quad \forall i$ , the formula can be rewritten as $p_a^{(2)} = \sum_i d_{i*}[a]$. In directed networks let the $s^{(2)}_{ab}$ strength of link$(a,b)$ be the maximum of the overlap of the modules *a* and *b* over the nodes, divided by the activity of module *a*, that is:

$$s^{(2)}_{ab} := \frac{\max_i\{T[a][b][i]\}}{p_a},$$

where $T[a][b][i]$ is the overlap ('common area') of node *i* between module *a* and *b*:

$$T[a][b][i] := \min\{d_i[a], d_i[b]\}.$$

The overlap of a node between two modules increases linearly proportional to its assignment-strength to each of the two given modules. Now, it is clear, that $s^{(2)}_{ab}$ scales linearly as the function of the community landscape height of the *i* node.

In undirected networks, we may use other definitions according to the model of the interactions used in the influence function calculation procedure. We now present a simple definition, in which the overlap between two modules is summarized over all nodes. For the node *i* let the overlap be propotional to its assignment-strength to both modules. The expression $d_i[a]d_i[b]$ scales as quadratic instead of linear in the function of the community landscape height of the node, therefore, the desired strength between modules *a* and *b* can be defined as

$$s^{(2)}_{ab} := \sum_i \frac{2d_i[a]d_i[b]}{m_i}.$$

The constant multiplier of 2 in the formula is a consequence of our convention, regarding an undirected link to two directed links with opposite directions.

In the undirected case, the activity of a module (where the module is regarded as a node of the second – or any higher – hierarchical level) is also given by the sum of the activities of the nodes belonging to that module, weighted by the module assignment-strength to that module, so $p_a^{(2)} = \sum_i p_i d_i[a]$. While $p_i \equiv 1 \quad \forall i$ for the networks considered in our current study, the above formula can be rewritten as $p_a^{(2)} = \sum_i d_i[a$ . We note that in the undirected case, $s^{(2)}_{ab}$ naturally equals $s^{(2)}_{ba}$ .

It is important to note, that the strength of the link between two modules already comprises the effect of direct and indirect interactions between the nodes of the original network, too. In other words this means that the higher level link strengths contain already the indirect impact between the nodes of the lower level network.

## *2. The modularization of a higher hierarchical level*

The structure of the next hierarchical level of the original network reflects the essential structure of the original network, making it easier to visualize, overview, describe and understand the original network. If the network of the next hierarchical level is still too complex, the construction of even higher hierarchical levels may be required. This is done by finding the modules of the representation of the network at the next hierarchical level. This process can be continued until the resulting network has no links and colaesces to individual single nodes, which is – obviously – the top level network. The number of top level nodes depends on the applied influence function calculation algorithm, among others. For example applying the PerturLand or LinkLand algorithm yields a number of top level nodes equal to the connected components in the original network, while in the case of the NodeLand algorithm more top level nodes than the number of original connected components may appear as a final result.

Technically it is important to remark, that each node without links of the network is defined as a separate module. According to this definition such a node makes a module containing one single node on every higher hierarchical levels of the network.

The link strengths between the modules can be directly used as community landscape heights, without any influence function calculation method, due to the fact, that the higher level link strength already comprises the effect of direct and indirect interactions between the nodes of the original network. As a consequence, the link strengths between the modules can be directly used as community landscape heights for the network of modules. However generally, one may be interested in types of interections between the modules other than the type of interaction between the original nodes. For example the interaction between groups of people (e.g. cities or countries) may be of quite different nature than the interactions between the constituent individuals. In this case one should apply an influence function model selected and/or adjusted according to the type of the interaction and re-run a modularization protocol at the second (or higher) level of the hierarchical structure in order to get the third (or, generally 'higher+1') level community landscape.

## *3. Projection of the modules of higher hierarchical levels*

Not only the nodes of the (*n-1*)th hierarchical level can be assigned to the nodes of the (*n*)th level as shown in the previous sections, but also nodes of any lower, (*n-2*)th, etc. hierarchical levels can be directly assigned to the nodes of the (*n*)th level. For example nodes of the (*n-2)*th level can be assigned to the nodes of the (*n*)th level (distributing the assignment-strengths of the (*n-2)*th level nodes to the (*n-1)*th level nodes between the (*n*)th level nodes). Since in practice the importance of this direct assignment procedure is used 'in reverse', that is to show how many nodes of e.g. the original network are beloning to very high hiearchical representation of this network having only a few modules already, we call this process as the 'projection' of the (*n)*th hierarchical level to the (*n-2)*th (or any lower) hierarchical level.

The following steps have to be taken to calculate the assignment-strength of node *i* of the (*n-2)*th hierarchical level to node *j* of the (*n*)th hierarchical level: For any node *k* of the (*n-1)*th hierarchical level, let us multiply the assignment-strength of node *i* to node *k* (note that node *k* is not on the same hierarchical level as node *i*, but at the hierarchical level of node *i* node *k* is actually one of the modules, where node *i* belongs) with the proportion with which node *k* is assigned to node *j*. The sum of the results of these multiplications gives the module assignment-strength of node *i* to node *j*.

To achieve an easy generalization of this procedure let us introduce the $M^{(n+1)}_{(n)}$ matrix, where $M^{(n+1)}_{(n)}[i,j]$ is the proportion with which the node *i* of the (*n*)th hierarchical level is assigned to module *j* of the same hierarchical level (or, to the node *j* of the (*n+1*)th hierarchical level):

$$M_{(n)}^{(n+1)}[i][j] \equiv \frac{d_i^{(n)}[j]}{m_i^{(n)}},$$

$$m_i^{(n)} \equiv \sum_j d_i^{(n)}[j]$$

Now any $M^{(a+k)}_{(a)}$ $(k>1)$ matrix can be derived, which project the module structure of a higher hierarchical level to the nodes of a lower hierarchical level, using

$$M_{(a)}^{(a+k)} := M_{(a)}^{(a+1)} M_{(a+1)}^{(a+2)} (...) M_{(a+k-1)}^{(a+k)}$$

that is, multiplying the module assignment matrices of consecutive levels. Naturally, the number of columns of any left-side matrix (the number of modules) by definition equals the number of rows of the right-side matrix (nodes of the higher hierarchical level). The assignment of node $i$ of the ($a$)th level to the higher ($a+k$)th hierarchical level, node $j$ is given by the equation $d_i[j] = M^{(a+k)}_{(a)}[i][j]m^{(a)}_i$ .

Applying the projection described here, we can not only analyze nodes of the previous hierarchical level from the viewpoint of a given hierarchical level, but also nodes of any previous hierarchical levels, including the original network. This gives rise to an exquisite possibility of an easy analysis, visualization and comprehension of extremely large original systems, such as telecommunication networks, neuronal cells, the human proteome, to name only a few.

A simple case illustrating this scenario can be seen on Figure S6 showing the hierarchical levels of the network science collaboration network (Newman, 2006a). The upper part of the figure shows the different hierarchical levels of the network. The bottom part of the figure shows the community structure of the higher hierarchical levels projected to the original network. The top hierarchical level would consist of separate nodes. It can be seen that the projection of a higher and higher hierarchical levels generally yields larger and larger modules, as expected.

## VIII. ModuLand-based network visualization methods

The hierarchical network representation can be used for a fast visualization of extremely large networks, where conventional network visualization methods became too slow to apply. A visualization method based on the ModuLand method starts with the modularization of the network as described in this paper and treats the nodes of the top-network not only as a representation of the respective community of the bottom-network as described in Section VII., but also as the center of the community of the bottom-network in the 2D space. The ModuLand method family gives numerical values to characterize the contribution of network nodes to each network community, and assigns community landscape heights to each link (please note that from the community landscape heights of the links the community landscape heights of the nodes can be derived). This data can be used to calculate the relative position of the nodes of the bottom-network from the community-centers. The ModuLand-based visualization protocol can be combined with other, existing network visualization methods. It is important to note that the extra time to run the ModuLand program before the visualization is not a disadvantage in case of extremely large networks, since many visualization tools run slower than the expected polynomial complexity, $O(n^3)$ of the ModuLand programs (Walshaw, 2003), and often give overpacked final results, which are difficult to comprehend. As a further advantage of the hierarchical visualization based on the ModuLand method the user may zoom-in to deeper and deeper layers of the hierarchical complexity of the modular structure described in Section VII. just in the section of interest in a large network.

# Supplementary Tables

## Table S1. Definitions of network modules

| Name | Definition | References |
|---|---|---|
| **Definitions based on 'local' topology**[a-d] | | |
| **Clique** | a complete subgraph of size *k*, where complete means that any two of the *k* nodes are connected with each other | Luce and Perry, 1949; Wasserman and Faust, 1994 |
| ***k*-clan** | a maximal connected subgraph having a subgraph-diameter ≤ *k*, where the subgraph-diameter is the maximal number of links amongst the shortest paths *inside* the subgraph connecting any two nodes of the subgraph | Alba, 1973; Mokken, 1979; Wasserman and Faust, 1994 |
| ***k*-club** | a connected subgraph, where the distance between nodes of the subgraph ≤ *k*, and where no further nodes can be added that have a distance ≤ *k* from all the existing nodes of the subgraph | Alba, 1973; Mokken, 1979; Wasserman and Faust, 1994 |
| ***k*-clique** | a maximal connected subgraph having a diameter ≤ *k*, where the diameter is the maximal number of links amongst the shortest paths (including those *outside* the subgraph), which connect any two nodes of the subgraph | Luce, 1950; Alba, 1973; Mokken, 1979; Wasserman and Faust, 1994 |
| ***k*-clique community** | a union of all cliques with *k* nodes that can be reached from each other through a series of adjacent cliques with *k* nodes, where two adjacent cliques with *k* nodes share *k*-1 nodes *(please note that in this definition the term k-clique is also often used, which means a clique with k nodes, and not the k-clique as defined in this set of definitions; the definition may be extended to include variable overlap between cliques)* | Alba and Moore, 1978; Palla et al., 2005; Zubcsek et al., 2008 |
| ***k*-component** | a maximal connected subgraph, where all possible partitions of the subgraph must cut at least *k* links | Matula, 1972 |
| ***k*-plex** | a maximal connected subgraph, where each of the *n* nodes of the subgraph is linked to at least *n*-*k* other nodes in the same subgraph | Seidman and Foster, 1978; Wasserman and Faust, 1994 |
| **strong LS-set** | a maximal connected subgraph, where each subset of nodes of the subgraph (including the individual nodes themselves) have more connections with other nodes of the subgraph than with nodes outside the subgraph | Wasserman and Faust, 1994; Radicchi et al., 2004 |
| **LS-set** | a maximal connected subgraph, where each node of the subgraph has more connections with other nodes of the subgraph than with nodes outside of the subgraph | Wasserman and Faust, 1994; Flake et al., 2002; Radicchi et al., 2004 |
| **lambda-set** | a maximal connected subgraph, where each node of the subgraph has a larger node-connectivity with other nodes of the subgraph than with nodes outside of the subgraph (where node-connectivity means the minimum number of nodes that must be removed from the network in order to leave no path between the two nodes) | Borgatti et al., 1990; Wasserman and Faust, 1994 |
| **modified (weak) LS-set** | a maximal connected subgraph, where the sum of the inter-modular links of the subgraph is smaller than the sum of the intra-modular links | Wasserman and Faust, 1994; Radicchi et al., 2004 |

## Table S1. Definitions of network modules (continuation)

| Name | Definition | References |
|---|---|---|
| **Definitions based on 'local' topology (continuation)[a-d]** | | |
| ***k*-core** | a maximal connected subgraph, where the nodes of the subgraph are connected to at least $k$ other nodes of the same subgraph; alternatively: the union of all $k$-shells with indices greater or equal $k$, where the $k$-shell is defined as the set of consecutively removed nodes and belonging links having a degree $\leq k$ | Seidman, 1983; Wasserman and Faust, 1994; Radicchi et al., 2004; Carmi et al., 2007; Kitsak et al., 2010 |
| **Definitions based on 'global' topology** | | |
| **Modularity(Q)-based definitions** | a set of subgraphs, which displays a larger modularity (Q) measure, than the same set of subgraphs in an appropriate null-model (for the definition of link- or motif-based modularity measures and – usually randomized – null-models see the references and Table S3) | Girvan and Newman, 2002; Newman and Girvan, 2004; Reichardt and Bornholdt, 2006a; Arenas et al., 2008a; Fortunato and Castellano, 2009; Fortunato, 2010 |
| **Definitions based on node similarity** | | |
| **Node similarity definitions** | a subgraph containing node-pairs, which are similar to each other based on their distance, eigenvector components, etc.; similarity may also be functional similarity coming from the emergent network properties – this latter similarity measure, however, may add a node of redundancy to the definition, since the emergent function often emerges from the modular structure | Leicht et al., 2006; Fortunato and Castellano, 2009; Fortunato, 2010 |
| **Definitions based on information content** | | |
| **Information content-based definitions** | a set of subgraphs, which allow the greatest compression of network structure with a minimal loss of information (the Network Information Bottleneck) | Ziv et al., 2005; Rosvall and Bergstrom, 2007; 2008 |
| **Definitions based on information propagation** | | |
| **Information propagation-based definitions** | a set of nodes, displaying a larger communication among them than to the rest of the network | Flake et al., 2002; and all network walk-based methods of Table S2 |
| **Global influence** | a set of nodes, displaying a larger influence (sum of weighted paths) to each other than to the rest of the network | Ghosh and Lerman, 2008 |
| **Definitions based on network dynamics[e]** | | |
| **Co-Set** | a subgraph, where all nodes occur or change simultaneously | Papin et al., 2004 |

[a]The definitions based on 'local' topology are listed in the order of their approximate stringency starting with the most stringent. A summary of additional (less exact) local community definitions can be found in Hinne (2007).

[b]The most widely used module definition is the LS-set or sometimes the modified LS-set. However, all the other definitions (and possibly even more) are satisfying our common sense of community formation based on the higher relative frequency/strength of links between community members compared to non-members ('static definition') as well as on the larger closeness or reachability of community members compared to non-members ('dynamic definition').

[c]In all the 'local' topology-based definitions no loops or multiple links were allowed.

[d]Most of the 'local' topology-based definitions do not include link weights or directedness.

[e]While there are plenty of mathematically explicit 'static' definitions, there is a remarkable paucity of explicit definitions, which include a dynamic network property.

## Table S2. Comparison of network module determination methods

| Name of method | Complete Data-set[a] | Weighted graph[b] | Directed graph[b] | Number of modules[c] | Assignment to modules[d] | Overlaps[e] | Polynomial complexity (speed)[f] | Test-modules[g] | Zachary-network[h] | References |
|---|---|---|---|---|---|---|---|---|---|---|
| **Agglomerative methods** | | | | | | | | | | |
| Hierarchical agglomeration (clustering) | + | + | – | parameter dependent | yes/no | – | N.A. | N.A. | N.A. | Johnson, 1967; Aldenfelder and Blashfield, 1984; Wasserman and Faust, 1994; Slater 2008; Rivera et al., 2010 |
| Clustering with previous overlap 'distribution' | + | – | – | parameter dependent | refined | + | ~O($n^3$), local version: O($n$ log$n$) | N.A. | N.A. | Gregory 2009 |
| Shortest path-similarity and hierarchical agglomeration | + | – | – | parameter dependent | yes/no | – | N.A. | N.A. | N.A. | Rives and Galitsky 2003; Arnau et al., 2005 |
| Neighborhood similarity and hierarchical agglomeration | + | – | – | parameter dependent | yes/no | – | N.A. | N.A. | N.A. | Andreopoulos et al., 2007 |
| Topological overlap methods | + | + | – | parameter dependent | yes/no | + | ≥O($n^3$) | N.A. | N.A. | Ravasz et al., 2002; Zhang and Horvath, 2005; Li and Horvath, 2007; Yip and Horvath, 2007 |
| Restricted neighborhood search/flow Markov clustering | + | + | – | parameter dependent | yes/no | – | N.A. | N.A. | N.A. | Enright et al., 2002; Spirin and Mirny, 2003; Dorow et al., 2004; King et al., 2004 |
| Clustering and a random walk process | + | + | + | parameter dependent | yes/no | – | ~O($n^2$) | N.A. | 1 | Delvenne et al., 2010; E et al., 2008; Li et al., 2008b |
| Clustering and a flocking process | + | + | + | parameter dependent | yes/no | – | N.A. | N.A. | N.A. | Li et al., 2008c |
| Markov clustering with noise | + | – | – | parameter dependent | yes/no | + | N.A. | N.A. | (1) | Gfeller et al., 2005; 2007 |

**Table S2. Comparison of network module determination methods (continuation)**

| Name of method | Complete Data-set[a] | Weighted graph[b] | Directed graph[b] | Number of modules[c] | Assignment to modules[d] | Overlaps[e] | Polynomial complexity (speed)[f] | Test-modules[g] | Zachary-network[h] | References |
|---|---|---|---|---|---|---|---|---|---|---|
| **Agglomerative methods (continuation)** | | | | | | | | | | |
| Clustering similarity- and distance similarity-based method | + | – | – | parameter dependent | yes/no | + | N.A. | N.A. | N.A. | Poyatos and Hurst, 2004 |
| Dissimilarity matrix reordering-based visual clustering | + | – | – | parameter dependent | yes/no | – | $O(n^2)$ | N.A. | N.A. | Yang et al., 2006 |
| Multilevel clustering (coarsening/de-coarsening) | +, – | – | – | automatic | yes/no | – | N.A. | N.A. | N.A. | Oliviera and Seok, 2006 |
| Network Information Bottleneck (NIB) clustering | + | – | – | automatic | with 'soft clustering': refined | With 'soft clustering: + | N.A. | 0.4, 0.45 | N.A. | Ziv et al., 2005 |
| Bayesian clustering | + | – | – | automatic | yes/no | – | $O(n^2)$ | N.A. | 0 | Hofman and Wiggins, 2008 |
| Expectation-maximization, kernel-derived method | + | + | – | automatic | refined | + | N.A. | N.A. | 3 overlaps | Ren et al., 2009 |
| Potts-model | + | + | + | parameter dependent | yes/no | – | ~O(nlogn) | 0.88 | N.A. | Blatt et al., 1996; Spirin and Mirny, 2003; Guimera et al., 2004; Ronhovde and Nussinov, 2009, 2010 |
| Fuzzy Potts-model | + | + | – | parameter dependent | yes/no | + | parameter dependent | 0.7 | N.A. | Reichardt and Bornholdt, 2004; 2006b; Ispolatov et al., 2006; Heimo et al., 2008b |
| Informational coherence and fuzzy Potts-model | + | – | – | parameter dependent | yes/no | + | $\geq O(n^3)$ | N.A. | N.A. | Shalizi et al., 2007 |
| Laplacian clustering | + | + | – | parameter dependent | yes/no | – | N.A. | N.A. | N.A. | Kim et al., 2008 |

**Table S2. Comparison of network module determination methods (continuation)**

| Name of method | Complete Data-set[a] | Weighted graph[b] | Directed graph[b] | Number of modules[c] | Assignment to modules[d] | Overlaps[e] | Polynomial complexity (speed)[f] | Test-modules[g] | Zachary-network[h] | References |
|---|---|---|---|---|---|---|---|---|---|---|
| **Agglomerative methods (continuation)** | | | | | | | | | | |
| Enforced frustration method | + | + | – | automatic | yes/no | – | $O(n^{3.2})$ | N.A. | 3 | Son et al., 2006 |
| Dynamical clustering (sync-dependent hierarchy detection) | + | – | – | parameter dependent | yes/no | – | $O(n^2)$ | 0.45 | 0 (3 modules) | Arenas et al., 2006; Boccaletti et al., 2007; Pluchino et al., 2008 |
| Dynamical simplex evolution method | + | – | – | automatic | yes/no | – | $O(n^2)$ | 0.83 | N.A. | Gudkov et al., 2008 |
| Symmetric connectivity and noise/continuous dynamics | + | + | – | automatic | yes/no | – | $O(n^2)$ | N.A. | N.A. | Krawczyk and Kulakowski, 2008 |
| *k*-means (*k*-median, *p*-median) of cluster distance minimization | + | – | – | parameter dependent | yes/no | – | $\sim O(n^4)$ | 0.26, 0.9 | 0 | MacQueen, 1967; Gustafsson et al., 2006; Angelini et al., 2007a; Brusco and Köhn, 2008; Kumar and Kannan, 2010 |
| fuzzy *k*-means clustering | + | – | – | parameter dependent | yes/no | – | N.A. | N.A. | N.A. | Bezdek, 1981; Dunn, 1974; Schwammle, 2010 |
| Network vectorization clustering | + | + | + | parameter dependent | yes/no | – | N.A. | 0.87 | 0 | Ren et al., 2008 |
| Co-occurrence methods | + | – | – | automatic | refined | + | N.A. | N.A. | N.A. | Papin et al., 2004; Jin et al., 2008 |
| Highly connected subgraph spectral analysis method | –, + | – | – | automatic | yes/no | – | N.A. | N.A. | N.A. | Kleinberg, 1997; Gibson et al., 1998; Hartuv and Shamir, 2000; Bu et al., 2003 |
| Distance-based *k*-clique clustering methods | + | – | – | automatic | yes/no | – | $O(n^3)$ | N.A. | N.A. | Edachery et al., 1999 |
| Game-based clustering | + | + | + | parameter dependent | yes/no | – | N.A. | N.A. | N.A. | Li et al., 2010b |

## Table S2. Comparison of network module determination methods (continuation)

| Name of method | Complete Data-set[a] | Weighted graph[b] | Directed graph[b] | Number of modules[c] | Assignment to modules[d] | Overlaps[e] | Polynomial complexity (speed)[f] | Test-modules[g] | Zachary-network[h] | References |
|---|---|---|---|---|---|---|---|---|---|---|
| **Network walk-based agglomerative methods[i]** | | | | | | | | | | |
| Clique percolation methods | –, + | + | + | automatic | Refined | + | $O(exp(n))$ – in real applications it runs faster, $O(n^2)$ | N.A. | N.A. | Alba and Moore, 1978; Palla et al., 2005; Adamcsek et al., 2006; Zotenko et al., 2006; Farkas et al., 2007; Palla et al., 2007a; Du et al., 2008; Kumpula et al., 2008; Lehmann et al., 2008; Shen et al., 2008; Zubcsek et al., 2008 |
| Bipartite cliques | + | – | – | automatic | yes/no | – | N.A. | N.A. | N.A. | Tanay et al., 2004 |
| Local community-based methods (basins of attraction) | –, + | + | – | automatic | yes/no | + | $O(nlogn)$, $O(n^3)$ | 0.18 | 1 | Altaf-Ul-Amin et al., 2006; Luo et al., 2006; Bagrow, 2008; Carmi et al., 2008; Hu et al., 2008b |
| Local fitness optimization | + | – | – | automatic | refined | + | $>O(n^2)$, fast | 0.6 | 5 overlaps | Baumes et al., 2005a; 2005b; Lancichinetti et al., 2009 |
| Local community with fuzzy clustering | + | – | – | automatic | yes/no | – | N.A. | N.A. | N.A. | Hu et al., 2007 |
| *k*-core-based methods | – | – | – | automatic | yes/no | + | $\sim O(n^3)$ | N.A. | N.A. | Bader et al., 2003; Wuchty and Almaas, 2005; Alvarez-Hamelin et al., 2006; Dorogovtsev et al., 2006; Baskerville et al., 2007; Carmi et al., 2007 |
| Local communities with *l*-shell label propagation | –, + | – | – | automatic | yes/no | + | $>O(n^3)$ | N.A. | 3 | Bagrow and Bolt, 2005; Porter et al., 2007 |
| Local communities with *t*-shell (initial triangles) label propagation | –, + | – | – | automatic | yes/no | + | N.A. | N.A. | 4 overlaps | Eckmann and Moses, 2002; Kelsic, 2005 |
| Local communities with bridge-bounding | + | – | – | automatic | yes/no | – | $\sim O(n)$ | N.A. | N.A. | Papadopoulos et al., 2009 |

**Table S2. Comparison of network module determination methods (continuation)**

| Name of method | Complete Data-set[a] | Weighted graph[b] | Directed graph[b] | Number of modules[c] | Assignment to modules[d] | Overlaps[e] | Polynomial complexity (speed)[f] | Test-modules[g] | Zachary-network[h] | References |
|---|---|---|---|---|---|---|---|---|---|---|
| **Network walk-based agglomerative methods[i] (continuation)** | | | | | | | | | | |
| Local communities with (hub-based) label propagation/hub duplication | –, + | + | + | automatic | yes/no | – | ~O($n$) | N.A. | 1 | Gibson et al., 1998; da Fontoura Costa, 2004; Ucar et al., 2006; Zhang et al., 2008 |
| Similarity-based network evolution methods | + | + | + (signed network) | automatic | yes/no | – | ≤O($n^3$) | 0.82 | 1 | Yang, 2006; Xiang et al., 2009 |
| Maximum flow, minimum cut methods (electric circuit models w/o or with spectral analysis) | + | + | – | parameter dependent | yes/no | – | O($n$), >O($n^3$) | 0.57 | 0,1 | Elias et al., 1956; Ahuja et al., 1993; Flake et al., 2002; Eriksen et al., 2003; Simonsen et al., 2004; Wu and Huberman, 2004; Alves, 2007; Slater 2008 |
| Community profile plot assessment method | + | + | – | parameter dependent | yes/no | – | N.A. | N.A. | three modules | Leskovec et al., 2008 |
| Communication-related Green's function with spectral analysis | + | – | – | parameter dependent | yes/no | + | >O($n^3$) | N.A. | overlaps | Estrada and Hatano, 2008 |
| Communicability graph and clique identification | + | – | – | parameter dependent | yes/no | + | exp($n$) but in reality faster | N.A. | 0 or overlaps | Estrada and Hatano, 2009 |
| Random walk-based similarity distance and hierarchical agglomeration | + | + | + | automatic | yes/no | – | ~O($n^3$) | N.A. | N.A. | Pons and Latapy, 2005 |
| Diffusion-based hierarchical communities | + | + | – | automatic | yes/no | – | O($n^3$) | ~0.75 | 0 | Zhou 2003a; 2003b; Zhou and Lipowski, 2004 |
| Diffusion kernel-based similarity method | + | – | – | automatic | yes/no | – | N.A. | 0.68 | 1 | Zhang et al., 2007b |

## Table S2. Comparison of network module determination methods (continuation)

| Name of method | Complete Data-set[a] | Weighted graph[b] | Directed graph[b] | Number of modules[c] | Assignment to modules[d] | Overlaps[e] | Polynomial complexity (speed)[f] | Test-modules[g] | Zachary-network[h] | References |
|---|---|---|---|---|---|---|---|---|---|---|
| **Network walk-based agglomerative methods[i] (continuation)** | | | | | | | | | | |
| Agent propagation, voting, mergers | + | − | − | automatic | yes/no | − | N.A. | N.A. | 0 | Young et al., 2004 |
| Opinion propagation with decaying confidence | + | − | − | automatic | yes/no | − | N.A. | N.A. | 1 | Morarescu and Girard, 2009 |
| Autonomous agent-based method for dynamic networks | + | − | − | automatic | yes/no | − | >O($n^2$) | N.A. | 1 | Yang et al., 2010 |
| Belief propagation Bayesian method to solve the Potts-model | + | − | + | automatic | yes/no | − | O($nlog^{\alpha}n$) | 0.95 | 1 | Hastings et al., 2006; Sulc and Zdeborova, 2010 |
| Label propagation with random link removal | + | + | − | automatic | yes/no | + | ~O($n$) | N.A. | 3 solutions (1 correct) | Raghavan et al., 2007; Tibély and Kertész, 2008; Xiaodong et al., 2008; Barber and Clark, 2009; Leung et al., 2009; Gregory, 2009; Liu and Murata, 2009 |
| Affinity propagation methods | + | + | + | automatic | yes/no | − | >O($n^2$) | N.A. | N.A. | Frey and Dueck, 2007; 2008; Leone et al., 2008; Wang et al., 2007; Brusco and Köhn, 2008 |
| Information flow-based methods | + | + | − | automatic | refined | + | O($n^2logn$) | N.A. | N.A. | Cho et al., 2006; 2007; Hwang et al., 2006; 2008 |
| Signal propagation with F-statistics and fuzzy C-means clustering | + | + | − | automatic | yes/no | − | >O($n^2$) | 0.8 | 0 | Hu et al., 2008a |
| Neuronal activation propagation times with principal component analysis | + | − | + | automatic | yes/no | − | N.A. | N.A. | N.A. | da Fontoura Costa, 2008a; 2008b |

## Table S2. Comparison of network module determination methods (continuation)

| Name of method | Complete Data-set[a] | Weighted graph[b] | Directed graph[b] | Number of modules[c] | Assignment to modules[d] | Overlaps[e] | Polynomial complexity (speed)[f] | Test-modules[g] | Zachary-network[h] | References |
|---|---|---|---|---|---|---|---|---|---|---|
| **Network walk-based agglomerative methods[i] (continuation)** | | | | | | | | | | |
| All shortest paths with principal component analysis | + | – | – | automatic | yes/no | – | N.A. | N.A. | 0 | da Fontoura Costa and Rodrigues, 2008a; Zhang et al., 2008 |
| Message percolation method | + | + | – | automatic | refined | + | N.A. | 0.15 | N.A. | Meng Muntz and Rezaei, 2006 |
| Structure-connected clusters (common neighbors) | + | – | – | automatic | yes/no | yes/no | $O(n^2)$ | N.A. | N.A. | Mete et al., 2008 |
| ModuLand method family | + | + | + | automatic | refined | + | implementation-dependent, ≤O(*n3*) possible | N.A. | 3 modules with overlaps | Kovács et al., 2006; current paper |
| **Methods (both agglomerative and divisive) based on modularity (Q) optimization[j]** | | | | | | | | | | |
| Hierarchical agglomeration with (greedy) optimization | + | – | – | parameter dependent | yes/no | – | $\sim O(n^2)$, $O(n log^2 n)$ | 0.33 | 0,1 | Newman, 2004a; Clauset et al., 2004; Gustafsson et al., 2006 |
| Hierarchical agglomeration with a random walk | + | – | – | parameter dependent | yes/no | – | $\sim O(n)$ | N.A. | 1 | Pujol et al., 2006 |
| Multi-step greedy algorithm with vertex mover refinement | + | + | – | parameter dependent | yes/no | – | $\geq O(n log n)$ | N.A. | N.A. | Schuetz and Caflish, 2008; Noack and Rotta, 2009; Sun et al., 2009 |
| Extremal division optimization | + | + | + | parameter dependent | yes/no | – | $O(n^2 log n)$ | 0.82 | 5 modules | Duch and Arenas, 2005 |
| Simulated annealing | + | + | + | automatic | yes/no | – | $\geq O(n^3)$ | 0.9 | N.A. | Guimera and Amaral, 2005; Massen and Doye, 2006 |
| Mean field annealing | + | – | – | automatic | yes/no | – | $\geq O(n^2)$ | N.A. | N.A. | Lehmann and Hansen, 2007 |
| Basin hopping | + | – | – | automatic | yes/no | – | N.A. | N.A. | N.A. | Massen and Doye, 2005 |

## Table S2. Comparison of network module determination methods (continuation)

| Name of method | Complete Data-set[a] | Weighted graph[b] | Directed graph[b] | Number of modules[c] | Assignment to modules[d] | Overlaps[e] | Polynomial complexity (speed)[f] | Test-modules[g] | Zachary-network[h] | References |
|---|---|---|---|---|---|---|---|---|---|---|
| **Methods (both agglomerative and divisive) based on modularity (Q) optimization[j] (continuation)** | | | | | | | | | | |
| Convex optimization | + | – | – | automatic | yes/no | – | N.A. | 0.29 | 1 | Hildebrand, 2008 |
| Linear or vector programming relaxation methods | + | – | – | automatic | yes/no | – | $O(n^2)$, $O(n^3)$ | N.A. | 1 (4 modules) | Agarwal and Kempe, 2008 |
| Hierarchical community detection based on affinity matrices | + | – | – | automatic | refined | + | $>O(n^3)$ | N.A. | N.A. | Sales-Pardo et al., 2007 |
| Genetic algorithm method | + | – | – | automatic | yes/no | – | $O(n)$ | N.A. | 0,1 | Tasgin et al., 2007 |
| Mixed integer mathematical programming | + | – | – | automatic | yes/no | – | N.A. (high) | N.A. | N.A. | Xu et al., 2007 |
| Modularity preserving pre-modularization size-reduction | – | + | – | automatic | yes/no | – | speed increase up to a factor of 4.27 | N.A. | 0 | Arenas et al., 2007 |
| Local modularity algorithm | + | – | – | automatic | yes/no | – | N.A. | N.A. | N.A. | Hinne, 2007 |
| Multiscale Q-optimization Using self-loops | + | + | + | automatic | yes/no | – | N.A. | N.A. | 0 | Arenas et al., 2008b |
| Methods using spectral properties of the network | + | + | + | automatic | yes/no | – | $\geq O(n^2)$ | 0.57, 0.72 | 0 | Donetti and Munoz, 2004; 2005; White and Smyth, 2005; Newman, 2006a; 2006b; Barber, 2007; Leicht and Newman, 2008; Richardson et al., 2008; Ruan and Zhang; 2008 |

**Table S2. Comparison of network module determination methods (continuation)**

| Name of method | Complete Data-set[a] | Weighted graph[b] | Directed graph[b] | Number of modules[c] | Assignment to modules[d] | Overlaps[e] | Polynomial complexity (speed)[f] | Test-modules[g] | Zachary-network[h] | References |
|---|---|---|---|---|---|---|---|---|---|---|
| **Methods (both agglomerative and divisive) based on modularity (Q) optimization[j] (continuation)** | | | | | | | | | | |
| Combinatorial approach to Modularity | + | + | + | automatic | yes/no | − | N.A. | N.A. | N.A. | Radicchi et al., 2010 |
| Multilevel partitioning using minimum weight cut of a derived complete graph | + | − | − | automatic | yes/no | − | ≥ $O(nlog^2n)$ | 0.7 | 1 | Djidjev, 2008 |
| **Optimization methods using alternative modularity definitions** | | | | | | | | | | |
| Hierarchical agglomeration for heterogeneous modules | + | − | − | parameter dependent | yes/no | − | $O(nlog^2n)$ | >0.33 | 0 | Danon et al., 2006 |
| Cluster density-based optimization | + | − | − | parameter dependent | yes/no | − | N.A. | N.A. | N.A. | Spirin and Mirny, 2003 |
| Maximal clique-based Q optimization | + | − | − | parameter dependent | refined | + | N.A. | N.A. | 4 modules + overlaps | Shen et al., 2009 |
| Modularity for directed graph and overlaps | + | − | + | automatic | yes/no | + | N.A. | N.A. | 2 overlaps | Nicosia et al., 2009 |
| Modularity for positive and negative links | + | + | + (signed network) | automatic | yes/no | − | N.A. | N.A. | N.A. | Bansal et al., 2004; Gómez et al., 2009; Kaplan and Forrest, 2008; Traag and Bruggeman, 2009 |
| Local modularity optimization | −, + | − | + | automatic | yes/no | + | ~$O(n^2)$ | ~0.5 | N.A. | Clauset, 2005; Muff et al., 2005; Rodrigues et al., 2007 |
| Local modularity optimization and hierarchical agglomeration | + | + | − | automatic | yes/no | − | N.A. (fast) | 0.67 | 4 modules | Blondel et al., 2008; Wallace and Gingras, 2008 |
| Multi-scale modularity with combined resolution parameters | + | + | + | automatic | yes/no | − | N.A. | N.A. | N.A. | Lambiotte, 2010 |
| Multiscale (local → global) modularity refinement | + | + | − | automatic | yes/no | − | N.A. | N.A. | N.A. | Pons, 2006 |

**Table S2. Comparison of network module determination methods (continuation)**

| Name of method | Complete Data-set[a] | Weighted graph[b] | Directed graph[b] | Number of modules[c] | Assignment to modules[d] | Overlaps[e] | Polynomial complexity (speed)[f] | Test-modules[g] | Zachary-network[h] | References |
|---|---|---|---|---|---|---|---|---|---|---|
| **Optimization methods using alternative modularity definitions (continuation)** | | | | | | | | | | |
| Community strength modularity optimization | + | − | − | automatic | yes/no | − | N.A. | N.A. | 3 modules | Medus and Dorso, 2009 |
| Local influence-based modularity optimization | + | + | + | automatic | yes/no | − | N.A. | N.A. | 0 | Ghosh and Lerman, 2008 |
| Motif-based modularity maximizing methods | + | + | + | parameter dependent | yes/no | − | N.A. | N.A. | 0 | Arenas et al., 2008a |
| Triangle-based modularity optimization | + | + | + | automatic | yes/no | − | $O(n^3)$ | N.A. | 4 modules | Serrour et al., 2010 |
| Centrality-based modularity optimization | + | − | − | parameter dependent | yes/no | − | N.A. | N.A. | 4 modules | Ghosh and Lerman, 2009; Lerman and Ghosh, 2009 |
| Statistical distribution-based modularity | + | − | − | parameter dependent | yes/no | − | N.A. | N.A. | N.A. | Pei and Zhang, 2007 |
| Blockmodeling-based modularity | + | − | + | parameter dependent | yes/no | − | N.A. | N.A. | N.A. | Reichardt and White, 2007 |
| Partition-coverage starting modularity generalization | + | − | − | parameter dependent | yes/no | − | N.A. | N.A. | N.A. | Gaertler et al., 2007 |
| Mutual information-based modularity maximizing methods (link to hierarchical clustering) | + | + | + | parameter dependent | yes/no | − | N.A. | N.A. | N.A. | Angelini et al., 2007b; Bickel and Chen, 2009 |
| Facility location theory-based method using strongly local modularity | + | − | − | automatic | yes/no | − | $O(n\log^2 n)$ | N.A. | 0 | Berry et al., 2007 |
| Fuzzy modularity optimization using *c*-means clustering | + | + | − | automatic | refined | + | $\sim O(n)$ | N.A. | 4 modules | Zhang et al., 2007a |

**Table S2. Comparison of network module determination methods (continuation)**

| Name of method | Complete Data-set[a] | Weighted graph[b] | Directed graph[b] | Number of modules[c] | Assignment to modules[d] | Overlaps[e] | Polynomial complexity (speed)[f] | Test-modules[g] | Zachary-network[h] | References |
|---|---|---|---|---|---|---|---|---|---|---|
| **Optimization methods using alternative modularity definitions (continuation)** | | | | | | | | | | |
| Fuzzy c-means modularity optimization based on improved Shared Nearest Neighbor method | + | + | – | Automatic | Refined | + | N.A. | N.A. | 2 modules | Xie et al., 2009 |
| Markov random walk and oscillator-synchronization-based modularity | + | + | + | automatic | yes/no | – | N.A. | N.A. | N.A. | Lambiotte et al., 2008; Li et al., 2008a |
| K-means clusters of node synchronization correlation matrix | + | + | + | automatic | yes/no | – | N.A. | N.A. | 2 modules | Li et al., 2010a; Shen et al., 2010 |
| Synchronization-based evolutionary subnetworks | + | – | – | automatic | yes/no | + | $O(n^2)$ | N.A. | 4 modules | Li et al., 2009 |
| Random walk link partition on weighted line graph (link-to-vertex dual) | + | + | + | automatic | yes/no | + | N.A. | N.A. | 4 modules | Evans and Lambiotte 2009a |
| Time-dependent Q optimization (multislice networks) | + | – | – | parameter dependent | yes/no | – | N.A. | N.A. | 2–4 modules | Mucha et al., 2010 |

## Table S2. Comparison of network module determination methods (continuation)

| Name of method | Complete Data-set[a] | Weighted graph[b] | Directed graph[b] | Number of modules[c] | Assignment to modules[d] | Overlaps[e] | Polynomial complexity (speed)[f] | Test-modules[g] | Zachary-network[h] | References |
|---|---|---|---|---|---|---|---|---|---|---|
| **Optimization methods using alternative modularity definitions (continuation)** | | | | | | | | | | |
| Q optimization on weighted line graph (link-to-vertex dual) | + | + | − | automatic | yes/no | + | N.A. | N.A. | N.A. | Evans and Lambiotte 2009b |
| Community detection based on community extraction criterion | − | + | + | automatic | yes/no | − | N.A. | N.A. | 3 | Zhao et al., 2010 |
| Self-organization-based modularity | + | − | − | automatic | yes/no | − | N.A. | N.A. | N.A. | Ahnert et al., 2009 |
| **Other divisive methods** | | | | | | | | | | |
| Centrality-based methods (betweenness random walk, current-based and information centralities) | + | − | − | parameter dependent | yes/no | − | $O(n^3)$, $O(n^4)$ | 0.3, 0.16 | 0, 1,3 | Girvan and Newman, 2002; Fortunato et al., 2004; Newman, 2004c; Newman and Girvan, 2004; Andrade et al., 2009 |
| Fuzzy betweenness centrality methods | + | − | − | parameter dependent | refined | + | $\leq O(n^3)$ | N.A. | N.A. | Wilkinson and Huberman, 2004; Pinney and Westhead, 2006; Gregory, 2007 |
| Conductance optimization | + | − | − | parameter dependent | yes/no | − | (NP-hard) | N.A. | N.A. | Bollobas, 1998; Cheng and Shen, 2010 |
| Cut-size optimization | + | − | − | parameter dependent | yes/no | − | (NP-hard) | N.A. | N.A. | Wei and Cheng, 1989 |
| Division optimization with branch and bound method | + | + | − | automatic | yes/no | − | N.A. | N.A. | N.A. | Hager et al., 2009 |
| Division optimization based on link-clustering (loops) | + | + | + | parameter dependent | yes/no | − | ≥O(n2) | 0.82 | 5 modules | Radicchi et al., 2004; Vragović and Louis, 2006 |

## Table S2. Comparison of network module determination methods (continuation)

| Name of method | Complete Data-set[a] | Weighted graph[b] | Directed graph[b] | Number of modules[c] | Assignment to modules[d] | Overlaps[e] | Polynomial complexity (speed)[f] | Test-modules[g] | Zachary-network[h] | References |
|---|---|---|---|---|---|---|---|---|---|---|
| **Other divisive methods (continuation)** | | | | | | | | | | |
| Mutual information-based methods | + | – | – | automatic | yes/no | – | >O(n3) | N.A. | 0 | Strehl and Ghosh, 2002; Rosvall and Bergstrom, 2007; 2008; Sun et al., 2007; Zhang et al., 2008; Kraskov and Grassberger, 2009 |
| Bootstrap resampling and significance clustering | + | – | – | automatic | yes/no | – | N.A. | N.A. | N.A. | Rosvall and Bergstrom, 2010 |
| Maximum likelihood method | + | + | + | automatic | refined | + | fast | N.A. | 0,1 | Čopič et al., 2009; Newman and Leicht, 2007; Mitrovic and Tadic, 2008; Mungan and Ramasco, 2010; Ramasco and Mungan, 2008; Vazquez, 2008a; 2008b; Wang and Lai, 2008; Zanghi et al., 2008 |
| Dendrogram maximum likelihood method | + | – | – | automatic | yes/no | – | N.A. | N.A. | N.A. | Clauset et al., 2008 |
| Tree mapping | + | + | – | automatic | yes/no | – | N.A. | N.A. | N.A. | da Fontoura Costa and Rodrigues, 2008b |
| Similarity optimization with simulated annealing | + | + | – | automatic | refined | + | N.A. | N.A. | N.A. | Nepusz et al., 2008 |

**Table S2. Comparison of network module determination methods (continuation)**

| Name of method | Complete Data-set[a] | Weighted graph[b] | Directed graph[b] | Number of modules[c] | Assignment to modules[d] | Overlaps[e] | Polynomial complexity (speed)[f] | Test-modules[g] | Zachary-network[h] | References |
|---|---|---|---|---|---|---|---|---|---|---|
| **Other divisive methods (continuation)** | | | | | | | | | | |
| Hypergraph mutual information | + | – | – | automatic | yes/no | – | N.A. | N.A. | N.A. | Strehl and Ghosh, 2002 |
| Random walk-based heat kernel pagerank and Cheeger inequality local cut | – | – | – | parameter dependent | yes/no | – | ~O($n^{1.5}$) | N.A. | N.A. | Chung, 2007 |
| Random walk-based LinkRank | + | + | + | parameter dependent | yes/no | – | N.A. | N.A. | N.A. | Kim et al., 2010 |
| Clustering (eigenvector) centrality and repetitive matrix bipartition | + | + | + (signed network) | automatic | yes/no | – | ~O($n$) | N.A. | N.A. | Yang and Liu, 2007 |
| Methods using spectral properties of the network or its local communities | –, + | + | + | automatic | yes/no | – | ≥O(n) | N.A. | N.A. | Barnes, 1982; Donath and Hoffman, 1972; Kleinberg, 1997; Gibson et al., 1998; Hartuv and Shamir, 2000; Capocci et al., 2005; Tibély et al., 2006; Heimo et al., 2008a; Sahai et al., 2009 |
| Division optimization based on community sub-matrix eigenvalue maximization | + | – | – | automatic | yes/no | – | N.A. | N.A. | N.A. | Chauhan et al., 2009 |
| Truncated singular value decomposition of the modular contribution matrix | + | – | – | automatic | yes/no | – | N.A. | N.A. | N.A. | Arenas et al., 2010 |

## Table S2. Comparison of network module determination methods (continuation)

| Name of method | Complete Data-set[a] | Weighted graph[b] | Directed graph[b] | Number of modules[c] | Assignment to modules[d] | Overlaps[e] | Polynomial complexity (speed)[f] | Test-modules[g] | Zachary-network[h] | References |
|---|---|---|---|---|---|---|---|---|---|---|
| **Other divisive methods (continuation)** | | | | | | | | | | |
| Matrix factorization (+ semi supervised clustering) | + | – | – | automatic | yes/no | – | N.A. (slow) | 0.53 | 3 | Zhang et al., 2007c; Wang et al., 2008b; Ma et al., 2010 |
| Spectral properties of the complement graph | + | + | + | automatic | yes/no | – | N.A. | N.A. | N.A. | Zarei and Samani, 2009; Zarei et al., 2009 |
| Symmetric community measurement (on graph and its complement) | + | – | – | automatic | yes/no | – | N.A. | N.A. | 3 modules | Wang and Lai, 2009 |
| **Methods for finding overlaps after any given non-overlapping clustering method** | | | | | | | | | | |
| Find overlappings by paralel genetic algorithms | + | + | – | automatic | yes/no | + | N.A. | N.A. | 2 modules with overlaps | Carchiolo et al., 2009 |
| Overlapping modularity measurement | + | – | – | automatic | yes/no | + | N.A. | N.A. | N.A. | Lázár et al., 2010 |

Data-assembly of the Table was closed on July 29th 2010. While we have taken a considerable effort to detect and read a large number of modularization methods, obviously the above list is extremely far from being complete. We would like to deeply apologize to all respectful colleagues, whose methods and significant efforts have been inadvertedly omitted from this list in this voluminous and extremely fast-growing field. Comparison of the methods was helped by the reviews of Newman (2004b), Danon et al. (2005), Fortunato and Castellano (2009), Habib and Paul (2010), Li et al. (2010c) and Fortunato (2010). 35 clustering algorithms have been nicely reviewed and clustered to a network by Jain et al. (2004). Several algorithms were compared by Lancichinetti and Fortunato (2009b), Leskovec et al. (2010) and Tibély et al. (2010). The modularity maximization methods were critically assessed by Good et al. (2010). Where more than one references are given, notes and numbers in the columns may refer to only one or a few of them. In case of multiple values separated with commas, each of them is referring to different reference(s). N.A. = data not available.

[a]At the column “Complete data-set” a “+” sign means that the method used all data from the original data set. The “–” sign denotes that some of the original data were deleted or comprised (like at coarse-graining methods), or the analysis was a fully local method using only a sub-segment of the original network. If both signs are present, some of the methods used all original data, while others not. Note, that many currently available network data are, in fact, only samples of a larger data-set, and in this sense, even the “+” methods use only a partial information. A detailed elucidation of the effects of sampling biases on modularization awaits further analysis (for an initial study see Lusseau et al., 2008).

[b]“Weighted graph” and “Directed graph” notes, if the method uses the additional information of weights or directedness for determining the modules. Some of the methods marked as “–” in these columns were applied to weighted and directed networks, but did not use these properties for the refinement of the modular structure.

[c]“Number of modules” is “parameter dependent”, if the maximal number of modules does not derived “automatic”-ally from the method.

[d]"Assignment of modules" is "refined", if the method gives a continuous scale for all links and nodes as their assignment to various modules (fuzzy clustering/partition), and "yes/no", if only a decisive "yes" or "no" answer (hard/crisp community clustering/partition) is given.

[e]"Overlaps" notes, if the method calculates overlaps between modules. If the method had a "yes/no" assignment, the existence of overlaps means that certain nodes/links were assigned to multiple modules simultaneously with an equal weight.

[f]"Speed" refers to the computational speed (computational complexity) of the method, where symbol $n$ denotes the number of nodes. To simplify the formulas and help comparison, many times it was assumed that $n$ is roughly equal to the number of links, i.e. the method is applied to sparse networks.

[g]Numbers in the "Test modules" column refer to the fraction of correctly identified nodes in the model network proposed by Girvan and Newman (2002) having four communities with 32 links each having 16 number of neighbors in average, half of them being intra-modular and the other half of the nodes pointing towards other modules.

[h]The "Zachary network" column gives the number of misplaced members of Zachary's karate club (Zachary, 1977), where the hidden modular division was later exposed by a real split, which made this network a gold-standard for the assessment of module determination methods (Girvan and Newman, 2002). Values in parentheses refer to the number of overlapping values, which are not real misplacements.

[i]Network walk-based methods also contain a large variety of methods, where the mutual information of the local network environment has been assessed and used for module determination.

[j]The modularity function (Q) has been suggested by Newman (2004a) as a measure of the "statistically surprising fraction of the links in a network fall within the chosen communities" (Leicht and Newman, 2008) meaning that the link-density larger than that of an appropriate model system. The exhaustive optimization of this function is an NP-complete problem (Brandes et al., 2007), which makes this straightforward method computationally untractable with larger size real networks. Therefore, several heuristic, optimum-search strategies have been applied to find the global optimum and to circumvent the traps of local optima using an algorithm with a reasonably low computational complexity.

## Table S3. Comparison of modularization methods

| Name of parameter, tool or method | Description of the parameter, tool or method helping the comparision of modularization methods | References |
|---|---|---|
| **Modularity (Q)** | Difference between the fraction of links within modules and the expected number of such links under an appropriate null model (like the one, which preserves the degree sequence of the original network, but otherwise randomizes link positions producing an ensemble of networks) – shown to be a necessary but not sufficient condition, since large random graphs may also have a division resulting in a high modularity and smaller modules may escape detection. | Newman and Girvan, 2004; Guimera et al., 2004; Reichardt and Bornholdt, 2006a; Fortunato and Barthélemy, 2007; Pei and Zhang, 2007; Good et al., 2010 |
| **Weighted modularity (Q)** | The same as above for weighted networks, where in the null model the node strength and not only the node degree is preserved. | Newman, 2004c |
| **Maximal clique-based modularity-like function ($Q_c$)** | Uses a continuous scale for module assignment and the assumption that a highly coeherent maximal clique is usually found only in one module. | Shen et al., 2009 |
| **Benchmark graphs** | Here a test-graph is examined and the recovery of the pre-set modular structure is tested for divisive methods or – more recently – for overlapping modules. | Zachary, 1977; Newman and Girvan, 2004; Lancichinetti et al., 2008, 2009; Reichardt and Leone, 2008; Lancichinetti and Fortunato, 2009a, 2009b; Sawardecker et al., 2009 |
| **Network Information Bottleneck (M)** | Perceives modularization as a coarse-graining process, which finds an optimum between minimizing the number of relevant modules and preserving the emergent information of the whole network and defines M as the area under the 'information curve', which is the function of the output information (after modularization) as the function of the input information (before modularization) of the network. | Ziv et al., 2005 |
| **Local modularity scores** | Utilizes a network walk and determines a local community around the starting node resulting in a large number of 'stopping criteria', where the local community is said to be complete; defines 'outwardness' of a node as the number of its neighbors outside the community minus the number of neighbors inside normalized by the degree and follows the quality of community-growth as the change in the number of extra-modular links. | Bagrow, 2008 |
| **Pairwise membership probability-based consistency** | The module memberships of network nodes are compared after different modularization methods, and a consistency measure of all comparisions for the whole network is defined. | Kwak et al., 2009 |
| **Clustering stability during a Markov chain random walk** | The autocovariance of the network partition during a Markov chain random walk is characterizing the stability of the clusters. If the clustering shows a high stability during the Markov process, it is good representation of the real network communities. | Delvenne et al., 2010 |
| **Statistical significance** | The statistical significance of communities is calculated using Extreme Statistics, and artificial communities arising as structural fluctuations of random graphs are detected. | Lancichinetti et al., 2010a |

## Table S3. Comparison of modularization methods (continuation)

| Parameter | Description | References |
|---|---|---|
| **Eigenvector stability** | The statistical significance of communities is calculated using the stability of the eigenvectors of the Laplacian matrix. | Hu et al., 2010 |
| **Perturbation resilience** | Altering the position of a few links will not change the modular structure (random graphs have multiple competing modularity maxima, while real networks typically have a single global maximum). | Massen and Doye, 2006; Karrer et al., 2008; Yip and Horvath, 2007; Hu et al., 2009 |
| **Resolution limit** | Optimization of global modularity measures may not detect small modules, if the size distribution of modules is large in the network (this problem is also known ad the "giant-component problem" referring to the fact, that the detection of small modules is often possible by the concurrent coalescence of larger modules to a giant component), these problems also warn that many quality functions of modularization work best only, if results with the same number of modules are compared. | Fortunato, 2007; Fortunato and Barthélemy, 2007; Kumpula et al., 2007; Berry et al., 2009 |
| **Computational time** | This parameter is usually given as the polynomial complexity of the modularization algorithm, where the exponent of the number of nodes and (especially) links of the starting network is critical. | Newman, 2004b; Dannon et al., 2005; Gustaffson et al., 2006; Fortunato and Castellano, 2009; Fortunato, 2010 |

The Supplementary Table lists the parameters, which have been designed so far to help the comparison of modularization methods judging their efficiency, reproducibility and usefulness. Most of the measures however have not been accommodating link weights (positive and negative) and directedness (+ link colors), which are highly significant parameters of module membership assignment in real-world networks.

**Table S4. Comparison of the modular assignment of the cAMP-dependent protein kinase family using four modularization methods on a high-fidelity yeast protein-protein interaction network**

| Methods and modules[b] | Modular assignment of the cAMP-protein kinase family member[a] | | | |
|---|---|---|---|---|
| | **TPK1 catalytic subunit** | **TPK2 catalytic subunit** | **TPK3 catalytic subunit** | **BCY1 regulatory subunit** |
| **ModuLand method** | | | | |
| Nuclear pore | 0[c] | 2 | 8 | 0[c] |
| Proteasome | 12 | 13 | 43 | 17 |
| Protein phosphatase 2A complex | 10 | 14 | 36 | 18 |
| Actomyosin complex | 11 | 13 | 36 | 39 |
| Ribosome small subunit assembly complex | 4 | 3 | 11 | 5 |
| Ribosome large subunit assembly complex | 5 | 6 | 8 | 30 |
| Hsp60 complex | 13 | 5 | 23 | 0[c] |
| CCR4/NOT transripctional control complex | 8 | 10 | 29 | 10 |
| Protein phosphatase 1 complex | 8 | 8 | 16 | 38 |
| ER-Golgi transport complex II | 27 | 8 | 44 | 4 |
| **ModuLand method – neighbors[d]** | | | | |
| Cell cycle regulatory complex | YDR212W (5)<br>YMR319C (26)<br>YFL033C (4) | 0[c] | YGL137W (76)<br>YDR490C (13)<br>YLR216C (25)<br>YPR160W (11)<br>YHR033W (41) | YBL106C (2) |
| Spliceosome | 0[c] | 0[c] | YNL227C (1)<br>YGR240C (2) | 0[c] |
| PCNA-loading DNA replication complex | YDR212W (1) | 0[c] | 0[c] | 0[c] |
| Golgi-vesicular transport complex | 0[c] | 0[c] | 0[c] | YBL106C (4) |
| **InfoMap method** | | | | |
| Signaling complex | + | + | + | + |
| **InfoMap method – neighbors[d]** | | | | |
| Hsp60 signaling complex | YDR212W | | | |
| Signaling complex (membrane traffic + pheromone) | YMR319C | | YDR490C | |
| ER-Golgi transport complex | | | YGL137W | |
| Golgi-vesicular transport complex | | | YML001W<br>YNL093W[d] | |
| Chaperone signaling complex | | | YLR216C | |

**Table S4. Comparison of the modular assignment of the cAMP-dependent protein kinase family using four modularization methods on a high-fidelity yeast protein-protein interaction network (continuation)**

| Methods and modules[b] | Modular assignment of the cAMP-protein kinase family member[a] | | | |
|---|---|---|---|---|
| | TPK1 catalytic subunit | TPK2 catalytic subunit | TPK3 catalytic subunit | BCY1 regulatory subunit |
| **InfoMap method – neighbors[d] (continuation)** | | | | |
| Translational initiation complex | | | YGR240C | |
| Signaling complex (cell cycle + translation) | | | YPR160W | |
| Complex with unknown function | | | YHR033W | |
| Spindle orientation + nuclear migration complex | | | | YBL106C |
| Protein phosphatase 1 complex | | | | YER133W |
| Casein kinase II complex | | | | YIL035C |
| **Louvain method** | | | | |
| Vesicular traffic complex | + | + | + | + |
| **Louvain method – neighbors[d]** | | | | |
| Hsp60 signaling + peroxisome complex | YDR212W | | | |
| Cell cycle regulation complex | YFL033C | | YGR240C | |
| ER-Golgi transport complex + RNA splicing | | | YGL137W | |
| Signaling complex | YMR319C | | YDR490C YLR216C[d] | |
| Transcription + replication complex | | | YPR160W | |
| Actomyosin complex | | | YHR033W | |
| Protein phosphatase complex + RNA maturation complex + ubiqutin/DNA-repair complex | | | | YER133W |
| Ribosome assembly + translation initiation complex | | | | YIL035C |
| **CFinder method** | | | | |
| cAMP kinase complex | + | + | + | + |

**Table S4. Comparison of the modular assignment of the cAMP-dependent protein kinase family using four modularization methods on a high-fidelity yeast protein-protein interaction network (continuation)**

| Methods and modules[b] | Modular assignment of the cAMP-protein kinase family member[a] | | | |
|---|---|---|---|---|
| | TPK1 catalytic subunit | TPK2 catalytic subunit | TPK3 catalytic subunit | BCY1 regulatory subunit |
| **CFinder method – neighbors[d]** | | | | |
| ER-Golgi transport complex | | | YGL137W | |
| ER-Golgi transport (late Golgi) complex | | | YNL093W | |
| Hsp90 complex | | | YLR216C | |
| Nucleosome | | | | YIL035C[e] |
| Ribosome large subunit assembly complex | | | | YIL035C[e] |
| Casein kinase II complex | | | | YIL035C[e] |

[a]Modularization of the high fidelity yeast protein-protein interaction network of Ekman et al. (2006) has been achieved using the NodeLand influence function determination method together with the ProportionalHill module membership assignment method with merging highly correlated modules using an arbitrary chosen correlation threshold of 0.9 (see Section VI.1.) as described in the present paper, and by freely available softwares of the InfoMap (Rosvall and Bergstom, 2008; http://www.tp.umu.se/~rosvall/code.html), the Louvain (Blondel et al., 2008; http://sites.google.com/site/findcommunities) and the CFinder (Adamcsek et al., 2006; Palla et al., 2005; http://CFinder.org) methods using k=4 cliques at the latter. In the ModuLand method the modular assignment values of the Table show the fraction of the total community landscape height of cAMP kinase catalytic (TPK1, TPK2, TPK3) or regulatory (BCY1) subunits proportional to the assignment-strength of the given node to the respective module. Instead of processing the modular assignment of the listed nodes into all modules, for each node $k$ only the $n_k$ most significant module assignments having the highest assignment values were considered, where $n_k$ is the effective number of the module assignments of the node $n$ as defined in Section V.6.bthe yes-no answers of the other methods are marked by a '+' sign in case the respective cAMP family member (or by the names of its neighbors) belonging indeed to the module. At the neighbors only their participation in novel modules was included to the Table.

[b]The names reflecting the possible functions of modules have been determined by comparing the functions of the proteins in the respective module and forming a consensus. Such a consensus was easy to achieve with the modules of the ModuLand method, since the community landscape height of the proteins automatically gives their 'centrality', i.e. importance in the module. Therefore, with the modules of the ModuLand method it was enough to determine the function of the module-core, i.e. compare the functions of the most central proteins. In all cases the functions of other, less central proteins confirmed the assignment of the module core of the ModuLand method (data not shown). In case of the other 3 methods, however, functions of all members had to be compared, which was a rather tedious job, especially with the large modules of the Louvain method often having more than a hundred members.

[c] Zero-s denote cases, where the assignment strength-related rank of the given module for node $k$ was lower than the threshold of the effective number of the module assignments, $n_k$ as described in more detail in note "a".

[d]Names of different neighbors of the same cAMP-family member are shown, The number in parentheses denotes the modular assignment strength of the respective neighbor in case of the ModuLand method.

[e]All these 3 complexes are overlapping complexes of the same neighbor, casein kinase 2.

**Table S5. Alignment of the modules of the cAMP-dependent protein kinase family obtained by four modularization methods on a high-fidelity yeast protein-protein interaction network**

| Aligned modules of the cAMP-kinase family by various modularization methods[a] | | | | References showing experimental evidence for module-related biological function[b] |
|---|---|---|---|---|
| ModuLand | Infomap | Louvain | CFinder | |
| Nuclear pore (956) | | | | Bharucha et al., 2008 |
| Proteasome (1366) | | | | Carlucci et al., 2008 |
| Protein phosphatase 2A complex (126) | | | | Müller et al., 2000 |
| Actomyosin complex (973) | | Actomyosin-complex (189) | | Leadsham et al., 2009; Legesse-Miller et al., 2006 |
| Ribosome small subunit assembly complex (618) + ribosome large subunit assembly complex (725) | | Ribosome assembly + translation initiation complex (174) | Ribosome large subunit assembly complex (29) | For translation initiation: Ashe et al., 2000 |
| Hsp60 complex (361) | Hsp60 signaling complex (25) | Hsp60 signaling + peroxisome complex (88) | | Not found |
| CCR4/NOT transripctional control complex (343) | | Transcription + replication complex (117) | | Lenssen et al., 2002 |
| Protein phosphatase 1 complex (94) | Protein phosphatase 1 complex (16) | Protein phosphatase complex + RNA maturation complex + ubiqutin/DNA-repair complex (70) | | De Wever et al., 2005 |
| ER-Golgi transport complex II (85) + Golgi-vesicular transport complex (45) | Golgi-vesicular transport complex (26) | Vesicular traffic complex (91) | ER-Golgi transport (late Golgi) complex (6) | Gelperin et al., 1995 |
| Cell cycle regulatory complex (901) | | | | Songyang et al., 1994 |
| Spliceosome (326) | | | | Not found |
| PCNA-loading DNA replication complex (39) | | | | Gupte et al., 2005 |

**Table S5. Alignment of the modules of the cAMP-dependent protein kinase family obtained by four modularization methods on a high-fidelity yeast protein-protein interaction network (continuation)**

| Aligned modules of the cAMP-kinase family by various modularization methods[a] | | | | References showing experimental evidence for module-related biological function[b] |
|---|---|---|---|---|
| ModuLand | Infomap | Louvain | CFinder | |
| | Signaling complex (18)[c] | | cAMP kinase complex (5) | Not found |
| | ER-Golgi transport complex (26) | ER-Golgi transport complex + RNA splicing (110) | ER-Golgi transport complex (12) | Gelperin et al., 1995 |
| | Signaling complex (membrane traffic + pheromone) (11) + chaperone signaling complex (15) | Signaling complex (158) | Hsp90 complex (5) | Alaamery and Hoffman, 2008 |
| | Translational initiation complex (14) | | | Ashe et al., 2000 |
| | Signaling complex (cell cycle + translation) (10) | | | Ashe et al., 2000; Songyang et al., 1994 |
| | Complex with unknown function (10)[c] | | | Not found |
| | Spindle orientation + nuclear migration complex (7) | | | Not found |
| | Casein kinase II complex (14) | | Casein kinase II complex (6) | Not found |
| | | Cell cycle regulation complex (79) | | Songyang et al., 1994 |
| | | | Nucleosome (8) | Not found |

[a]Modularization of the high fidelity yeast protein-protein interaction network of Ekman et al. (2006) has been achieved using the NodeLand influence function determination method together with the ProportionalHill module membership assignment method with merging highly correlated modules using an arbitrary chosen correlation threshold of 0.9 (see Section VI.1.) as described in the present paper, and by freely available softwares of the InfoMap (Rosvall and Bergstom, 2008; http://www.tp.umu.se/~rosvall/code.html), the Louvain (Blondel et al., 2008; http://sites.google.com/site/findcommunities) and the CFinder (Adamcsek et al., 2006; Palla et al., 2005; http://CFinder.org) methods using k=4 cliques at the latter. Names reflecting the possible functions of modules have been determined as described in the legend of Suppl. Table S4. Alignment of modules belonging to the cAMP family members has been performed by comparison of their members. Two modules were considered 'aligned' and were written in the same line, if the majority of

the members of the smaller module could be found in the larger module as well. (The size of modules is given as a number of their members in parentheses. In case of the ModuLand method this size reflects the effective number of members – for the definition of the term, 'effective number' see Section V.6.b. –, since otherwise all 2,444 members of the giant component of the protein-protein interaction network should have been listed in case of all ModuLand modules.)

[b]References for experimental evidence of the modular assignment were sought in the >19 million records of www.pubmed.com using the search term <cAMP kinase AND (yeast OR cerevisiae)> AND the description of the respective function.

[c]In case of "signaling" or "unknown function", the search for experimental evidence of the identified novel cAMP kinase family-related function was obvious or impossible, respectively. In these cases we searched www.pubmed.com for the interacting partners listed.

# Supplementary Discussion

The network approach, i.e. the description of complex systems as an assembly of their connected nodes, became an increasingly useful method to describe, understand and visualize the enormous data sets of current science and everyday life from cells to society (Boccaletti et al., 2006; Barabasi and Oltvai, 2004; Csermely, 2006; Newman, 2003; Strogatz, 2001; Watts, 1999). The importance of network modules has been recognized rather early (Homans, 1950; Rice, 1927; Simon, 1962; Weiss and Jacobson, 1955), and their determination drew more and more interest, since modules are key players to understand the functional organization and evolution of networks (Danon et al., 2005; Fortunato, 2010; Fortunato and Castellano, 2009; Hartwell et al., 1999; Newman, 2004b; Porter et al., 2009). Modular overlaps and their key function in social organization have been described by Georg Simmel a long ago (Simmel, 1922) and gained recently a growing attention and interest (Palla et al., 2005).

The ideal network modularization method delivers both meaningful partitions and community overlaps. Additionally, an ideal method describes both the connected hierarchy and the disjoint embeddedness of the autonomous sub-networks. Finally, an ideal method is both fast and accurate at the same time. The simultaneous optimization of the above requirement-pairs is impossible, therefore a novel method can not be 'better' than the previous methods, it may only be different in the ratio how it fulfills one or another ascpect of the above, diverse requirements. There are several modularization methods, which explore the local topology of hubs (da Fontoura Costa, 2004), extend the communities by label propagation (Raghavan et al., 2007; Leung et al., 2009), employ a highly efficient optimization of the Potts-model (Ronhovde and Nussinov, 2009), or a gradual hierarchical agglomeration process (Pujol et al., 2006), and thus achieve a computational time up to nearly linear with the number of network nodes. These methods are particularly suitable to assess the modular structure of extremely large networks. However, they use only a subset of the available structural information and, therefore, in many cases have only a limited accuracy. A large number of methods exist, which provide a rather accurate representation of network communities (Table S2). However, most of these accurate methods are computationally expensive. Similar statements apply for handling the modular overlaps and network hierarchy. The ModuLand method family includes both accurate (but somewhat slower) and fast (but relatively inaccurate) methods, as well as methods handling variable extent of modular overlaps at multiple hierarchical levels. Thus the ModuLand method family is not 'better' than the previous methods, but – as a significant and important novelty – provides the experimenter a general framework to respond to the otherwise intractably extreme requirements. This general framework allows an easy shift of emphasis from one optimization parameter to another thus finding an optimal method for the analysis of the given network or data structure. Moreover, the general framework of the ModuLand family also gives a unified background and tool for the easy comparision and evaluation of various modularization attempts, techniques and ideas.

**Comparison of the ModuLand method family with existing methods**

As we have summarized in the main text, the ModuLand method family is novel, since (a) it includes an unparalleled variety of integrated influence function-determination methods; and (b) uses the hills of community landscapes as a basis of module determination.

(a) The various influence function calculation algorithms of the ModuLand family resemble to many local methods described in Table S2. One of the closest methods is that of Bagrow and Bollt (2005), who not only define local communities by the spreading of *l*-shells from the nodes of the network, but also include a process aiming to achieve a community-assignment 'consensus' of the *l*-shells including the given node.

(b) Previous network landscape methods used clustering coefficients (Eckmann and Moses, 2002), link number per visualized network unit area (Ramani et al., 2005), loop-coefficients (Vragovic and Louis, 2006) or degrees (Axelsen et al., 2006) to define the landscape-height. These definitions utilize mainly local nodes of topology, while the ModuLand method assesses a wide range of structural information. Moreover, none of the previous authors used their landscapes for module determination. Recently a number of publications showed a 'hidden metric space' behind network topologies, which links network structure to a landscape-type representation (Krioukov et al., 2008; Narayan et al., 2009). The 'consensus'-building approach of Bagrow and Bollt (2005) resembles to the construction of the community landscape of our method. The recent work of Roswall and Bergstrom (2008) published during the course of the current study (Kovacs et al., 2006) uses the probability flow of random walks to construct a map of scientific communication. This method is similar to our PerturLand influence function calculation algorithm, but its application yields non-overlapping modules. Moreover, none of the methods mentioned above and listed in Table S2 use the hills of a community landscape-type network representation to determine the modular structure. The hill-finding approach, which is the second phase of the ModuLand methods, gives an additional layer of flexibility where the relatively inaccurate results of simpler 'consensus'-building algorithms and the large computational costs of accurate optimization processes can be tailored to the network and to the experimenter's needs and possibilities.

In the following, we compare the ModuLand algorithm with a few important, selected existing method in more detail.

The „*leading eigenvector method*" of Newman (2006b) is able to divide the network in two non-overlapping communities maximizing the modularity measure Q, dividing nodes based on the sign of the respective components of the leading eigenvector of the appropriately crafted modularity matrix. It is possible to divide the network into multiple modules by applying the above division recursively. The magnitude of components of the leading eigenvector can serve as a kind of measure of the respective nodes being central in their own community. However, we do not know about practical applications of this centrality measure either for the bi-partitioning or the recursive division case. The "*vector partitioning algorithm*" of Newman (2006b) assigns a so-called community vector to each node of the network by taking into account multiple leading eigenvectors of the modularity matrix. Non-overlapping communities of the network maximizing modularity Q can then be sought by solving the classical vector partitioning problem. Both the ModuLand framework and modularity-based methods let their users adapt to the specificities of the analyzed network, for the ModuLand framework *via* the choice of its sub-steps (like the community landscape construction method), while modularity-based methods have a null-model to be chosen to reflect our expectations about the network. The paper of Newman (2006b) also introduces the "*community centrality*" measure quantifying the contribution of each node to the modularity Q. As any centrality measure, this community centrality may also be used to form a ModuLand community landscape, therefore making it possible to include the modularity-based method of Newman (2006b) into the ModuLand framework.

The paper of Evans et al. (2009b) shows that meaningful modules can be found in networks by finding modules of links instead of nodes, so that nodes can trivially belong to multiple modules, if their links do. However, the above method is without the fine information about the membership strength of the nodes to different modules as can be uncovered with the ModuLand framework. The link module approach of Evans et al. (2009b) is similar to that of the ModuLand framework, which by default constructs the community landscape over links of the network and calculates node assignment to modules based on link assignment.

The method described in Lancichinetti et al. (2009) is similar to the ModuLand framework in that it is able to find overlapping modules and determine a module hierarchy. For determining the modules instead of executing the local module finding procedure for each node of the network, it is only executed for nodes not contained in any local module yet. Therefore, this local module finding step can be inserted into the ModuLand framework as an influence determination method. Note that executing the influence determination method only for a fraction of nodes is a possible valid approximation method inside the ModuLand framework too. Also, the method of Lancichinetti et al. (2009) does not yield fine information about the membership strength of the nodes to different modules as the ModuLand framework does, but yield binary containment information instead. Lancichinetti et al. (2009) define the term 'cover' as a set of clusters, where each node is assigned to at least one cluster. By tuning the resolution parameter of the local module finding method modules of different size can be obtained, constituting the various hierarchies of network covers. The authors validate the meaningful covers by inspecting the stability of their modules regarding to the variation of the resolution parameter. Covers stable over a wider range of the parameter are said to be meaningful, and these covers are candidates for forming a series of hierarchical levels. However, the constraint that a modules of a lower hierarchical level must entirely be included in a module of a higher hierarchical level is furthermore imposed by the authors. The ModuLand framework has a different approach for determining hierarchical levels, so validation of meaningful resolution parameters is not required. In fact, the resolution parameter for constructing the original influence zones (and thus the community landscape) may be set to yield small modules, and after that the hierarchical levels are automatically determined. Moreover, modules of the higher hierarchical levels of the ModuLand framework do not enforce a strict relation with modules of the lower levels, but the higher level modules can also be continuously overlapping over the lower level modules. However, it would be interesting to investigate the relation between the meaningful covers identified by the method of Lancichinetti et al. (2009) and the hierarchical levels uncovered by its implementation in the ModuLand framework.

**Scale-free distribution of module parameters**

The community size-distribution, the community degree distribution, the community overlap-size and the node membership number distribution of the University of South Florida word association network (Nelson et al., 1998) are shown on Figure S8. We have used cumulative distributions, since they were shown more accurate, than frequency-distributions (Tanaka et al., 2005). The distributions recover the formerly observed, highly heterogenous patterns (Arenas et al., 2004; Clauset et al., 2004; González et al., 2006; 2007; Guimera et al., 2003; Lancichinetti et al., 2009; Palla et al., 2005; Pollner et al., 2006; Radicchi et al., 2004) but significantly deviate from the expected linearity on the log-log scale at both low and high values. This apparent discrepancy has two major reasons.

1.) While many of the previous distribution patterns of module parameters were similar to a scale-free distribution, quite often rather significant deviations from linearity on the log-log scale could also be observed. For example a non-linear log-log distribution pattern was quite obvious at the community degree ditribution of the South Florida work association network (Palla et al., 2005), at the community size distribution of the amazon.com network (Clauset et al., 2004), the Add-Health school firendship network (González et al., 2006) and biological networks (Lancichinetti et al., 2010b) as well as at the node membership number distribution of the Add-Health school friendship network (González et al., 2007).
2.) The properties of the ModuLand method may also influence a scale-free distribution of modular parameters to the fashion observed on Figure S8.
    - Those ModuLand method versions, which apply a a local maxima-based hill determination do not require any previous knowledge about the possible number of network modules. However, due to the fact that these methods only investigate the community landscape without further information about the underlying influence function and influence zone structure, these methods are almost totally blind regarding the size of the underlying influence zones. In other words, the local maxima-based hill determination methods can not discriminate between a community landscape built up from many small influence zones (contaning only a few nodes) and a community landscape built up from a few large influence zones. Even if the original influence zones are small, the modules can still be of arbitrary size almost independently from the size of the influence zones, thus significantly lowering the number of small modules.
    - Noise-like errors (coming either from the inaccuracy of the data or from the approximative nature of the influence function calculation algorithms) may cause the appearance of small local maxima on the community landscape. Thus, by the introduction of 'artificial' local modules, even a very low level of noise may significantly fragment large modules. The effect of additional small modules will become rather noticeable in the case of the TotalHill module membership assignment method, where the size of modules assigned to the noise-induced small local maxima may become larger than that of the ProportionalHill or GradientHill module membership assignment methods. The module merging method described in Section VI.1. addresses this problem, but still residual discrepancies may remain.

    The above two properties shift both the small and large modules towards intermediate size entities causing the curvilinear pattern of Figure S8a.
    - The deviations of the effective module degree distribution on Figure S8b from linearity may be explained partly by the reasons listed before. We would like to add here that the preferential attachment model of modules (Pollner et al., 2006) takes into account a direct interaction of modules and constituent network nodes. However, the influence function calculation step of the ModuLand method is based on the effective, indirect impact of the starting node to the rest of the network (see Section III.1.), which again diminishes the number of both low-degree and extremely high-degree entities contributing to the curvilinear pattern of Figure S8b.
    - The above considerations also influence the overlap size and membership number distributions (Figures S8c and S8d).

    At higher hierarchical levels deviations from the ideal scale-free distribution of modular properties largely come from the definition of link weights at these levels reflecting the overlaps of primary modules (see Section VII.1.).

**Robustness of the results obtained by the ModuLand method**

In our work we applied the benchmark graph generation method published by Lancichinetti et al. (2008), which produces non-overlapping modules, in order to check the correspondence of our highly overlapping modules with the surely known partitions of the benchmarks. The comparison of our results with the recently published updated benchmark graph generation method by Lancichinetti and Fortunato (2009), which also supports overlapping modules, is an interesting and challenging problem of its own, and would require a study which is beyond the scope of our current paper.

In Figures S13a and S13b we show that the identified modules correspond consistently to the modules of the benchmark graph of Lancichinetti et al. (2008) over a range of parameter settings, where modules can be defined in the strong sense. Strong sense means here, at least the half of the neighboring nodes are assigned to the same module as the given node, see Lancichinetti et al. (2008).

**Overlapping word-association modules**

The inter-modular, central position of multiple meaning words (heteronyms, antagonyms and homophones) shown by the ModuLand analysis of the University of South Florida word association network (Nelson et al., 1998) gives a further support of the earlier findings of Sigman and Cecchi (2002) showing that word ambiguity greatly improves the small-world character of the Word-net. The multiple meaning words of 'bright' and 'focus' shown on Figure S9 are homophones, where the multiple meanings are neither extremely disjoint (as would be the case for heteronym words with a different pronunciation) nor opposing (as would be the case for antagonym words). As expected, the analyzed homophone words are located in densely interconnected modules. Interestingly, our representation of the associative neighborhood of 'bright' does not show words referring to the intellect, which has been successfully picked up by the k-clique method of Palla et al. (2005). This shows the importance of the similarity threshold, which we set to 13% making the image clearer, by cutting all words having less than 13% modular similarity than the modular content of the examined word, 'bright'. Words related to the intellect were less than this threshold in this case.

Interestingly, words with the highest community landscape height representing centrally important nodes of American thinking based on word-associations (as determined by the NodeLand algorithm yielding almost a thousand modules total) are the following in decreasing order of their importance: money > cold > car > water > food > tree > book > church > dinner. The list represents well the major human needs and circumstances (water, food, dinner, cold, tree), values of the American society (book, church) and major icons of the American lifestyle (money, car).

**Comparison of the hierarchical network representation of the ModuLand method family with that of other methods**

The complexity of most networks exceeds the cognitive limits of the 'logical', left hemisphere of the brain, which is able to handle a handful of separate pieces of information only. This necessity of data-reduction together with the early recognition of the embedded structure of networks, where nodes of complex networks are networks themselves (such as the cells of our brain can be represented as networks of proteins, etc.) made the hierarchical representation of networks a centerpiece of network studies. A number of hierarchy-representations came from the tree-structures of clustering schemes, or utilized the development of various renormalization-type processes, where the information content of network topology was gradually reduced utilizing the 'fractal-like' behavior of many real networks (Guimerá et al., 2003; Hartwell et al., 1999; Ravasz et al., 2002; Song et al., 2005; Wiuf et al., 2006). Recently several powerful methods have been published to extract network hierarchy (Clauset et al., 2008; Pan and Sinha, 2009; Sales-Pardo et al., 2007) and/or to detect the overlapping and hierarchical structure of network modules (Ahn et al., 2010; Lancichinetti et al., 2009; Shen et al., 2008). The exposed hierarchy can be used for the simplified visualization of large networks (Walshaw, 2003) demonstrated by the example of GenPro (Vlasblom et al., 2006), a visualization subprogram of Cytoscape (Shannon et al., 2003).

The ModuLand method treats primary network modules as nodes of the next hierarchical layer of the network, thus gives a hierarchical representation of the network community structure (Section VII.). The hierarchy of the network science collaboration network (Newman, 2006b) is shown on Figure S6, while that of Community-44 of the Add-Health school friendship network (Moody, 2001; González et al., 2007) is presented on Figure S10 and (partially) on Figure S12. The hierarchical network representation can be used for a fast visualization of extremely large networks, where conventional network visualization methods became too slow to apply (see Section VIII.). The hierarchical representation of modules can also be a highly efficient method for efficient packet delivery (Danon et al., 2008; Palotai, 2008; Palotai and Csermely, 2009).

**Modular hierarchy of a school friendship network**

The hierarchical module structure of Community-44 of the Add-Health database (Moody, 2001; González et al., 2007) obtained by various methods of the ModuLand method family is shown on Figs. S10-S12. Most modularization methods obtain several or all of the 4 major modules at higher hierarchical levels. However, the methods also show a highly refined modular structure at lower levels up to 232 modules. Interestingly, if we decreased the $X$ value of the PerturLand algorithm, thus decreased the putative exchange of information between the schoolchildren of the network, their modules have been better separated (cf. panels of Figure S12) – which is in agreement with our expectations.

**Comparison of the community landscape height with other centrality measures**

The vertical scale of the community landscape represents the combination (in its simplest form: sum) of the individual influence functions for the given link or node. Since the influence functions correspond to the indirect impact of their start-sites to all links of the network, their sum gives a centrality-type measure, which is high, if the overall impact of all network segments (nodes, links or groups) is large. However, combining the influence functions is not the only option for constructing the community landscape. The concept of centrality has a long-standing tradition in network science, and any centrality measure defined in the past or to be defined in the future can serve as a basis of a community landscape. The seminal work of Linton C. Freeman (1978/79) clarified three structural measures of topological centrality: degree (the number of direct neighbors), betweenness centrality (the number of shortest paths traversing through the node) and closeness (the sum of shortest paths leading to all other nodes). Later a number of other centrality measures have been defined, which take into account more, or different details of the overall topology of the network. These measures include

- the lobby index, or Hirsch-index, which is the largest integer $k$ of a node $x$ fulfilling the criterion that node $x$ has at least $k$ neighbors with a degree of at least $k$ (Korn et al., 2009),
- the eigenvector centrality, or α-centrality, which is the principal eigenvector of the adjacency matrix related to the combined degree of the node and its neighbors (Bonacich, 1972),
- PageRank, which is the damped random-walk-based prestige-measure of Google related to the principal eigenvector of the transition matrix describing the damped random walk (Brin and Page, 1998; Perra et al., 2009),
- approximation of PageRank-type measure by local node properties (in- and out-degrees) and by the centrality of the module the nodes maximally belongs (Masuda et al., 2009),
- Katz centrality, which is the total number of paths linking the given node to other nodes in the network exponentially weighted by the length of the path (Katz, 1953), or Bonacich centrality, where the total number of paths is attenuated by two factors α or β for indirect and direct links, respectively (Bonacich, 1987),
- subgraph centrality, which is related to the closed walks starting and ending at the given node (Estrada and Rodríguez-Velázquez, 2005) or centrality measures based on biased random walks (Estrada 2010; Estrada et al., 2008; Lee et al., 2009),
- information centrality, which is the drop of graph performance removing the given node or link (Latora and Marchiori, 2007),
- dynamic centrality, or influence centralitry, which measure the influence of a node based on dynamic network datasets (Klemm et al., 2010; Lerman et al., 2010),
- information-flow score, which is a complex measure of information centrality modelling the network as an electrical circuit (Missiuro et al., 2009) and
- a large variety of centrality-type measures used in the network descriptiptions of ecosytems (for their summary see Jordán and Scheuring, 2004).

Another set of centrality measures can be derived from the community structure of the network. Measures, like the bridgeness, defined differently by Nepusz et al. (2008) and this paper, take into account the inter-modularity of a node. On the contrary, the height of the community landscape may merge the local (degree-, or eigenvector centrality-based), mesoscopic (community-based) and global centrality measures in different ratios, depending on the exact method of the ModuLand method family. The complexity of centrality measures is further substantiated by the findings of Pollner et al. (2008), who demonstrated that the inter-modular position is resulting in high centralities by a number of conventional centrality measures. Similarly, the grossly inter-modular 'creative nodes' defined by Csermely (2008) display an extremely high centrality in a large variety of networks. The community landscape height could combine both the 'traditional', local centrality increments (mostly characteristic to the cores of well-defined network modules) and the 'bridge-related' centrality increments typical to key information channels of modular overlaps. Different realizations of the ModuLand method family give different weight to these two types of centrality-increments, and thus regulate the ‘roughness’ of the community landscape, i.e. the overall extent of modular overlaps. (A rough community landscape has rather well defined hills, where overlaps are minor. On the contrary, a flat community landscape provides large overlaps and less well-defined individual modules.)

**Central nodes of power-grid a network**

We have tested the centrality measures of the ModuLand method by removing the most central nodes of the Western Power Grid network of the USA (Watts and Strogatz, 1998, Figure 4 of the main text) and calculating the efficiency of the network as defined by Latora and Marchiori (2001) after the removal. When removing nodes in the order of their decreasing centrality from the network, we did not re-calculate the centrality values after each removal. While this may sound as a large simplification at the first time, it is worth to consider that in a real case, a rather voluminous dynamics can be observed in the real network after each removal, which rearranges the links of the network rather significantly. Thus, lacking simulations reflecting the possible network dynamics, the starting model itself is already simple enough to accommodate this further simplifying assumption for comparative purposes.

The bridgeness centrality measure – for its definition, see Section V.6.d. – of the ModuLand method defined more important nodes of the power-grid network than either the degree or the betweenness centrality measures (Figure 4).

**Discrimination between date- and party-hubs**

We have analyzed the modular structure of the yeast proteome using the high-confidence protein-protein interaction data of Ekman et al. (2006). Our analysis using either the LinkLand or the NodeLand algorithms (Figure 5A of the main text, Figure S14) uncovered a number of major modules with a well-known biochemical function such as the proteasome cell cycle regulation, ribosome biogenesis, nuclear pore complex, actomyosin, RNA splicing, etc. Importantly, those modules, which were functionally related, such as the proteasome and cell cycle regulation as well as the ribosome biogenesis, nuclear pore complex and RNA splicing showed a large overlap. The functional analysis of the modules showed several, novel interesting features. As an example of this, rather surprisingly, the superoxide dismutase has been recovered as a part of the mitochondrial ribosome (Kovacs et al., 2006; Mihalik et al., 2008). It turned out from a literature search that, in fact, a functional relationship has been uncovered between this enzyme and the mitochondrial ribosome before by Zielinski et al. (2002), which gives an experimental evidence for the functional meaning of the ModuLand-based module membership assignment in a biological network. We have also identified characteristic changes in the yeast proteome after stress (Mihalik et al., 2008; Palotai et al., 2008), which were in agreement with the expectations (Szalay et al., 2007).

The dissection of date- and party-hubs of protein interaction networks, i.e. proteins sequentially or simultaneously interacting with a large number of neighbors, has been proved notoriously difficult (Ekman et al., 2006; Han et al., 2004; Kim et al., 2006; Komurov and White, 2006) and has been intensively debated (Batada et al., 2006; 2007; Bertin et al., 2007; Agarwal et al., 2010). We were curious, if our modularization method may help to discriminate between these different hub classes. In agreement with earlier findings (Han et al., 2004, Yu et al., 2007; Yu et al., 2008) we assumed that date-hubs should have a more inter-modular position than party-hubs. Date-hubs had a larger bridgeness than party-hubs having a similar centrality (Figure 5 of main text). Similarly, date-hubs had a larger modular overlap than party-hubs of the same degree. We have obtained similar results, if we made the modularization using the NodeLand, LinkLand or PerturLand algorithms of the ModuLand method family (Kovács et al., 2006). As we have described in the main text the bridgeness-based dissection of date- and party hubs misclassified only a single date-hub (of 201 total) and 28 party-hubs (of 318 total) of the consensus-based identification of these proteins. The large variability of date- and party-hub identification may come in part by the differences of the initial datasets. As noted by Yu et al. (2008) the two-hybrid methods pick up more inter-modular contacts (date hubs), while the immunprecipitation-based methods enrich intra-modular interactions (party hubs). Our result almost exclusively identifying date-hubs as inter-modular proteins poses this important class as multifunctional, ‘moonlighting’ proteins (Gianchandani et al., 2006), mediators of cross-modular effects (Andreopoulos et al., 2007), non-hub bottlenecks (Yu et al., 2007) and creative proteins (Csermely, 2008) helping to survive unprecedented, novel challenges, and playing a key role in the development and evolvability of complex systems. The modular analysis is an important tool to identify these latter, creative nodes of complex systems (Csermely, 2008; Kovacs et al., 2006).

## Conclusions and perspectives

As a summary, the ModuLand module determination method is a powerful and novel modularization tool, which recovers many results of earlier determinations, uncovers a rich hierarchy of complex networks, provides a set of novel centrality and other measures to characterize network nodes and is able to predict the dynamic behavior of network nodes from their topology. The method has an important potential to assess network dynamics and evolution (Palla et al., 2007b), to predict missing nodes or to identify hidden or mislabeled nodes or links (Clauset et al., 2008), to design efficient packet routing algorithms (Danon et al., 2008; Palotai, 2008; Palotai and Csermely, 2009), or to assess the roughness of various landscapes including fracture surfaces (Haavig Bakke and Hansen, 2007) or energy landscapes. The ModuLand method family may also help us to identify the inter-modular contacts and increasing overlaps of the developing brain (Fair et al., 2008) as well as to predict bankruptcies of financial networks (Fujiwara and Aoyama, 2008) as a few of the myriads of exciting applications.

We invite our colleagues to explore these possibilities and offer the method as a freely downloadable software at our web-site, <www.linkgroup.hu/modules.php>. We are happy to list, link and accommodate the upgrades and versions of this method developed by other groups at this platform. Moreover, we highly welcome any results of direct comparison of the more than hundred different modularization techniques described in Table S2 (and obviously those, which we may have left out for which we deeply apologize) using the unifying platform of the ModuLand method for their conversion and direct comparison as described in Section IV.4.

## Supplementary References